\documentclass[fleqn,usenatbib]{mnras}

\usepackage{newtxtext,newtxmath}
\usepackage{listings}
\usepackage{color}
\usepackage{nicefrac}
\usepackage{subcaption}

\definecolor{dkgreen}{rgb}{0,0.3,0}
\definecolor{gray}{rgb}{0.5,0.5,0.5}
\definecolor{mauve}{rgb}{0.58,0,0.82}
\definecolor{golden}{rgb}{0.86,0.65,0.01}

\usepackage[T1]{fontenc}

\DeclareRobustCommand{\VAN}[3]{#2}
\let\VANthebibliography\thebibliography
\def\thebibliography{\DeclareRobustCommand{\VAN}[3]{##3}\VANthebibliography}

\usepackage{graphicx}	

\usepackage{amsmath}	

\title[Spurious astrometric solutions]{A classifier for spurious astrometric solutions in Gaia eDR3}

\author[Jan Rybizki et al.]{
Jan Rybizki$^{1}$$^{\dagger}$,
Gregory M. Green$^{1}$\thanks{E-mail: green@mpia.de}$^{\dagger}$,
Hans-Walter Rix$^{1}$,
Kareem El-Badry$^{1}$,
\newauthor
Markus Demleitner$^{2}$,
Eleonora Zari$^{1}$,
Andrzej Udalski$^{3}$,
Richard L. Smart$^{4}$,
\newauthor
and Andrew Gould$^{1,5}$
\\
$^{\dagger}$These authors contributed equally\\
$^{1}$Max Planck Institute for Astronomy,
	K\"onigstuhl 17, D-69117 Heidelberg, Germany\\
$^{2}$Astronomisches Rechen-Institut, Zentrum f{\"u}r Astronomie der Universit{\"a}t Heidelberg, M{\"o}nchhofstrasse 12-14, D-69120 Heidelberg, Germany\\
$^{3}$Astronomical Observatory, University of Warsaw, Al. Ujazdowskie 4, 00-478 Warszawa, Poland\\
$^{4}$INAF - Osservatorio Astrofisico di Torino, via Osservatorio 20, 10025 Pino Torinese (TO), Italy\\
$^{5}$Department of Astronomy, Ohio State University, 4055 McPherson Laboratory, 140 West 18th Avenue, Columbus, Ohio 43210, USA
}

\date{Accepted 06-Dec-2021. Received 10-Sep-2021}

\pubyear{2020}

\begin{document}
\label{firstpage}
\pagerange{\pageref{firstpage}--\pageref{lastpage}}
\maketitle

\begin{abstract}
The Gaia early Data Release 3 has delivered exquisite astrometric data for 1.47 billion sources, which is revolutionizing many fields in astronomy. For a small fraction of these sources, the astrometric solutions are poor, and the reported values and uncertainties may not apply. Before any analysis, it is important to recognize and excise these spurious results -- this is commonly done by means of quality flags in the Gaia catalog. Here, we devise a means of separating ``good'' from ``bad'' astrometric solutions that is an order of magnitude cleaner than any single flag: 99.3\% pure and 97.3\% complete, as validated on our test data.
We devise an extensive sample of manifestly bad astrometric solutions, with parallax that is \textit{negative} at $\ge 4.5\sigma$; and a corresponding sample of presumably good solutions, including sources in HEALPix pixels on the sky that do not contain such negative parallaxes, and sources that fall on the main sequence in a color--absolute magnitude diagram. We then train a neural network that uses 17 pertinent Gaia catalog entries and information about nearby sources to discriminate between these two samples, captured in a single ``astrometric fidelity'' parameter. A diverse set of verification tests shows that our approach works very cleanly, including for sources with positive parallaxes. The main limitations of our approach are in the very low-SNR and the crowded regime. Our astrometric fidelities for all of eDR3 can be queried via the Virtual Observatory, our code and data are public.
\end{abstract}

\begin{keywords}
Galaxy: stellar content, Galaxy: kinematics and dynamics, software: public release, space vehicles: instruments, virtual observatory tools
\end{keywords}


\section{Introduction} \label{sec:intro}
Parallax measurements contain information about the distance to astrophysical objects, and are critical to anchoring the cosmic distance ladder. At the same time, kinematic measurements -- proper motions and radial velocities -- provide phase-space information that is key to understanding Milky Way dynamics and external galaxies. The 1.47 billion astrometric measurements reported in \textit{Gaia} Early Data Release 3 \citep[``eDR3'']{EDR3Summary} constitute the largest astrometric dataset ever produced.

While this astrometric catalog is of extremely high quality \citep{2020arXiv201203380L}, a significant fraction of astrometric solutions are spurious \citep{EDR3CatalogueValidation}. Spurious astrometric solutions are a distinct issue from negative parallaxes, which are an expected outcome of the normally distributed parallax measurement \citep{2015PASP..127..994B,2018A&A...616A...9L}. Spurious solutions result from specific failure modes \citep{EDR3CatalogueValidation}, and as an ensemble, have true parallaxes that are incompatible with the reported catalog values (and their uncertainties). This can be a particular concern when looking at sparsely populated portions of the color-magnitude diagram, or at extreme objects, such as the nearest or fastest-moving stars (i.e., those with the largest parallaxes or proper motions, respectively). For example, naively selecting all objects with measured parallaxes greater than 10~mas (corresponding to a distance of less than 100~pc) yields a catalog with an estimated 50\,\% of spurious parallax measurements. \textit{Gaia} eDR3 provides a number of astrometric quality parameters that can be used to exclude such spurious solutions. The ``Gaia Catalogue of Nearby Stars'' \citep[``GCNS'']{EDR3GCNS} uses a combination of these parameters to filter out spurious sources, obtaining a highly complete and pure subset of \textit{Gaia} eDR3 sources lying within 100~pc. In this paper, we use a similar approach to extend this work to the entire \textit{Gaia} eDR3 catalog.

\textit{Gaia} eDR3 provides 1.47 billion astrometric measurements containing of a two-dimensional position on the sky, a two-dimensional proper motion and a parallax (in addition, a 7.2M subset also has radial velocity measurements). There are many possible sources of excess noise in these astrometric measurements. Some error modes, such as unmodeled acceleration caused by an unresolved binary companion typically introduce small residuals into the astrometric solution, which will usually be accounted for in the parallax uncertainty estimate \citep{2020arXiv201203380L}. However, other error modes, such as incorrect epoch cross-matches with background or spurious sources and also close source pairs, which might be partially resolved \citep{EDR3CatalogueValidation}, can introduce very large residuals, scattered around the true parallax \citep{EDR3GCNS}, which are unaccounted for in the reported parallax uncertainty. Spurious astrometric solutions mainly happen in very dense parts of the sky \citep{EDR3CatalogueValidation}. It is this latter class of ``catastrophic'' errors in the astrometric solutions (leading to errors in excess of the stated uncertainties) that we will attempt to detect. We will ignore the fact that the reported parallax uncertainties in eDR3 are slightly underestimated \citep{2021AJ....161..214Z} and can be significantly underestimated for specific classes of sources \citep{El-Badry2021c}.

One can try to mitigate these spurious astrometric solutions with cuts on \texttt{ruwe}, \texttt{visibility\_periods\_used} and apparent $G$ magnitude. These cuts are known to exclude many valid sources and also bias the sky coverage. In the GCNS, the approach was to use several astrometric quality indicators to train a random forest classifier on a good and a bad training sample. Since the GCNS only contains sources with observed parallax of greater than 8~mas, it primarily covers the extremely high parallax signal-to-noise ratio (SNR) regime. For the ``bad'' examples, sources with measured parallax less than -8~mas were used, exploiting the fact that spurious astrometric solutions can be expected to scatter equally above and below the true parallax. Sources with extremely negative measured parallaxes can therefore be taken to be representative of sources with spurious, positive measured parallaxes. For the ``good'' examples, sources in low-density regions of the sky that were cross-matched to 2MASS and that showed consistent absolute magnitudes in Gaia and 2MASS bands with the locus of the main stellar populations (e.g. main-sequence or white-dwarfs) were used.

When trying to classify the spurious parallax solutions for the whole Gaia eDR3 catalog, one also needs to make informed decisions for low parallax SNR, as 85\,\% of sources fall into this regime (for $\left| \mathrm{SNR} \right| < 4.5$). Since low-SNR parallaxes less stringently constrain distance, it is more difficult to establish a genuine difference between valid and spurious astrometric solutions. We aim to mitigate this difficulty by training specialised models for the high- and low-SNR regimes.

In the following, we will attach a single scalar measure of ``astrometric fidelity'', bound between 0 and 1, to all sources in eDR3. This will prove useful when, for example, trying to clean CAMDs (see Section~\ref{sec:camd_slices}). Contamination in CAMDs can also result from spurious colors, which our astrometric fidelity measure is not intended to flag. However, we will show in Section~\ref{sec:cmd_fidelity} how these can be filtered out as well.

\section{Open science approach}

We previously submitted a preprint to \url{arxiv.org}, with a first version (\texttt{v1}) of our astrometric fidelities uploaded to the German Astrophysical Virtual Observatory (GAVO), in order to collect feedback from the community. Now, in the updated version (\texttt{v2}) of this work, we have implemented improvements that we will highlight, together with the general procedure in the following sections. We will also compare \texttt{v1} to \texttt{v2} and look at the improvements.

All of the work in this paper, including the training of the classifier and the various validation tests, can be replicated using a \texttt{Python} notebook and data that we have made available.\footnote{The notebook can be found at \url{https://colab.research.google.com/drive/1lPzhGSSIjx2nQ7XM2v8bQZtkf0Atrk0z?usp=sharing}, while the necessary training and validation data is stored at \url{https://keeper.mpdl.mpg.de/d/21d3582c0df94e19921d/}}. This notebook should have legacy value for the community, as our analysis can be easily repeated for upcoming data releases. Additionally, others can augment our training data with extra sets of good or bad astrometric solutions, in order to improve the classifier or to produce special purpose classifiers for specific stellar populations or regions of the sky with specific properties (e.g. high density areas).

\section{Training data}
\label{sec:training-data}

In order to train a classifier that distinguishes between sources with reliable and unreliable astrometric solutions, we first must construct a training sample of sources that we can identify as falling into each category. We will refer to spurious (valid) astrometric solutions as ``\texttt{bad}'' (``\texttt{good}''). In the following, we describe our method of identifying both \texttt{bad} and \texttt{good} training sets.

In all of the following, we neglect the Gaia parallax zero-point offset \citep{2020arXiv201201742L}, which should be safe for most of the sources, as its magnitude is on the order of 20$\mu$as.

\subsection{Spurious sources}
\label{sec:bad-training-sources}

\subsubsection{Large negative parallaxes}

We obtain the bulk of our \texttt{bad} training sample by selecting sources with $\mathtt{parallax\_over\_error} < -4.5$. We use the following query:
\begin{lstlisting}
SELECT *
FROM gaiaedr3.gaia_source
WHERE parallax_over_error < -4.5
\end{lstlisting}
This returns 4.18 million sources. If all of the 1.47 billion sources in Gaia eDR3 with measured parallaxes had a true parallax of zero, and all of the measurement errors were Gaussian, then we would expect approximately 5000 stars -- nearly three orders of magnitude fewer -- to satisfy the above cut. Since in reality, sources have positive true parallaxes, the discrepancy is even larger. Thus, even with the most pessimistic assumptions, the contamination rate of our \texttt{bad} training sample by sources with good astrometric solutions is $\sim 0.1\%$.

\begin{figure}
	\includegraphics[width=\linewidth]{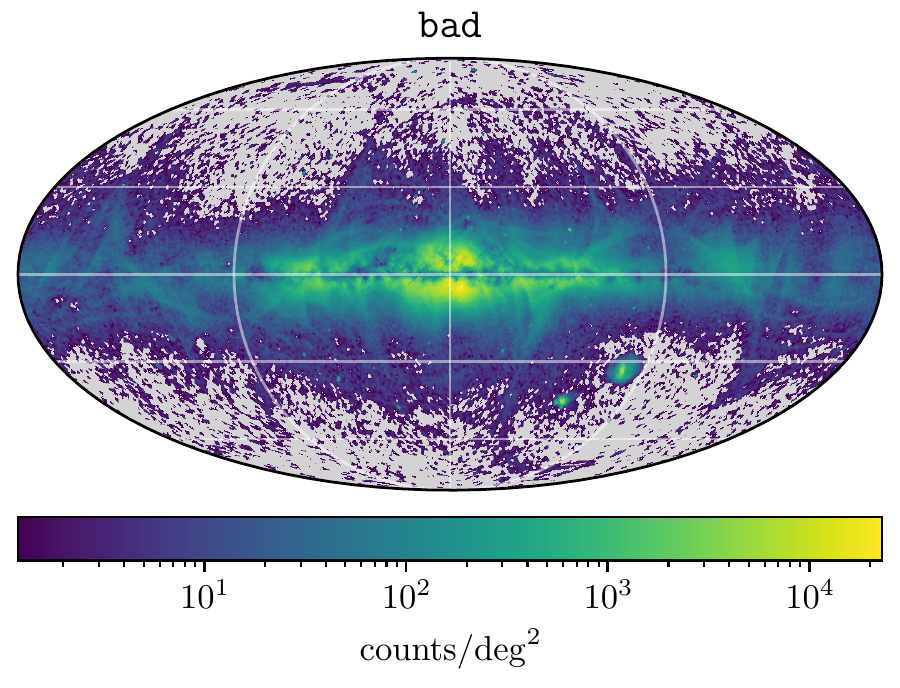}
	\caption{Density distribution of the sources identified as \texttt{bad} for our training sample by their highly negative ($<-4.5\sigma$) parallaxes, shown using a Mollweide projection in Galactic coordinates, with a logarithmic color scale.}
	\label{fig:skydist_bad}
\end{figure}

Fig.~\ref{fig:skydist_bad} shows the distribution of our \texttt{bad} sources over the sky. Crowded areas such as the bulge, disc and the Magellanic clouds stand out, but scanning-law patterns are also visible, with regions of the sky that are scanned most often (notably the two rings along $\mathrm{ecliptic\ latitude} \approx \pm 45^{\circ}$) having higher densities of spurious astrometric solutions. We conjecture that this is due to the many scans along a similar scanning angle, which on the one hand could increase the probability of spurious detections ocurring at the same place and therefore reduce the probability of these sources being filtered out in the downstream process \citep{2020arXiv201206420T} and on the other hand could amplify the disturbance from a close neighbor.

\subsubsection{Iterative identification of spurious parallaxes}
\label{sec:iterative-identification-spurious}

\begin{figure*}
    \includegraphics[width=0.45\textwidth]{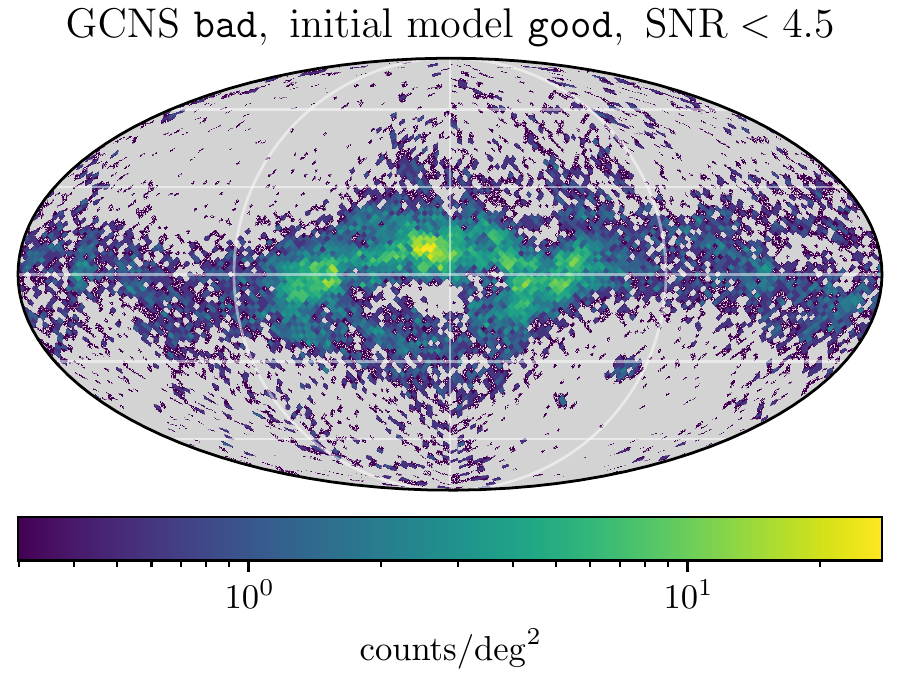}
    \hfill
    \includegraphics[width=0.45\textwidth]{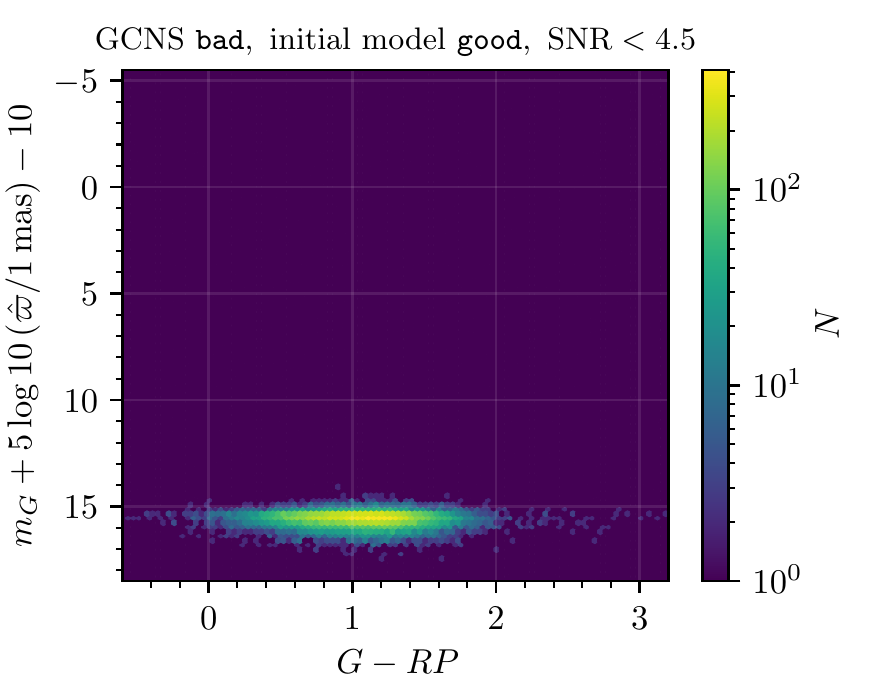}
	\caption{Sky distribution (left panel, using a Mollweide projection of Galactic coordinates) and CAMD (right panel) of the low-SNR GCNS sources ($\hat{\varpi} \ge 8\,\mathrm{mas}$) identified as \texttt{good} by our initial model but as \texttt{bad} by the GCNS, using a logarithmic color scale. These sources are concentrated in the Galactic plane, and even show overdensities in the LMC and SMC. We use these sources to create an additional \texttt{bad} dataset, on which our final model is trained.}
	\label{fig:gcns_bad}
\end{figure*}

Thus far, all of our \texttt{bad} training examples are in the high-SNR regime. We follow an iterative approach to obtain a sample of sources in the low-SNR regime with unreliable parallaxes. We first train our model without any low-SNR \texttt{bad} training examples. We then classify all GCNS sources (i.e., all source with observed parallax greater than 8~mas) and compare to the GCNS classifications. We are particularly interested in the sources that we classify as \texttt{good} and that the GCNS classifies as \texttt{bad}. In Fig.~\ref{fig:gcns_bad}, we show the sky distribution and CAMD of such sources in the low-SNR regime. Only a negligible fraction of such mismatched sources are in the high-SNR regime. In the low-SNR regime, the sky distribution of these sources is highly concentrated in the Galactic disk, and even contains overdensities in the directions of the Magellanic clouds. This indicates that a large fraction of these sources are likely spurious (cf. Fig.~\ref{fig:skydist_bad}).

We therefore add the $\sim$23k low-SNR sources that the GCNS classifies as \texttt{bad} and our initial model classifies as \texttt{good} to our \texttt{bad} training sample, and then retrain the model. Because of the relatively small size of this sample, we assign 10 times the normal weight to these sources during training.

\subsection{Good sources}
\label{sec:good-training-sources}

\begin{figure}
	\includegraphics[width=\linewidth]{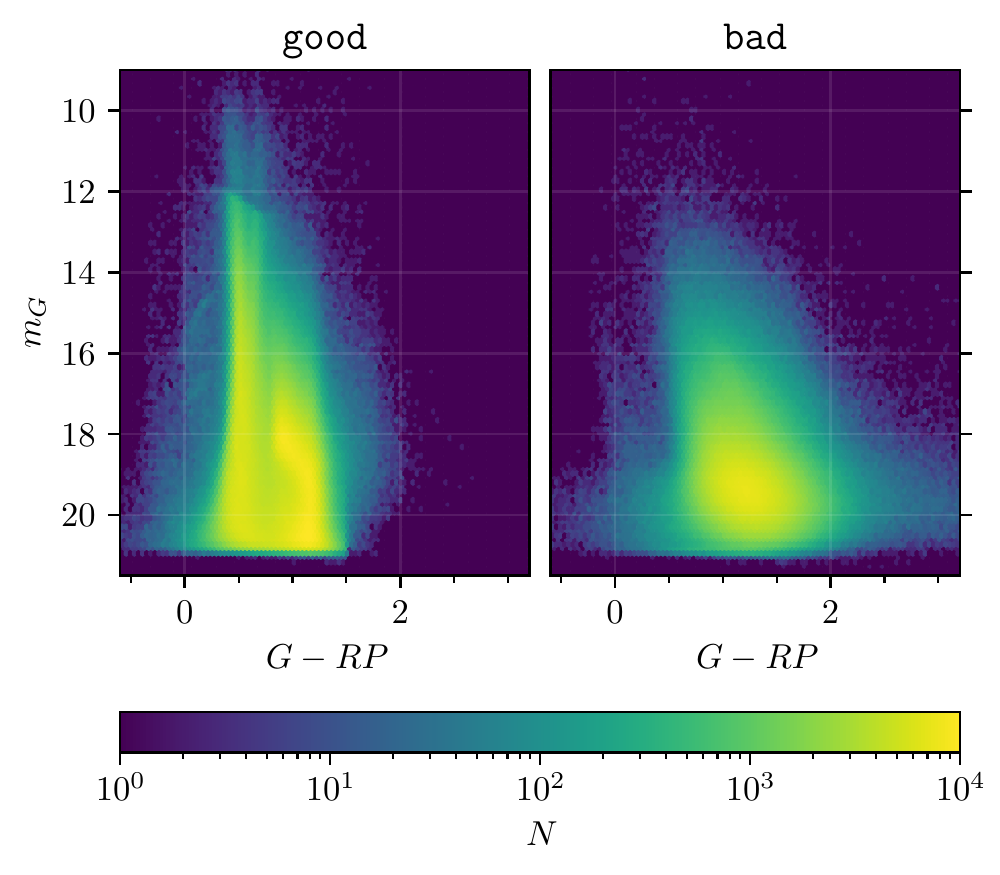}
\caption{Color-magnitude diagram (CMD) distribution of the \texttt{good} (left panel) and \texttt{bad} (right panel) training samples, on a logarithmic color scale.
	The left panel shows all 6.25 million \texttt{good} sources (Section~\ref{sec:good-training-sources}), while the right panel shows all 4.20 million \texttt{bad} sources (Section~\ref{sec:bad-training-sources}). The left panel shows the well-established and expected bimodal structure in the color distribution that arises from the main sequence turn-off and the red giant branch. The right panel shows not a hint of such structure, implying that the colors of these bad sources may  commonly be spurious, too.}
	\label{fig:CMD_training}
\end{figure}

We combine three different datasets in order to construct the \texttt{good} training set. The primary dataset consists of sources that lie in regions of the sky devoid of \texttt{bad} sources (i.e., regions that lack sources with significantly negative parallaxes). The second dataset consists of sources near the Galactic plane that lie on the main sequence in Pan-STARRS~1 \citep[][``PS1'']{2016arXiv161205560C} color-color and color-absolute-magnitude (using Gaia parallaxes) space. The third dataset consists of globular cluster members that have measured parallaxes consistent with the globular cluster parallax. By combining these different datasets, we hope to obtain a sample of \texttt{good} sources from a range of different regimes (e.g., source density and sky position). Additionally, in constructing the \texttt{good} training set, we do not cut on any astrometric flags that we use later as potential features in the classifier -- doing so would render the classifier's job trivial. Similarly, as can be seen in Fig.~\ref{fig:CMD_training}, where the CMDs of the \texttt{good} and the \texttt{bad} training set are shown, both training sets have a large overlap in magnitude and color space\footnote{In \texttt{v1}, when we crossmatched to 2MASS instead of PS1, we had a deficit of blue, faint \texttt{good} training sources. Explicitly below the line connecting $\left(G-RP,m_G\right) = \left(0.5,18\right)$ and $\left(1.2,20\right)$, we had no \texttt{good} training sources. The drop in density for sources brighter than $G = 12\,\mathrm{mag}$ is due to the saturation limit of PS1.}, with an exception for very red colors. These considerations should allow our classifier to more effectively generalize from our training dataset to the entire Gaia eDR3 catalog.

These three datasets are described in detail below.

\subsubsection{Regions without bad sources}

In order to construct our primary \texttt{good} training set, we select all sources with $\mathtt{parallax\_over\_error} > -3$ in regions of the sky that do not contain any sources with significantly negative parallaxes. This means that our primary \texttt{good} training dataset and our \texttt{bad} training dataset come from disjoint regions of the sky. This \texttt{good} sample does not come from the Galactic plane (i.e., $|b| > 19^\circ$ for all \texttt{good} sources). In detail, we separately query each HEALPix \citep{2005ApJ...622..759G} level-6 (i.e., $\mathtt{nside} = 64$) pixel that contains no sources with $\mathtt{parallax\_over\_error} < -3.6$. We use a stricter cut in sparser regions of the sky, omitting HEALpix pixels with a source count below 3500 that contain any sources with $\mathtt{parallax\_over\_error} < -2.7$. In all, we use 1614 out of 49152 HEALPix level-6 pixels on the sky to construct our primary \texttt{good} training set. Our query for a single pixel is as follows:
\begin{lstlisting}
SELECT *
FROM (SELECT dr3.*, ps1.original_ext_source_id as obj_id,
FLOOR(dr3.source_id/140737488355328) as hpx6
FROM gaiaedr3.gaia_source as dr3
JOIN gaiaedr3.panstarrs1_best_neighbour AS ps1
USING (source_id)
WHERE source_id BETWEEN 0 AND 562949953421311) AS subquery
-- Query only first HEALpix of level 6
JOIN gaiadr2.panstarrs1_original_valid AS ps
USING (obj_id)
-- only sources with a crossmatch to PS1 are queried
\end{lstlisting}
We obtain a total of 4.85 million sources. The requirement that the source also be visible in PS1 ensures that we do not include spurious Gaia sources (though the PS1 xmatch does not help in filtering spurious astrometric solutions).

We split our primary \texttt{good} training set into two subsets: a high-SNR subset with $\mathtt{parallax\_over\_error}\ (\mathrm{SNR}) > 4.5$ and a low-SNR sample with $-3.0 < \mathrm{SNR} < 4.5$. In order to further purify our primary \texttt{good} training set, we require $G-RP < 1.8~\mathrm{mag}$ (1.5~mag) for the high-SNR (low-SNR) subset. The cut on $G-RP$ requires $RP$ photometry and removes about 1\,\% of unphysically red sources. This extremely red color sometimes coincides with nearby sources and/or high \texttt{phot\_bp\_rp\_excess\_factor}. After these photometric cuts, our primary \texttt{good} training set contains 4.82 million sources.

Of course, our \texttt{bad} and primary \texttt{good} training sets probe quite different regimes, with the primary \texttt{good} training set coming mostly from low-extinction regions and sparse fields at high Galactic latitudes. We therefore add in two additional \texttt{good} training sets, described in the following subsections, which are tailored to cover crowded regions of the sky. We hope (and later verify) that in the space of astrometric parameters and quality flags, our training sets cover the relevant feature space and will allow our classifier to have discriminative power over the entire sky, as was the case for the GCNS.

\subsubsection{Sources lying on the main sequence}

\begin{figure*}
	\includegraphics[width=0.8\linewidth]{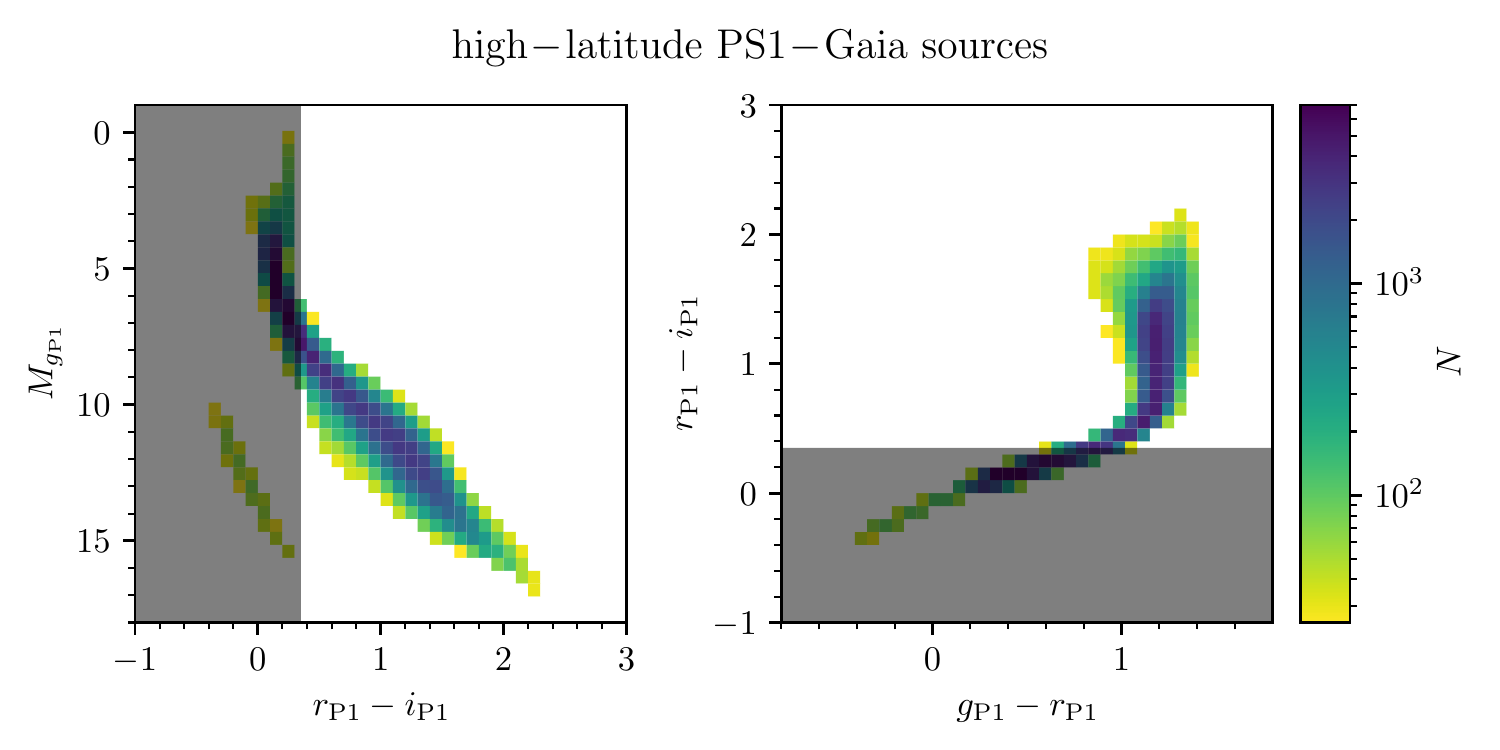}
	\caption{The left panel shows a PS1-Gaia CAMD (at high Galactic latitudes), using PS1 photometry (de-reddened using Bayestar19) and high-SNR Gaia parallaxes. We use stars with $r_{\mathrm{P1}}-i_{\mathrm{P1}} > 0.35$ (the unshaded region) to define a main sequence mask on the CAMD. We additionally construct a PS1 color-color mask (right panel). We then use these CAMD and color-color templates to select stars in the Galactic plane with good Gaia astrometry.
	}
	\label{fig:good_camd}
\end{figure*}

In order to obtain sources with well measured parallaxes nearer to the Galactic plane, we make use of a combination of PS1 photometry and Gaia parallaxes to select sources that lie on the stellar main sequence. We make use of PS1 photometry, instead of Gaia photometry, due to the increased contamination of Gaia $BP-RP$ colors in crowded regions of the sky.

On a CAMD, stars with PS1 $r-i > 0.35$ that have well measured, high-SNR parallaxes should mostly lie on the main sequence, while stars with spurious parallax measurements will be scattered off of the main sequence. Using stars with high-SNR parallax measurements near the Galactic North Pole ($\left| b \right| > 80^{\circ}$), we construct a mask on the PS1 $M_g$ vs. $r-i$ CAMD, as well as a mask on the $g-r$ vs. $r-i$ color-color diagram, defining the main sequence, as illustrated in Fig.~\ref{fig:good_camd}. We then select stars in the region $0^{\circ} \leq \ell \leq 80^{\circ}$, $\left| b \right| < 20^{\circ}$ that have high-SNR parallax measurements ($\mathtt{parallax\_over\_error} > 4.5$) and $\mathtt{ruwe} < 1.2$, and which lie on our CAMD and color-color main-sequence masks, based on their measured Gaia eDR3 parallaxes and extinction-corrected PS1 magnitudes. For our extinction correction, we use the mean Bayestar19 \citep{Green2019-Bayestar19,Green2018dustmaps} extinction estimate at the source distance (computed naively as $1\,\mathrm{mas}\,\mathrm{kpc} / \mathtt{parallax}$). In order to limit the effect of the extinction correction on our selection, we exclude regions of the sky for which the \citet{SFD1998} reddening estimate exceeds $\mathrm{E} \left(B-V\right) = 0.5\,\mathrm{mag}$.

This procedure yields 980k sources, which we use as our second \texttt{good} dataset. This dataset covers crowded regions closer to the Galactic plane, helping to balance our \texttt{good} training set.

\subsubsection{Globular cluster sources}

\citet{VasilievBaumgardt2021-GCs} (``VB21'') found that the initial version (v1) of our classifier underperformed in globular clusters. This is most likely due to the fact that our primary \texttt{good} training examples come from sparse regions of the sky, while our \texttt{bad} training examples are drawn primarily from dense regions of the sky.

In order to obtain \texttt{good} training data in the most crowded regime, we make use of globular cluster members identified by VB21. In detail, we select sources identified by VB21 as having high cluster membership probability (>90\%), which additionally have measured parallaxes consistent with their respective host-cluster parallax (as determined by VB21, i.e., their $\mathtt{qflag} \ge 2$). This procedure results in 450k \texttt{good} sources.

\section{Training the classifier}
\label{sec:training-nn}

We train a neural network to distinguish between our \texttt{good} and \texttt{bad} training sets, using a set of features that come from the Gaia Archive, as well as information about the neighborhood of each source. In the following, we describe our training features, the architecture and training of the neural network, and finally compare our neural network to simpler classifiers.

\subsection{Training features}
Generally, we want a set of features that enable our classifier to distinguish between \texttt{good} and \texttt{bad} sources, while avoiding features that will allow our classifier to trivially distinguish between the training sets, but which would not generalize to the entire Gaia eDR3 catalog. As the GCNS obtained good classification results for nearby stars, we use a similar set of features as that work (marked in gray in Table A.1 of \citealt{EDR3GCNS}). In detail, we use the following features from the Gaia EDR3 \texttt{gaia\_source} table:\footnote{See the \href{https://gea.esac.esa.int/archive/documentation/GEDR3/Gaia\_archive/chap\_datamodel/sec\_dm\_main\_tables/ssec\_dm\_gaia_source.html}{Gaia eDR3 data model} for details on what each feature represents.}
\begin{itemize}
    \item \texttt{parallax\_error}
    \item \texttt{parallax\_over\_error}
    \item \texttt{pmra}
    \item \texttt{pmdec}
    \item \texttt{pmdec\_error}
    \item \texttt{pmra\_error}
    \item \texttt{astrometric\_sigma5d\_max}
    \item \texttt{astrometric\_excess\_noise}
    \item \texttt{visibility\_periods\_used}
    \item \texttt{ruwe}
    \item \texttt{astrometric\_gof\_al}
    \item \texttt{ipd\_gof\_harmonic\_amplitude}
    \item \texttt{ipd\_frac\_odd\_win}
    \item \texttt{ipd\_frac\_multi\_peak}
    \item \texttt{matched\_transits\_removed}
    \item \texttt{astrometric\_params\_solved}
    \item \texttt{astrometric\_excess\_noise\_sig}
\end{itemize}
Critically, we take the absolute value of \texttt{parallax\_over\_error} before passing it to the classifier, as our \texttt{bad} training contains only strongly negative \texttt{parallax\_over\_error} values. Giving the classifier information about the sign of the parallax measurements would render classification of the training set trivial, but would not generalize to the entire Gaia survey, in which spurious parallaxes can be both positive or negative. We additionally take the absolute value of \texttt{pmra} and \texttt{pmdec}, as the direction of the proper motions might carry information about the location of a star on the sky -- information we do not want to give the classifier, given the spatially disjoint nature of our \texttt{bad} and primary \texttt{good} training sets\footnote{The total proper motion would be a possible replacement for \texttt{pmra} and \texttt{pmdec}. However, following GCNS, we use the individual vector components.}. Finally, and perhaps trivially, the actual source sky position is of course not used in the classification.

From \citet{2020arXiv201201742L}, we know that astrometric solutions with different numbers of parameters have systematic differences. We therefore add \texttt{astrometric\_params\_solved} to the feature list. This feature was neither used in v1 nor in the GCNS paper. Similarly, \texttt{matched\_transits\_removed} is also new. Both parameters depend on information from Gaia DR2 that might not be available due to a new or changed \texttt{source\_id} (which can happen without any relation to the astrometric solution). Tests showed that this does not reduce the performance of our classification, probably because other features still indicate the quality of the astrometric solution and the training sets will contain enough examples such that the classifier can still be safely generalized.

In addition to the \texttt{gaia\_source} table columns, for each source, we calculate the distance to the nearest source that is at least $X\,\mathrm{mag}$ brighter, for various values of $X$. We call these features \texttt{dist\_nearest\_neighbor\_at\_least\_X\_brighter}, where $\mathtt{X} \in \left\{ 0, 2, 4, 6, 10 \right\}$. The presence of nearby, bright neighbours is plausibly one of the largest causes of bad astrometry or colors (see Section~\ref{sec:cmd_fidelity}),\footnote{One could add \texttt{phot\_bp\_rp\_excess\_factor} as a training feature, as it also correlates with nearby sources, but this would restrict the applicability of our model to eDR3 sources which have astrometric solutions \emph{and} both $BP$ and $RP$ photometry. We therefore use \texttt{dist\_nearest\_neighbor\_at\_least\_X\_brighter} as a proxy that does not need $BP$ and $RP$ photometry.} and is therefore potentially a powerful feature for identifying spurious astrometric solutions. These features were not used by the GCNS. In the development of our classifier, we additionally explored the use of fainter nearby neighbors (i.e., neighbors with $\mathrm{X} < 0$), but found that they do not significantly improve the overall performance of the classifier, beyond what we achieve with the above features. As information about faint neighbors might simply be a proxy for overall density of sources on the sky, we decided against providing it to the classifier. We truncate \texttt{dist\_nearest\_neighbor\_at\_least\_X\_brighter} at 5~arcsec, as distant neighbors should have no appreciable effect on parallax measurements, and any inferences drawn from the presence of more distant neighbors would likely reflect idiosyncrasies of our training data (e.g., most of our \texttt{good} samples come from low-density regions of the sky) and generalize poorly to the entire eDR3 catalog. Our table of neighbor distances is available together with our \texttt{v1} and \texttt{v2} fidelities from GAVO. We also calculate and report a metric of expected color contamination that accounts for both magnitude difference and angular separation, as described in Section~\ref{sec:cmd_fidelity}.

\subsection{Training for two different |SNR| regimes}
\label{sec:two-SNR-regimes}

We train two different classifiers, intended for use in the regimes $\left| \mathrm{SNR} \right| < 4.5$ and $\left| \mathrm{SNR} \right| > 4.5$ (where we use \texttt{parallax\_over\_error} as a proxy for SNR). We will refer to these classifiers as the ``low-SNR'' and ``high-SNR'' classifiers, respectively. The most important difference between these two classifiers is that the high-SNR classifier uses |SNR| as a feature, while the low-SNR classifier does not. Recall that our \texttt{bad} training set includes very few sources with $\left| SNR \right| < 4.5$ (i.e., those from our iterative retraining). If we were to allow the low-SNR classifier to take |SNR| into account, it would learn that there are very few \texttt{bad} sources with $\left| SNR \right| < 4.5$, which is simply an artifact of our method of identifying bad training data.

\subsection{Neural network training}
\label{sec:nn}

In contrast to the work on GCNS, where spurious sources were identified using a random forest, we employ a feed-forward neural network (NN) here. Our NN model consists of 4 hidden layers, each with 64 neurons and a Rectified Linear Unit (ReLU) activation. The final layer has a single neuron with a sigmoid activation, and represents the probability that a source belongs to the \texttt{good} class. We use the binary cross-entropy loss function \citep[e.g.][]{Goodfellow-et-al-2016}, which is closely related to the Kullback-Leibler divergence and which measures how much additional information would be needed, on average, to correct the classifier's prediction. Given input features $\vec{x}$, the classifier outputs a probability $P \left( \vec{x} \right)$ that the source belongs to the \texttt{good} class. Denote true class (the label) by $y \in \left\{ 0, 1 \right\}$ (where $y = 1$ signifies ``\texttt{good}''). The binary cross-entropy is then given by
\begin{align}
    \mathcal{H}
    &=
    - \left( 1-y \right) \ln \left[ 1 - P \left( \vec{x} \right) \right]
    - y \ln P \left( \vec{x} \right)
    \, .
    \label{eqn:binary-crossentropy}
\end{align}
As the binary cross-entropy is a measure of information, it can be expressed in units of bits or nats (by using natural logarithms above, we have chosen to use nats).

We implement our model in Tensorflow 2 \citep{Abadi2016Tensorflow} and Keras \citep{chollet2015keras}. We train for 150 epochs with an Adam optimizer \citep{Kingma2014}, using a learning rate of $10^{-3}$ in the first 50 epochs, a learning rate of $10^{-4}$ in the subsequent 50 epochs, and a learning rate of $10^{-5}$ in the final 50 epochs. During training, we apply a dropout rate of 0.1 after each hidden layer in order to avoid over-fitting. The features are normalised (to zero mean and unit variance) prior to the training.

We set aside 10\% of the data as a ``test'' dataset, which we use to assess the performance of our final, trained model. Of the remaining training data, we set 20\% aside as a ``validation'' dataset, which we use to assess progress during the training procedure. Thus, in total, 72\% of our data is used for training, 18\% is used for validation during training, and 10\% is used to test the performance of the final model.

We train our high- and low-SNR classifiers separately. For both classifiers, our validation loss is lower than our training loss (due to the use of dropout during training), indicating that we are not over-fitting. For both classifiers we do not restrict the good training sample to the respective SNR range. This reduces slightly the performance measured with the test set (that was drawn from the training sample), but improves performance when applying the classifier to independent validation sets, i.e. generalizes better to the whole eDR3.

As explained in Section~\ref{sec:iterative-identification-spurious}, after training our model, we find that low-SNR sources that GCNS classifies as \texttt{bad} but ours as \texttt{good} have an implausible on-sky distribution. We incorporate these sources into our \texttt{bad} training sample and re-train the model. In the remainder of the paper, we report the results with this retrained model.

We assess our final performance by applying the low-SNR classifier to our low-SNR test dataset, and our high-SNR classifier to our high-SNR test dataset. On this combined test dataset, we achieve a loss of 0.0571 nats of cross-entropy, with a purity of 99.3\% and a completeness of 97.3\%. On the high-SNR dataset, we achieve a binary cross-entropy of 0.0457 nats, with a purity of 98.9\% and a completeness of 97.2\%\footnote{For comparison, when using an ExtraTree \citep{10.1007/s10994-006-6226-1} classifier (which is closer in performance to the GCNS paper's random forest) in the high-SNR regime, we obtain a purity of 93.7\,\% and a completeness of 95.7\,\%, showing the superior performance of the neural net classifier.}. This performance is an upper limit for real world applications, as our training samples are fairly well separated in feature space. Particularly in the low-SNR regime, it is difficult to discern \texttt{good} from \texttt{bad} astrometric solutions.

\begin{figure}
	\includegraphics[width=\linewidth]{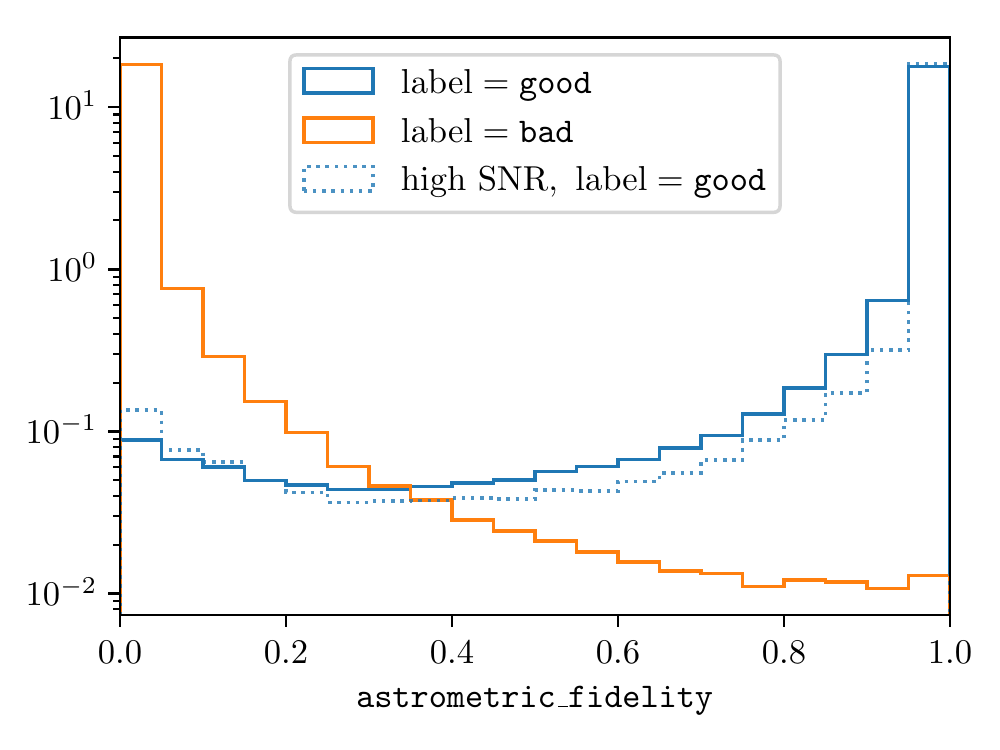}
	\caption{Histogram of the predicted classifier probabilities (of belonging to the \texttt{good} class) for sources in the test dataset, split by training label. As the \texttt{good} class contains both low- and high-SNR training data, we additionally show the classifier probabilities for the high-SNR \texttt{good} sources. The $x$-axis represents the probability output by the classifier that a given source is \texttt{good}, which we term the ``astrometric fidelity.''}
	\label{fig:classification_hist}
\end{figure}

In Fig.~\ref{fig:classification_hist}, we show histograms of the predicted astrometric fidelity for \texttt{good} and \texttt{bad} sources in the test dataset. In the rest of this work, we classify objects with $\mathrm{fidelity} > 0.5$ as \texttt{good}, though users can make stricter (or looser) cuts to improve purity at the expense of completeness (or completeness at the expense of purity).

\subsection{Comparison with simpler classifiers}

\begin{figure}
	\includegraphics[width=\linewidth]{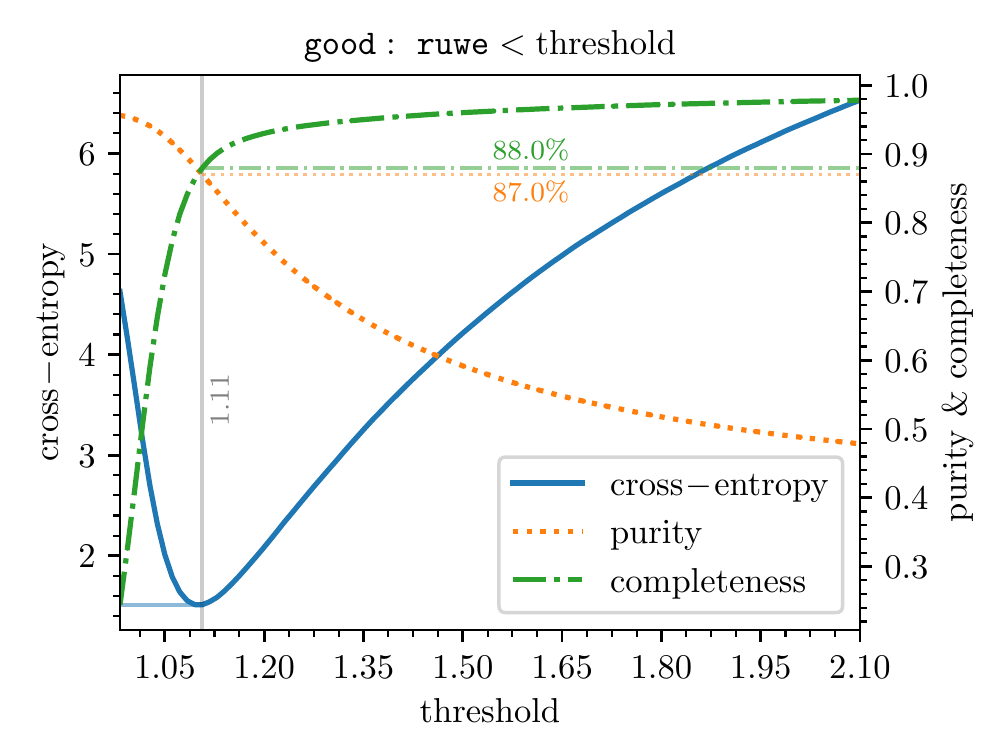}
	\includegraphics[width=\linewidth]{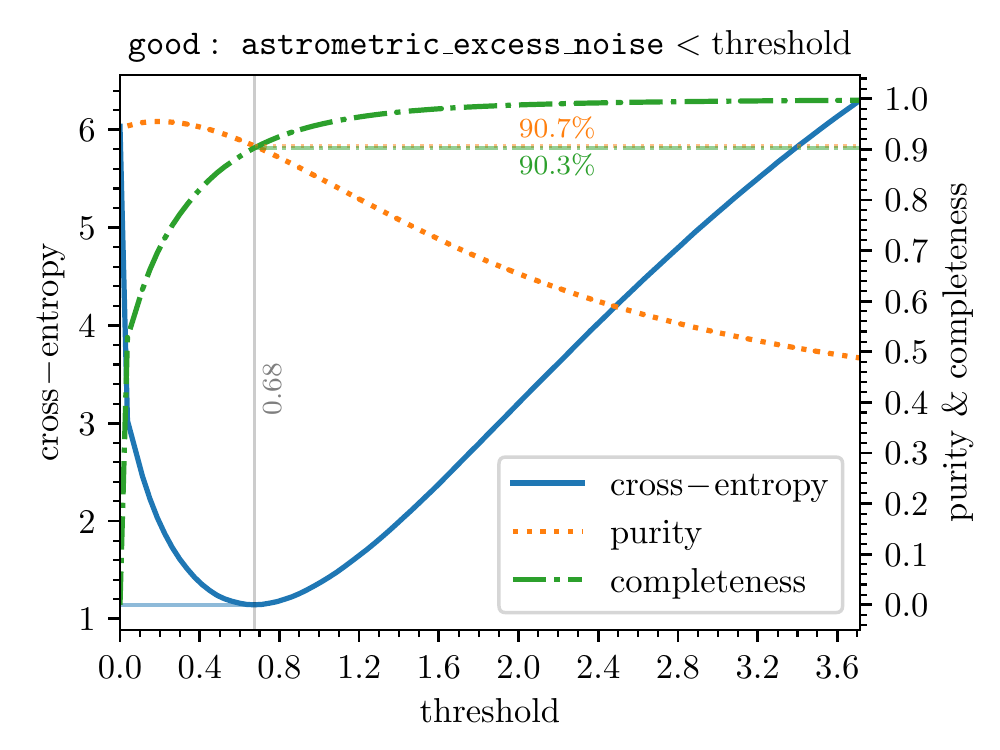}
	\caption{Performance of simple cuts on \texttt{ruwe} (top panel) and \texttt{astrometric\_excess\_noise} (bottom panel) in differentiating \texttt{good} and \texttt{bad} astrometric solutions. In each panel, we show how binary cross-entropy (Eq.~\ref{eqn:binary-crossentropy}), purity and completeness depend on the threshold chosen for the cut. Our neural network predicting astrometric fidelity achieves an order of magnitude less contamination (99.3\% purity and 97.3\% completeness -- see Section~\ref{sec:two-SNR-regimes}) than the optimal choices for these cuts (at minimal cross-entropy).}
	\label{fig:simple_cuts}
\end{figure}

Here, we compare the astrometric fidelity predicted by our neural network to analogous quantities obtained using simpler classifiers. First, we evaluate how cleanly simple cuts on \texttt{ruwe} and \texttt{astrometric\_excess\_noise} separate \texttt{good} and \texttt{bad} sources in the high-SNR test dataset. Fig.~\ref{fig:simple_cuts} shows the binary cross-entropy, purity and completeness of the cut as a function of the threshold value for each feature. For \texttt{ruwe}, we achieve a minimum binary cross-entropy of 1.51~nats using a cut of $\texttt{ruwe} < 1.11$, corresponding to a purity of 87.0\% and a completeness of 88.0\%. For \texttt{astrometric\_excess\_noise}, we achieve a minimum binary cross-entropy of 1.14~nats for a cut of $\texttt{astrometric\_excess\_noise} < 0.68$, corresponding to a purity of 90.7\% and a completeness of 90.3\%.

Next, we train a logistic model, which takes into account a linear combination of features. This model assigns a probability
\begin{align}
  P \left( \texttt{good} \mid \vec{x} \right) = 
    \left[
    1 + e^{-\left( \vec{w} \cdot \vec{x} + b \right)}
    \right]^{-1}
\end{align}
of belonging to the \texttt{good} class to each source, where $\vec{x}$ is a vector containing the features, $\vec{w}$ is a vector containing a weight for each feature, and $b$, the bias, is a scalar. We use the Adam optimizer to find the weights and bias that minimize the binary cross-entropy of the predictions. On the high-SNR test dataset, we obtain a binary cross-entropy of 0.100~nats, a purity of 95.7\% and a completeness of 95.9\%. This is better than what we achieve with simple cuts, but still represents nearly twice the binary cross-entropy we obtain with the full neural network model.

The full neural network is not significantly more difficult to implement than these simpler classifiers, and it achieves a far more complete and pure separation of the test dataset. For these reasons, we strongly favor use of the full neural network classification over simpler alternatives.

\section{Outside validation}
\label{sec:outside}
Our classifier performs very strongly on the test dataset, which is statistically identical to the training data. In this section, we perform several tests using a variety of outside datasets, in order to determine whether our classifier generalizes beyond our training data. Here, the validation revolves around external information on the  distance (e.g. membership in a cluster or the Large Magellanic Cloud), the plausibility of the distribution of sources on the sky or in the disk of the Milky Way, or simply external measurements (e.g., comparison of Gaia proper motions with those measured by other surveys).

\subsection{Gaia Catalogue of Nearby Stars}

All 1.2 million sources with parallax $> 8\,\mathrm{mas}$ from eDR3 have classifications from the GCNS \citep{EDR3GCNS}. Here, we measure our agreement with these classifications. As described in Section~\ref{sec:iterative-identification-spurious}, we train two iterations of our model, using a sample of low-SNR sources from the GCNS in our \texttt{bad} training sample in the second iteration. Thus, our final model is influenced by the GCNS classifications, and will -- by construction -- have good agreement.

We therefore report our agreement with GCNS both for our initial and for our final model. Taking the GCNS classifications as ground truth, our initial model achieves a purity of 96.0\,\% and a completeness of 99.4\,\%. Our final model achieves a purity of 99.7\,\% and a completeness of 99.3\,\%.

\subsection{Clusters}

\begin{figure}
	\includegraphics[width=\linewidth]{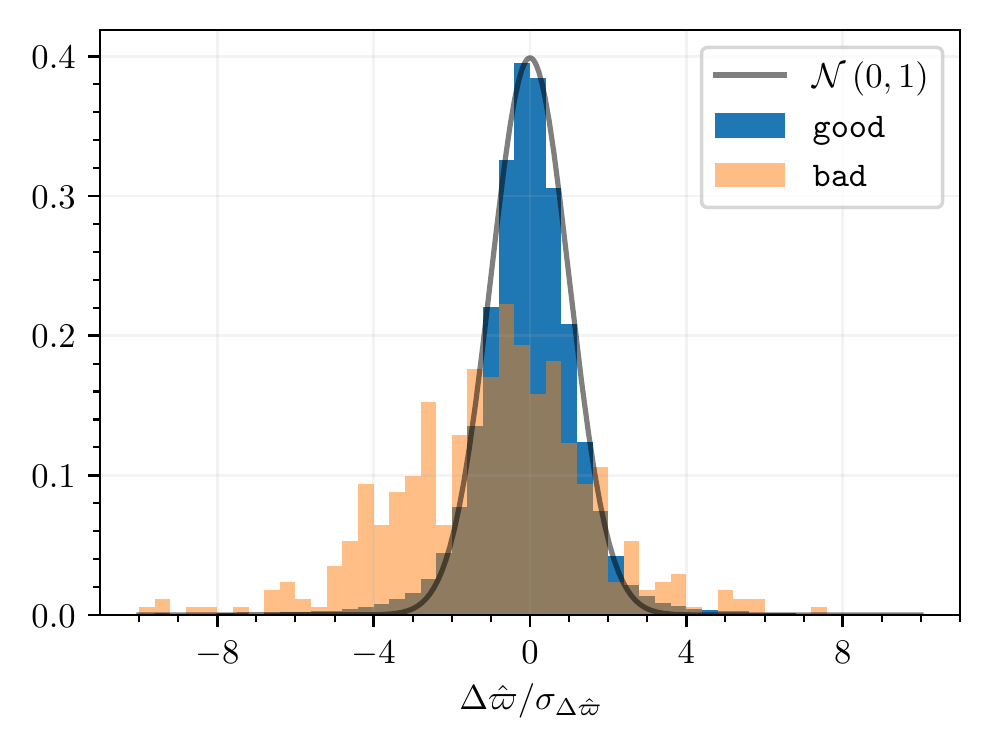}
	\caption{Validation of our astrometric fidelity prediction using open and globular clusters. The figure shows histograms of the standardized parallax residuals for \texttt{good} and \texttt{bad} sources. The true parallaxes are estimated using the variance-weighted mean of the parallaxes in each cluster. We restrict this comparison to clusters with distances determined to 20\% or better. The \texttt{good} sources closely follow the expected unit normal distribution, in marked contrast to the standardized residuals of the \texttt{bad} sources.}
	\label{fig:cluster_comparison}
\end{figure}

Open and globular clusters, and the ``prior'' information on the distance of their likely member stars, offer a great opportunity to validate our parallax classifier. We begin with a catalog of 162,484 sources assigned to 121 clusters, coming from \citet{2018AJ....156...58B}. This catalog was compiled using a method similar to that used in \citet{2018A&A...616A..10G}. For each individual cluster, we calculate the variance-weighted mean parallax of its member sources, as well as the corresponding uncertainty in the mean parallax:
\begin{align}
  \left< \hat{\varpi} \right> = \sum_i \frac{\hat{\varpi}_i}{\sigma_i^2}
  \, , \hspace{0.5cm}
  \sigma_{\left< \hat{\varpi} \right>} =
    \left(
      \sum_i \frac{1}{\sigma_i^2}
    \right)^{-\nicefrac{1}{2}}
  \, .
\end{align}
We then select the 41 clusters for which $\sigma_{\left< \hat{\varpi} \right>} / \left< \hat{\varpi} \right> < 0.2$ (i.e., for which parallax is determined to better than 20\%). For each source in each of these clusters, we then calculate a parallax residual, using the estimated cluster parallax as a reference:
\begin{align}
    \Delta \hat{\varpi} \equiv \hat{\varpi} - \left< \hat{\varpi} \right>
    \, , \hspace{0.5cm}
    \sigma_{\Delta \hat{\varpi}}
    &=
    \left(
        \sigma_{\hat{\varpi}}^2
        + \sigma_{\left< \hat{\varpi} \right>}^2
    \right)^{\nicefrac{1}{2}}
    \, .
\end{align}
The distribution of these parallax residuals (divided by the corresponding uncertainties) is shown in Fig.~\ref{fig:cluster_comparison}. Our classifier labels approximately 1\% of sources in these clusters as \texttt{bad}. The standardized residuals of sources classified as \texttt{good} roughly follow the expected unit normal distribution, while the distribution of standardized residuals of the sources classified as \texttt{bad} is shifted negative and has much longer tails.

\subsection{OGLE proper motions}

The Fourth Phase of the Optical Gravitational Lensing Experiment \citep[OGLE-IV,][]{OGLE-IV} began observing the bulge of the Milky Way in 2010. Here, we validate our classifier using sources with proper-motion measurements from both OGLE-IV (OGLE Uranus astrometry project, Udalski et al. 2021, in preparation) and Gaia eDR3. Our assumption is that objects with spurious parallax determinations in Gaia eDR3 are more likely to have spurious proper-motion determinations. This should be reflected in the proper-motion residuals between Gaia eDR3 and OGLE-IV, with sources classified as \texttt{bad} in Gaia eDR3 having systematically higher $\chi^2$ values in this comparison. We begin with a catalog of OGLE-IV sources with proper-motion measuresments, lying in a $0.15\,\mathrm{deg} \times 0.15\,\mathrm{deg}$ box centered on $\left( \alpha_{\mathrm{J2000}} ,\, \delta_{\mathrm{J2000}} \right) = \left( 271.761\,\mathrm{deg} ,\, -26.698\,\mathrm{deg} \right)$. Using a matching radius of $0.2^{\prime\prime}$, we obtain 14125 matching Gaia eDR3 sources with measured proper motions. Our classifier labels 2870 of these sources \texttt{good}.

\begin{figure}
	\includegraphics[width=\linewidth]{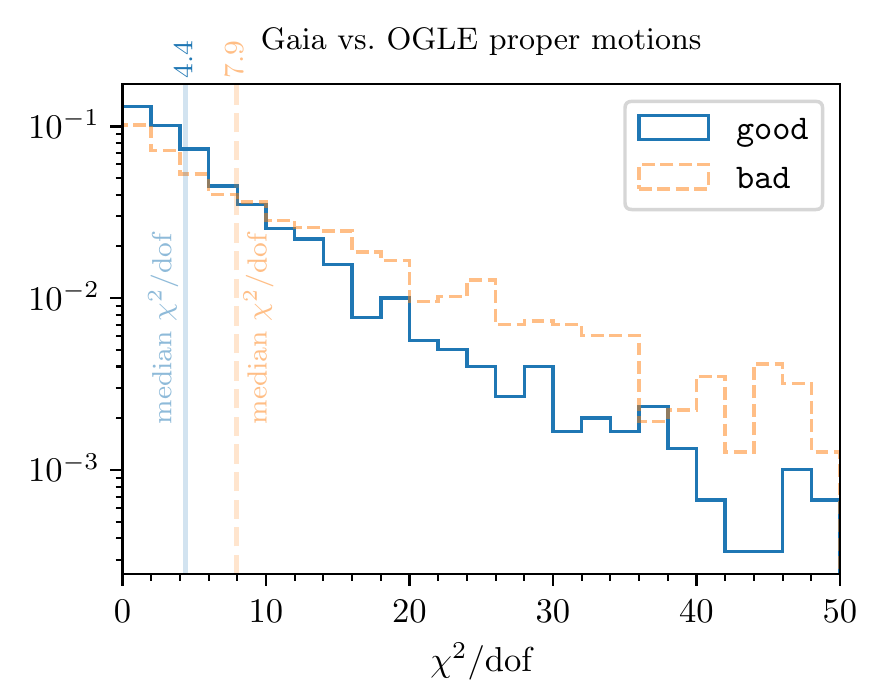}
	\caption{Validation of our astrometric fidelity classification through proper motion comparison between OGLE and Gaia eDR3. Shown is the distribution of $\chi^2 / \mathrm{dof}$, based on a comparison of Gaia eDR3 and OGLE-IV proper motions in the Galactic Bulge, for sources labeled \texttt{good} and \texttt{bad} by our classifier. For ideal data, the median $\chi^2 / \mathrm{dof}$ would be $\sim 0.69$. We find that that proper-motion uncertainties are underestimated for Gaia eDR3 and/or OGLE-IV, leading to larger $\chi^2 / \mathrm{dof}$ values. However, for sources labeled \texttt{good} by our classifier, Gaia eDR3 and OGLE-IV proper motions match significantly better, as indicated by the lower median $\chi^2 / \mathrm{dof}$ values (4.4 for the \texttt{good} subsample, vs. 7.9 for the \texttt{bad} subsample).}
	\label{fig:ogle_pm_comparison}
\end{figure}

We calculate the proper-motion residuals, $\Delta \vec{\mu} \equiv \vec{\mu}_{\mathrm{Gaia}} - \vec{\mu}_{\mathrm{OGLE}}$, as well as the covariance matrix of the residuals, $C_{\Delta \vec{\mu}} = C_{\mu, \mathrm{Gaia}} + C_{\mu, \mathrm{OGLE}}$. We then calculate $\chi^2 = \Delta \vec{\mu}^T C_{\Delta \vec{\mu}}^{-1} \Delta \vec{\mu}$ for each source. If the uncertainties are well estimated and the residuals follow a Gaussian distribution, then the $\chi^2$ values that we obtain should follow a $\chi^2$ distribution with two degrees of freedom. However, we find that the resulting $\chi^2$ values are significantly larger, on average, than expected, both for sources labeled \texttt{good} and \texttt{bad}, indicating that Gaia eDR3 and/or OGLE-IV proper-motion uncertainties are underestimated in the Galactic Bulge. One could attempt to address this problem by inflating the uncertainties by a constant factor or by introducing a systematic error floor. However, these different methods of ``correcting'' the proper-motion uncertainties impact the distributions of $\chi^2$ values obtained for the \texttt{good} and \texttt{bad} sources differently, as the \texttt{good} sources tend to have smaller estimated proper-motion uncertainties than the \texttt{bad} sources. In order to avoid these difficulties, we restrict our comparison to sources in a relatively small range of estimated proper-motion uncertainties, for which we assume the true proper-motion uncertainties to be similar. In particular, we select sources with $0.1 \, \mathrm{mas\,yr^{-1}} < \left| C_{\Delta \vec{\mu}} \right|^{\nicefrac{1}{4}} < 0.2 \, \mathrm{mas\,yr^{-1}}$, obtaining 1503 sources labeled \texttt{good} and 1657 sources labeled \texttt{bad}. The resulting distributions of $\chi^2$ values are displayed in Fig.~\ref{fig:ogle_pm_comparison}. We find that sources labeled \texttt{good} by our classifier tend to have significantly lower $\chi^2$ values than those labeled \texttt{bad}, with the median $\chi^2$ per degree of freedom (dof) for the \texttt{good} subsample being 4.4, and the median $\chi^2 / \mathrm{dof}$ of the \texttt{bad} subsample being 7.9.

\subsection{Large Magellanic Cloud}

In the direction of the Large Magellanic Cloud (LMC), the vast majority of sources should be at a distance of $\sim$50~kpc, corresponding to a parallax of 0.02~mas. This affords us another opportunity to validate our classifications, as almost all stars labeled \texttt{good} in this region of the sky should have reported parallaxes consistent with 0.02~mas. We expect the \texttt{bad} sources to have larger than reported residuals, and to scatter equally to positive and negative parallaxes, leading to a widened distribution of reported parallaxes centered on 0.02~mas.

We query a 0.25~deg cone in Gaia eDR3, centered on Galactic coordinates $\left( \ell, b \right) = \left( 280.47 \, \mathrm{deg}, -32.88 \, \mathrm{deg} \right)$, obtaining 252,115 sources, which we then run through our classifier. In this densely crowded region of the sky, only 16.05\% of all sources are classified as \texttt{good}, while only 0.7\% of high-SNR sources are classified as \texttt{good}. In order to model the small number of Milky Way foreground stars in this field, we compare to a control field of the same apparent size with the same Galactic latitude, and longitude reflected around $\ell = 0 \, \mathrm{deg}$. This control field has 1589 sources. In the control field, 85.3\% of the sources are classified as \texttt{good}.

\begin{figure}
	\includegraphics[width=\linewidth]{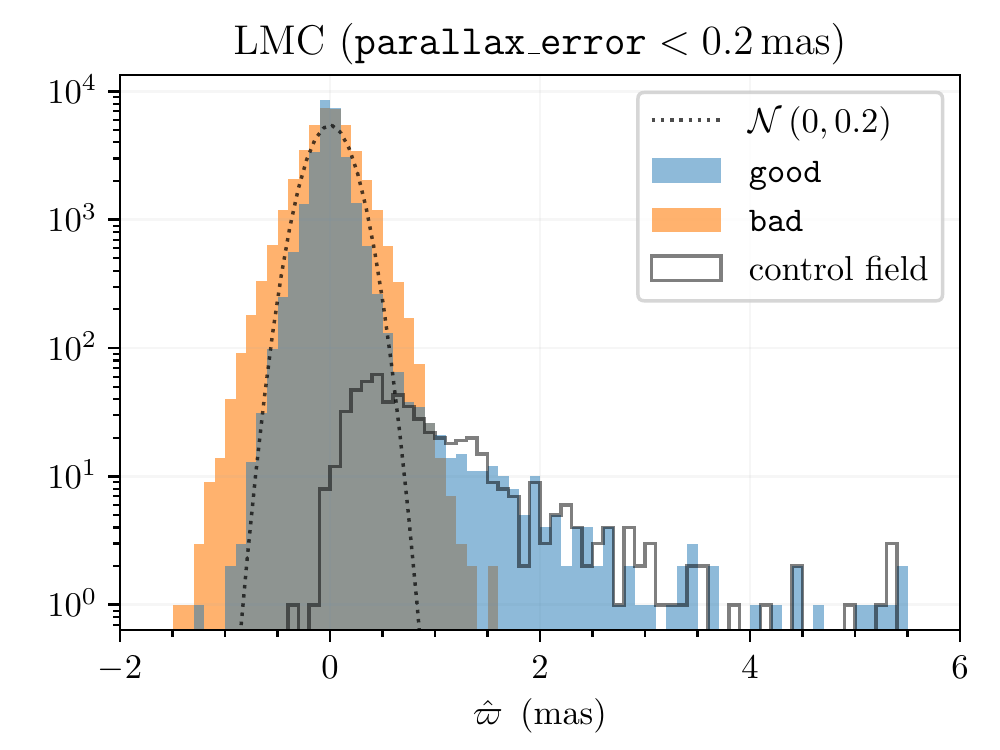}
	\caption{Validation of our astrometric fidelity classification through the parallax distribution of \texttt{good} and \texttt{bad} sources with small parallax uncertainties ($\sigma_{\hat{\varpi}} < 0.2 \, \mathrm{mas}$) towards the LMC. The parallax distribution of \texttt{good} sources is consistent with a large population of distant ($\hat{\varpi} \approx 0$) sources (approximated by a normal distribution with zero mean and a standard deviation of 0.2~mas), along with an expected population of foreground stars (matching the distribution of parallaxes in a control field). In contrast, the \texttt{bad} parallaxes are consistent with a distant population of stars with parallax uncertainties that are underestimated by $\sim$50\,\%. This indicates that in dense regions of the sky, our astrometric fidelities distinguish between sources with well-measured parallaxes, and sources with parallax residuals that are substantially larger than the reported uncertainties.}
	\label{fig:lmc}
\end{figure}

Fig.~\ref{fig:lmc} shows the parallax distribution of \texttt{good} and \texttt{bad} sources with small reported errors ($\mathtt{parallax\_error} < 0.2$) in our LMC field. The parallax distribution of \texttt{good} sources is consistent with a distant population of stars with well-measured errors, plus a small foreground population of Milky Way stars at larger parallax (matching the control field). The \texttt{bad} sources are consistent with a distant population of stars with significantly underestimated parallax errors. Our classifier is thus clearly identifying sources with excess parallax residuals, and even in this dense field, is still cleanly identifying foreground stars.

\subsection{The Structure of the Galactic Disk in OBA stars}

The catalog of O-, B-, and A-type (OBA) stars devised by \citet{Zari2021-hot-stars} (``Z21'') offers another opportunity to test our classifier with an ensemble of sources at {\it low Galactic latitudes}. Z21 selects stars brighter than $G = 16 \, \mathrm{mag}$,  with \textit{Gaia} eDR3 and 2MASS colors consistent with (reddened) OBA-type stars. Z21 does not apply any condition on the parallax error, as the sample was designed to be inclusive for spectroscopic follow-up.

\begin{figure*}
    \centering
    \includegraphics[width=0.48\hsize]{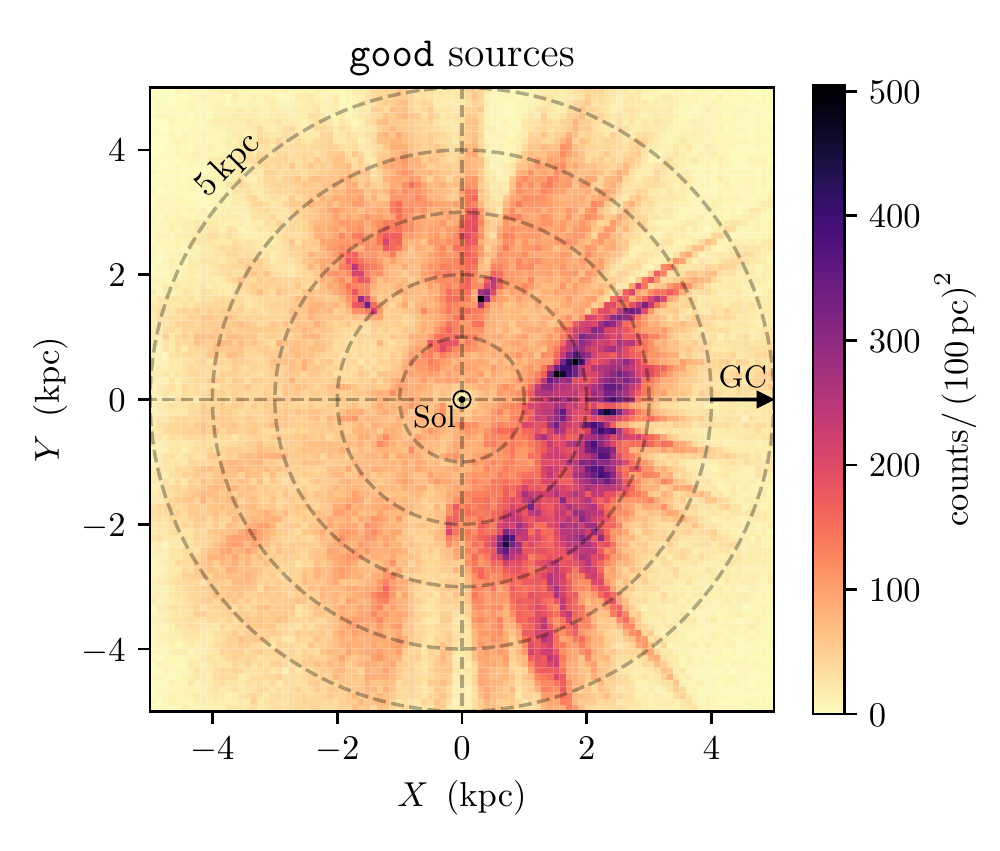}
    \includegraphics[width=0.48\hsize]{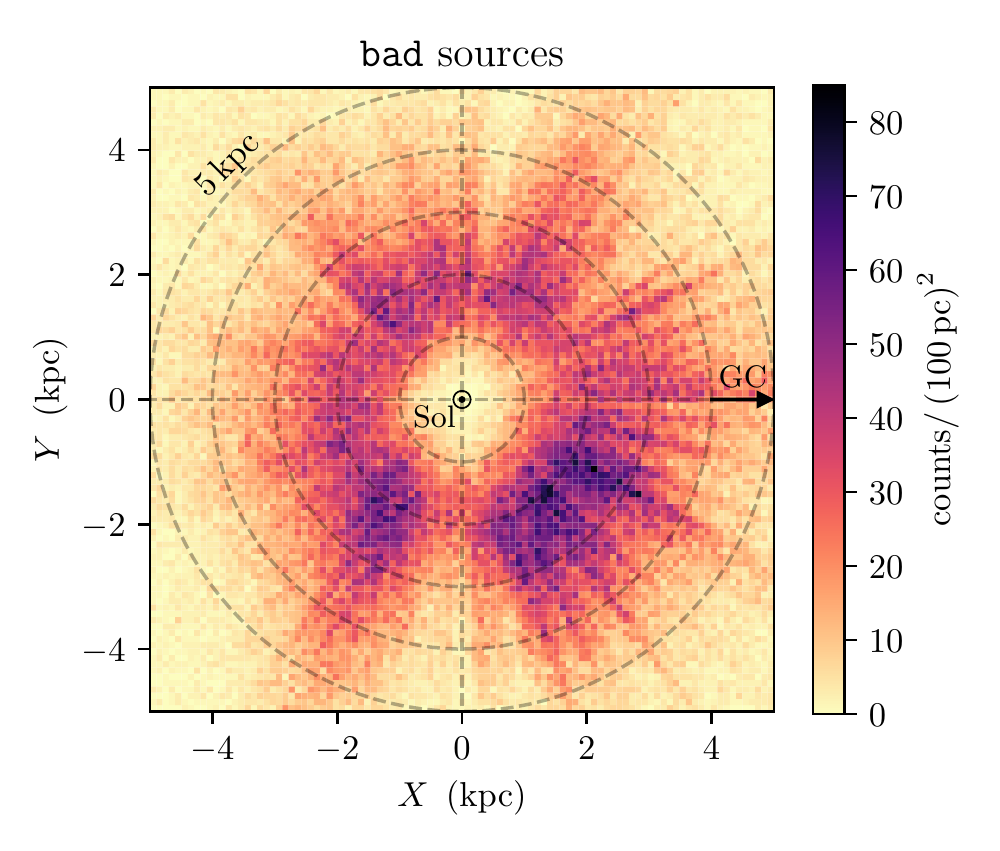}
    \caption{Validation of our astrometric fidelity classification through the astrophysical plausibility of the distribution of young stars in the Galactic plane, as identified by \citet{Zari2021-hot-stars}. The left (right) panel shows the spatial distribution of \texttt{good} (\texttt{bad}) OBA-star sources in the Galactic plane, with the Sun located at $\left( X, Y \right) = \left( 0 \, \mathrm{kpc}, 0 \, \mathrm{kpc} \right)$, and the Galactic center at $\left( 8.2 \, \mathrm{kpc}, 0 \, \mathrm{kpc} \right)$. We have divided the plane into pixels 100~pc on a side. The \texttt{good} sources show concentrations at many known locations of young stars, and show a spiral-arm-like morphology. The \texttt{bad} sources show a ring-like structure, exactly centered on the Sun and at the (seeming) distance of the most common parallax -- clearly, far too Ptolemaic a distribution to be real.}
    \label{fig:OBA}
\end{figure*}

We run our classifier on the resulting catalog, which contains $\sim$1 million stars. Fig.~\ref{fig:OBA} shows the distribution of Z21 sources classified as \texttt{good} (left, $\sim$80\% of the initial sample) and \texttt{bad} (right) in the Galactic plane ($|b| < 25^{\circ}$). The distribution of sources with \texttt{good} astrometric solutions shows known regions of young stars and traces the spiral arm structure of the Milky Way disk, as discussed in Z21 (cf. their Fig.~11). The distribution of sources with \texttt{bad} astrometric solutions shows a ring-like feature centered on the Sun, at a distance of 2 to 3~kpc, which is physically implausible and hence presumably spurious. This is expected, as the parallax distribution of all sources in the OBA catalog peaks at around 0.3~mas ($\sim$3~kpc).

A similar structure (called the ``bloody-eye effect'') was observed in Gaia DR2 by \citet[][e.g. their Fig. A2]{2019A&A...628A..94A}, in which the cleaning of spurious results was achieved by a quality flag definition (\texttt{STARHORSE\_OUTFLAG[0]}) based on a cut in the space of relative distance uncertainty vs. $\log g$. Our astrometric fidelity classifier confirms the presence of the effect in Gaia eDR3, and resolves the issue using Gaia astrometric data alone.

\subsection{White dwarf test}

We use a spectroscopically confirmed sample of white dwarfs (WDs) from \citet{2021arXiv210607669G} to determine how many presumably good WDs are excluded by our classifier. We only use WDs with \texttt{sdss\_clean1} == 1, together with \texttt{type} belonging to the set $\{$WD | DZ | DQ | DC | DB | DA$\}$. All 23k WDs should have well measured parallaxes, as their positions on the CAMD lie in the WD area. Whereas \texttt{v1} of our classifier marked 7.7\,\% of these WDs as \texttt{bad}, our updated \texttt{v2} classifier only classifies 2.6\,\% of the WDs as \texttt{bad}. This improvement is likely due to the fact that our \texttt{good} training data is cross-matched with PS1 (rather than with 2MASS, as in \texttt{v1}), which allows for fainter blue objects to enter the training sample. This test can serve as a simple estimate (upper limit) for the incompleteness that our classification may induce for high-latitude samples of sources across a wide range of apparent magnitudes.

\subsection{Source distribution in parallax bins}
\label{sec:parallax-bin-validation}

As a final approach to validation, we visually inspect the projected sky distribution and CAMD of Gaia eDR3 sources classified as \texttt{good} and \texttt{bad} in narrow bins of measured parallax. We additionally compare with mock observations generated using GeDR3mock (see in Appendix\,\ref{sec:gedr3mock_query} for the query, where we sample the uncertainties and also use HEALPix-specific $G$-band magnitude limits).

We refer to the parallax bins by their corresponding nominal distances. The 100~pc sample consists of the 1.2 million sources with $\hat{\varpi} > 8\,\mathrm{mas}$ (the GCNS sample). The 300~pc sample consists of the 1.3 million sources with $3.3\,\mathrm{mas} < \hat{\varpi} < 3.4\,\mathrm{mas}$, the 1~kpc sample ($1\,\mathrm{mas} < \hat{\varpi} < 1.01\,\mathrm{mas}$) contains 3.2 million sources, the 3~kpc sample ($0.333\,\mathrm{mas} < \hat{\varpi} < 0.334\,\mathrm{mas}$) contains 1.3 million sources, the 10~kpc sample ($0.1\,\mathrm{mas} < \hat{\varpi} < 0.101\,\mathrm{mas}$) contains 1.2 million sources, and the 30~kpc sample ($0.0325\,\mathrm{mas} < \hat{\varpi} < 0.034\,\mathrm{mas}$) contains 1.4 million sources. 

The figures we will show in the following subsections will all follow a similar pattern: The distance slices will increase from top to bottom and on the left side will be the GeDR3mock sources (our expectation for \texttt{good} sources), in the middle the eDR3 sources classified as \texttt{good} and in the right panel the ones classified as \texttt{bad}.

\subsubsection{Sky distribution}

\begin{figure*}
	\includegraphics[width=0.29\linewidth]{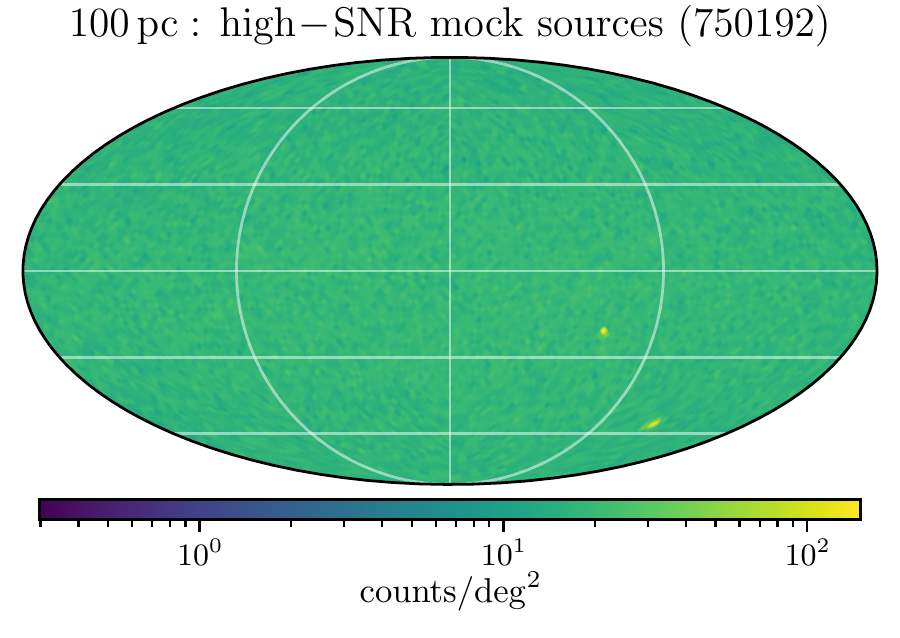}
	\hfill
	\includegraphics[width=0.29\linewidth]{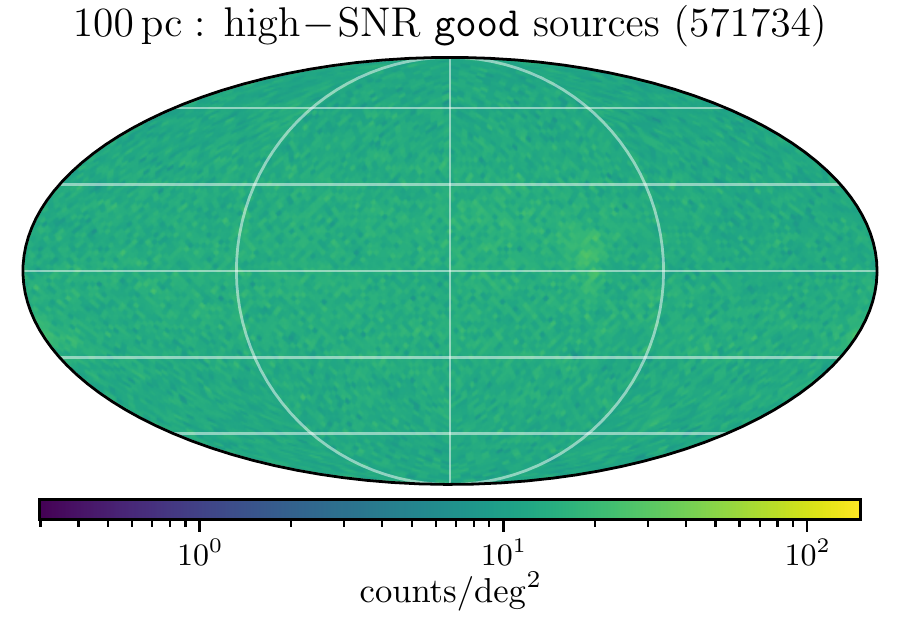}
	\hfill
	\includegraphics[width=0.29\linewidth]{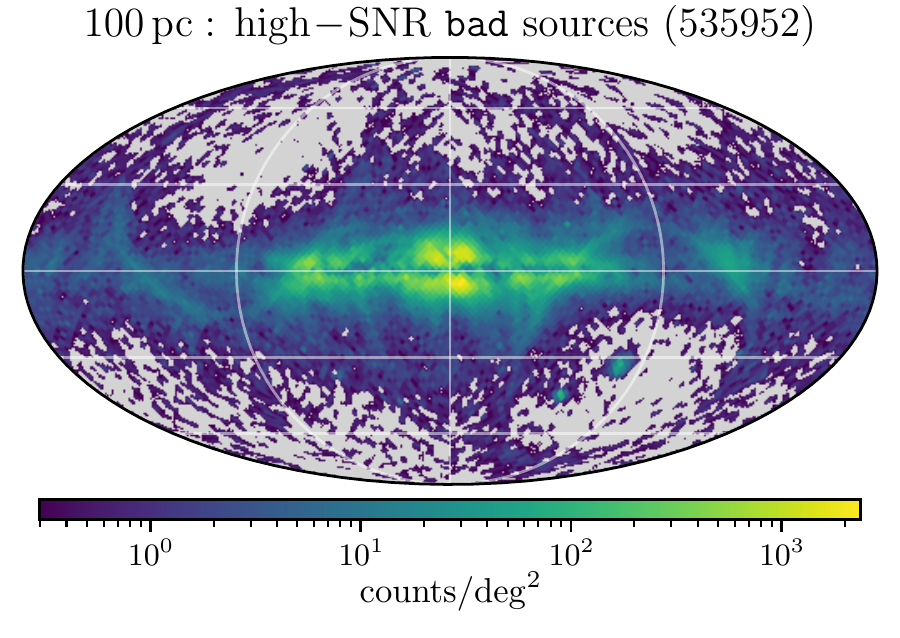}
	\includegraphics[width=0.29\linewidth]{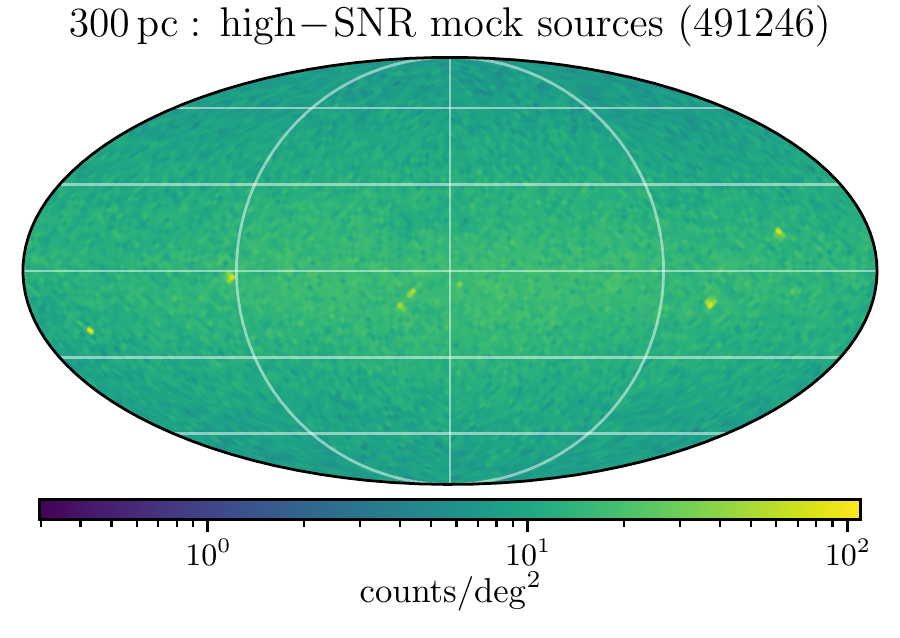}
	\hfill
	\includegraphics[width=0.29\linewidth]{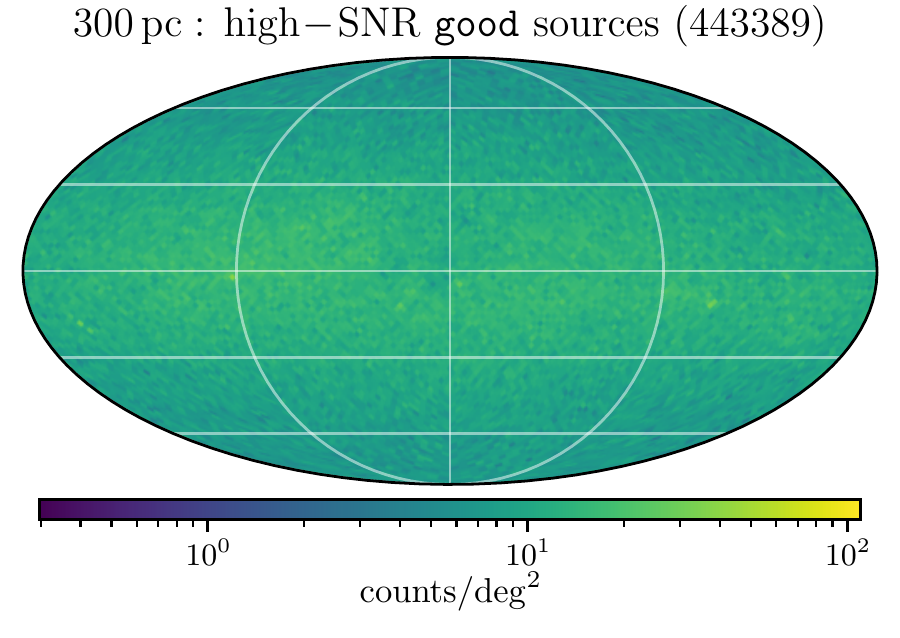}
	\hfill
	\includegraphics[width=0.29\linewidth]{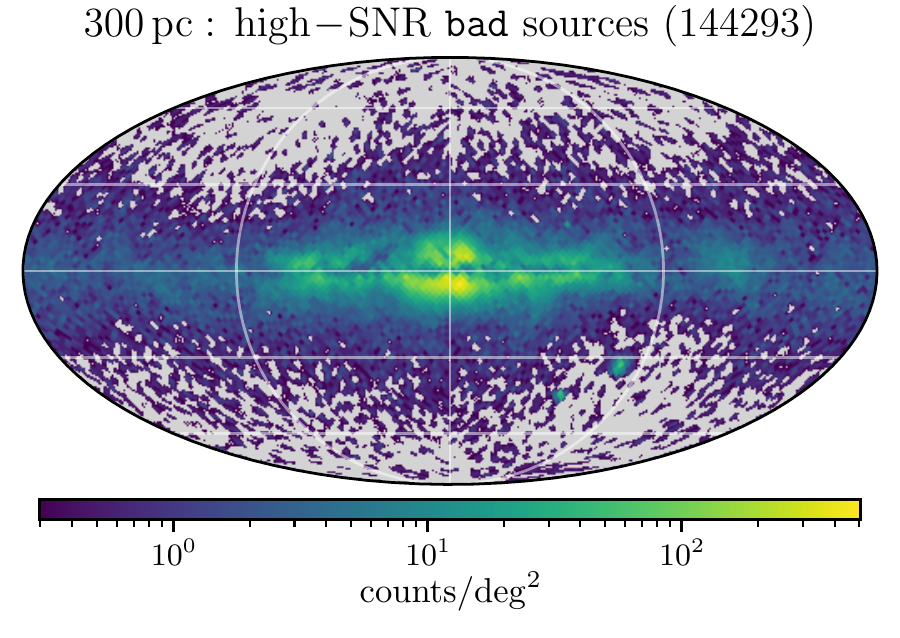}
	\includegraphics[width=0.29\linewidth]{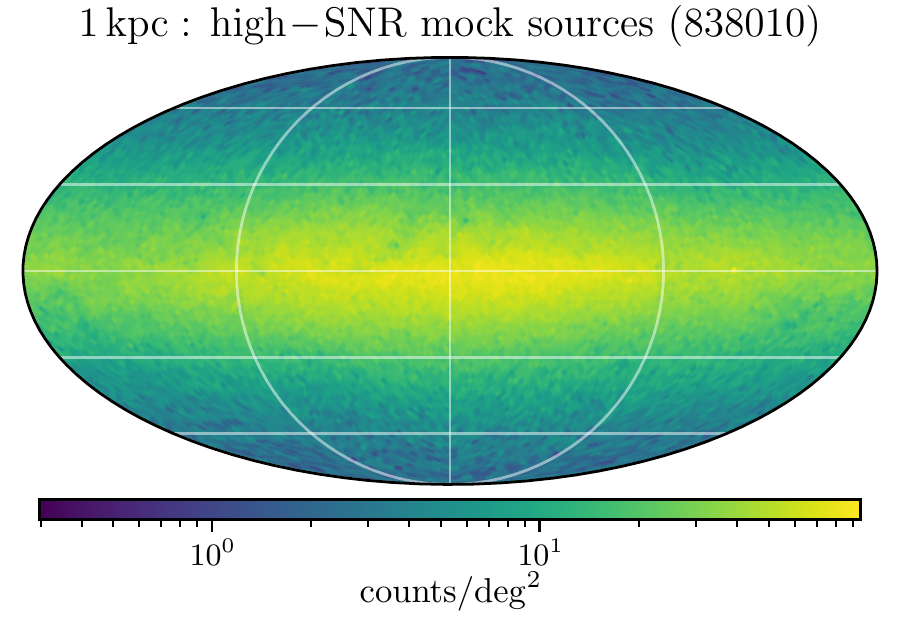}
	\hfill
	\includegraphics[width=0.29\linewidth]{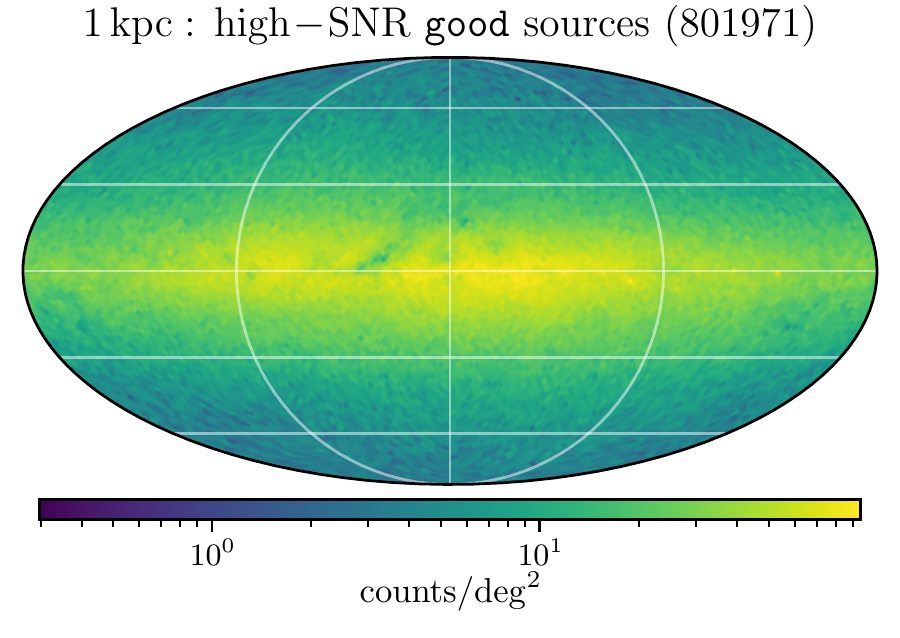}
	\hfill
	\includegraphics[width=0.29\linewidth]{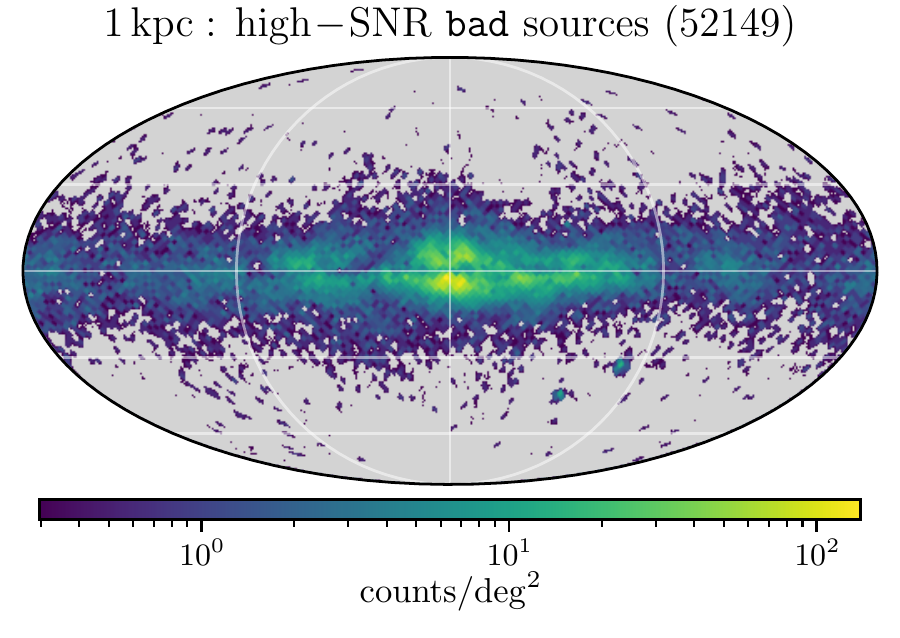}
	\includegraphics[width=0.29\linewidth]{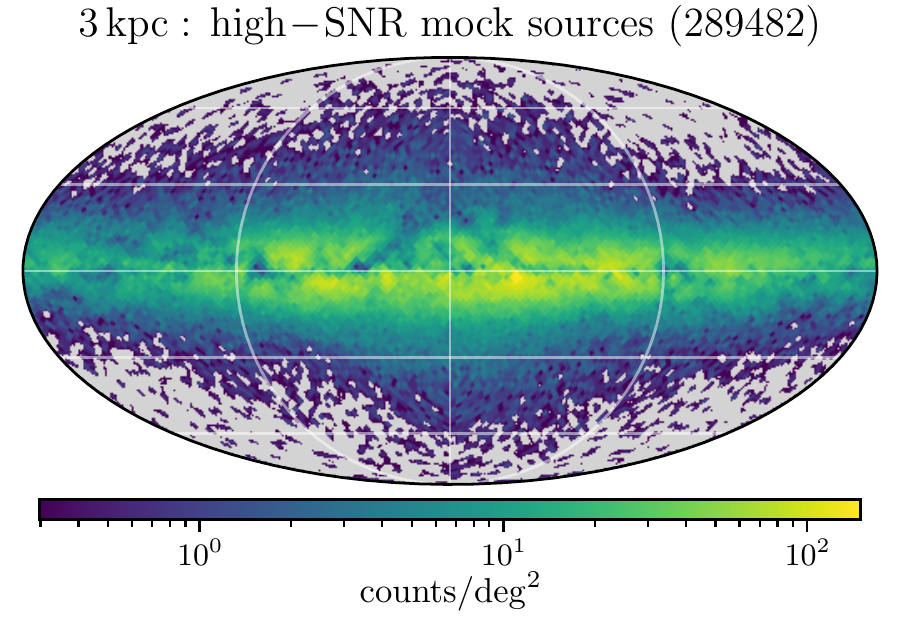}
	\hfill
	\includegraphics[width=0.29\linewidth]{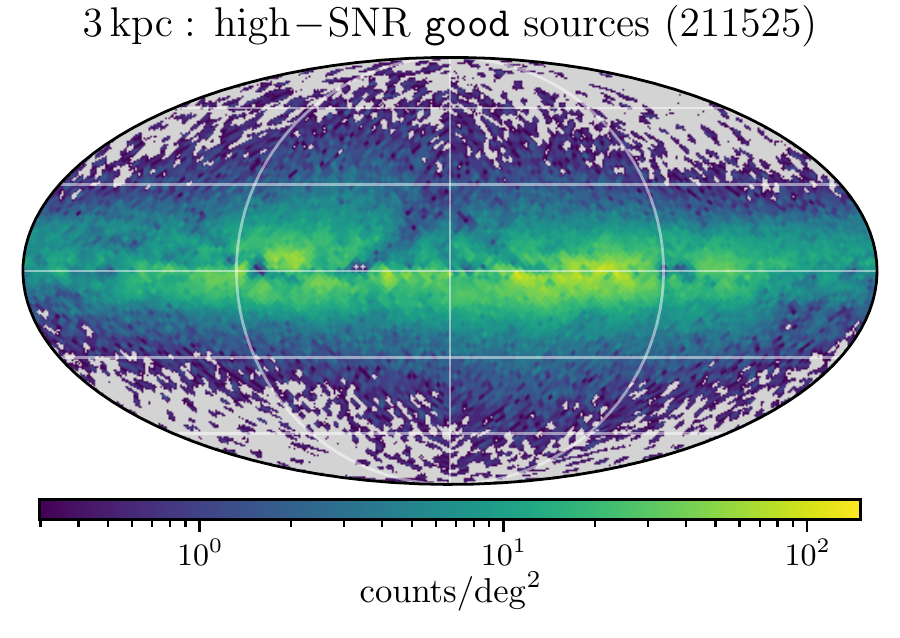}
	\hfill
	\includegraphics[width=0.29\linewidth]{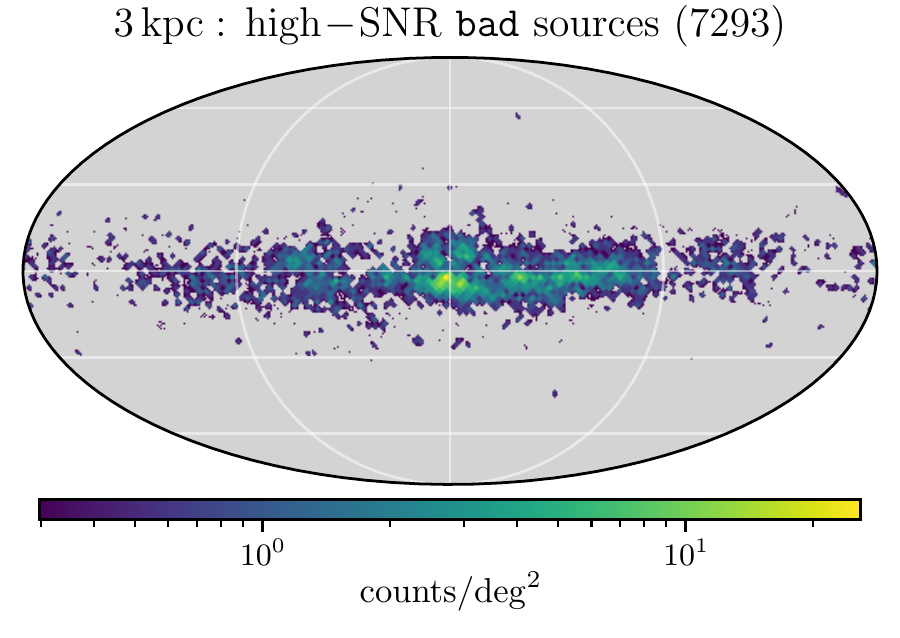}
	\includegraphics[width=0.29\linewidth]{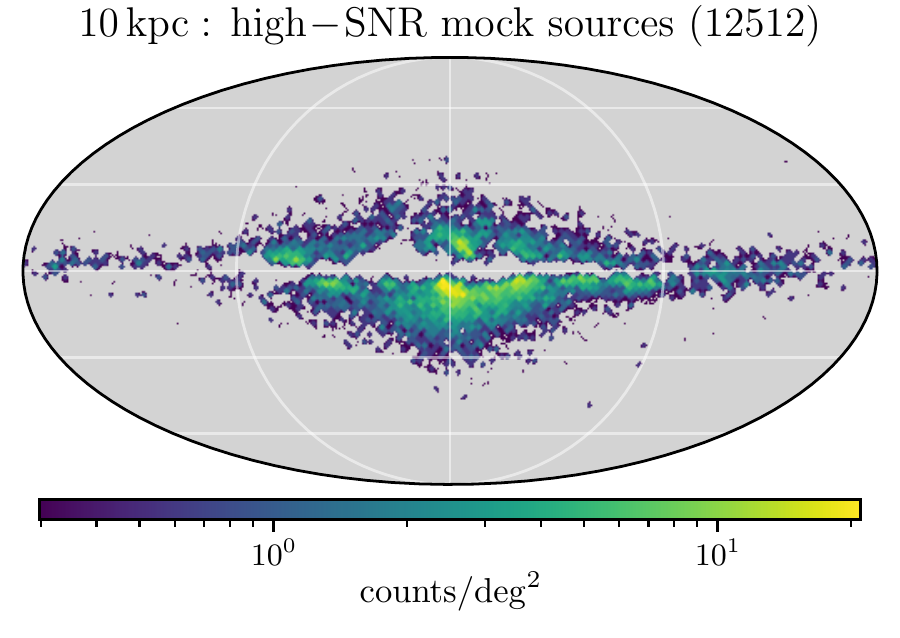}
	\hfill
	\includegraphics[width=0.29\linewidth]{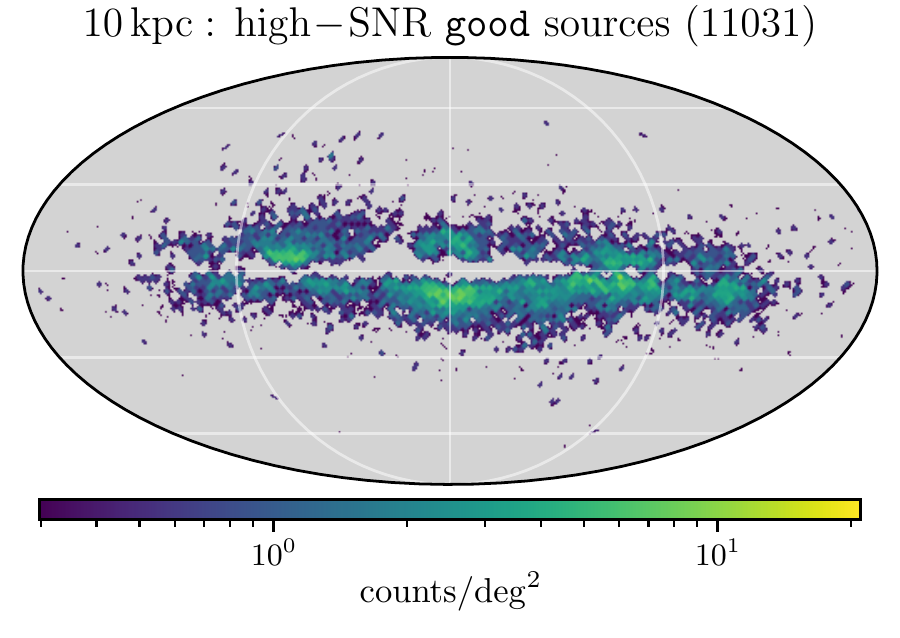}
	\hfill
	\includegraphics[width=0.29\linewidth]{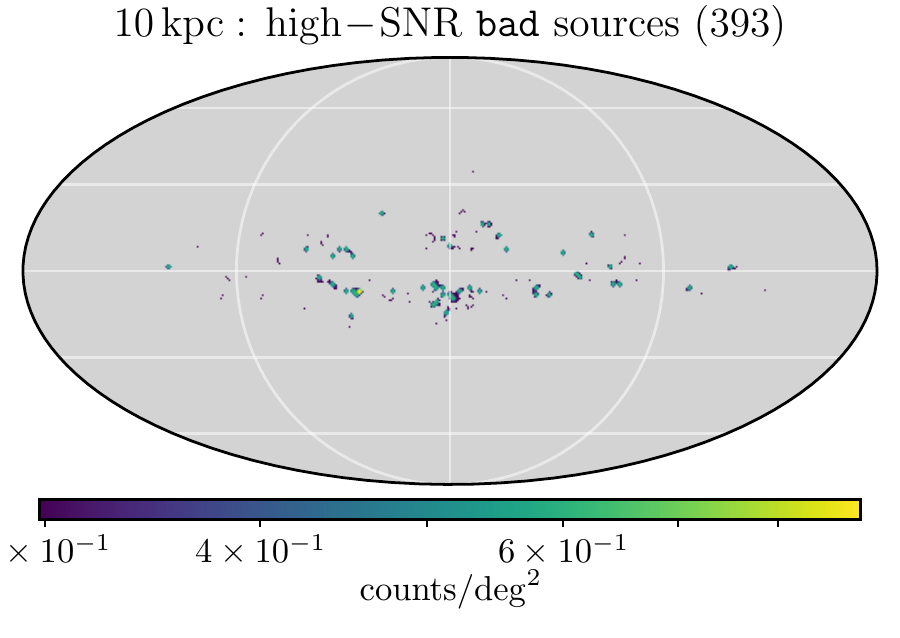}
	\caption{Sky distribution of high-SNR sources in GeDR3mock (left), in eDR3 classified as \texttt{good} (middle), and classified as \texttt{bad} (right column). We show the sky density of sources on a Mollweide projection of Galactic coordinates, with a logarithmic color scale (which is the same for the left and middle panels in each row). From top to bottom, the distance increases logarithmically from 100~pc to 10~kpc. We also list the number of sources in each panel, to illustrate the ratio of \texttt{good} to \texttt{bad} sources at the respective measured parallax and the number of expected sources according to GeDR3mock.}
	\label{fig:skydist_highsnr}
\end{figure*}

We begin by looking at the sky distributions of the high-SNR sources in Fig.~\ref{fig:skydist_highsnr}. The density is shown on a logarithmic color scale and the range of the left (GeDR3mock) and middle (eDR3 \texttt{good}) panel are set by the total range of the middle panel. In the middle panel we see that with increasing distance the structure of the Galactic disc is becoming more and more visible. At 10kpc the high-SNR sources peter out. Except for a higher normalization in the 100\,pc sample the GeDR3mock distribution on the left looks very similar to the eDR3 data\footnote{In the 100 and 300\,pc bin the open clusters are more pronounced in GeDR3mock than in the \texttt{good} sources. The reason is that known cluster's true sky positions were used while masses were assigned randomly when producing GeDR3mock clusters \citep{Rybizki2020-GeDR3mock}.}. The fraction of \texttt{bad} sources (right panel) decreases with increasing distance from 48\,\% at 100pc to 3\,\% at 3 and 10kpc. With increasing distance it also resembles less and less the distribution of \texttt{bad} training sources as shown in Fig.~\ref{fig:skydist_bad} and concentrates more towards the Galactic plane.

\begin{figure*}
	\includegraphics[width=0.29\linewidth]{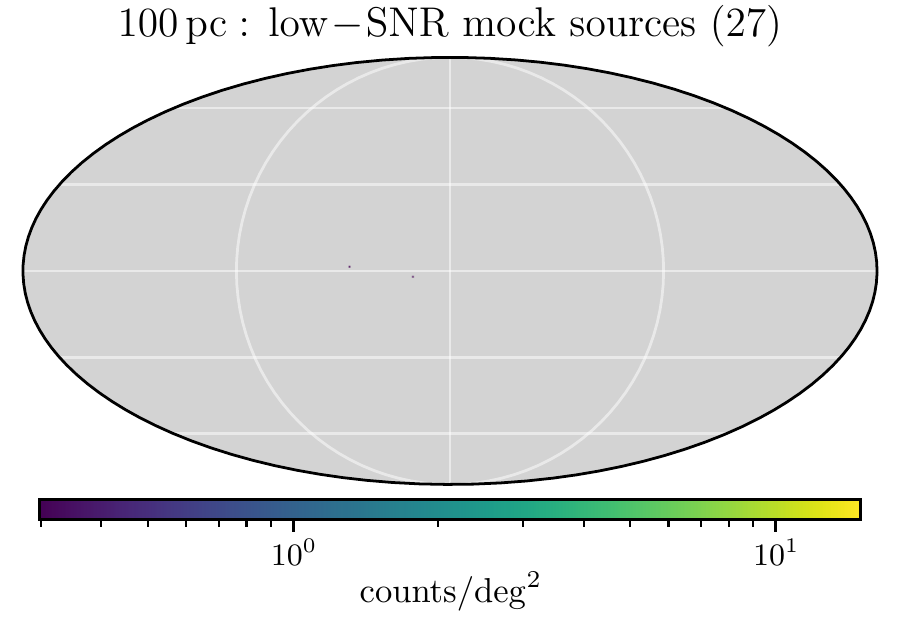}
	\hfill
	\includegraphics[width=0.29\linewidth]{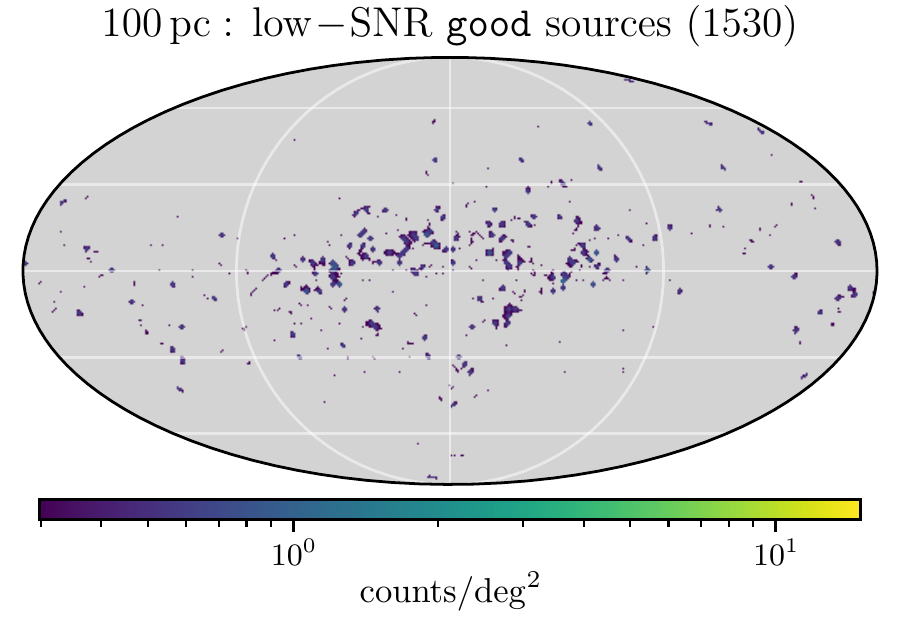}
	\hfill
	\includegraphics[width=0.29\linewidth]{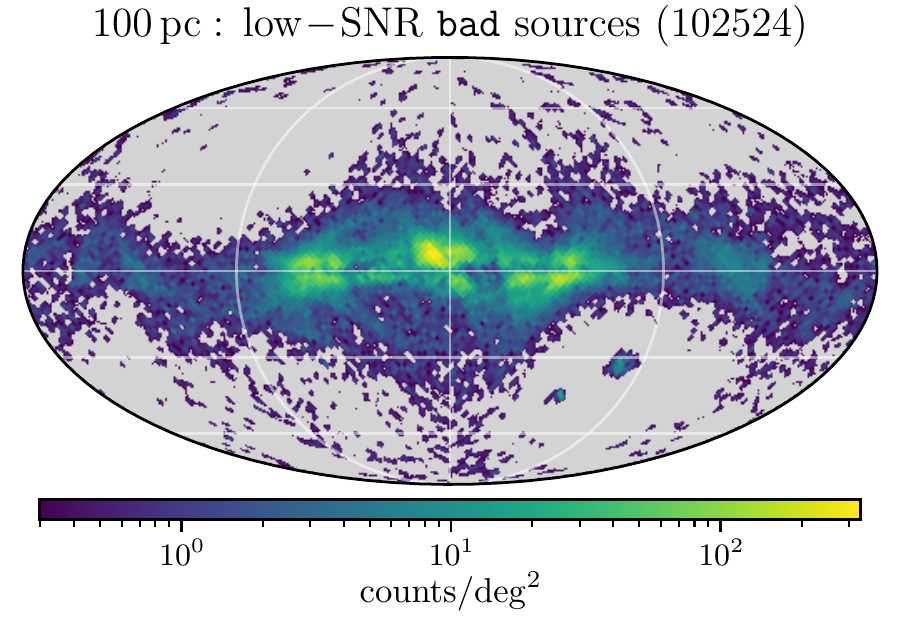}
	\includegraphics[width=0.29\linewidth]{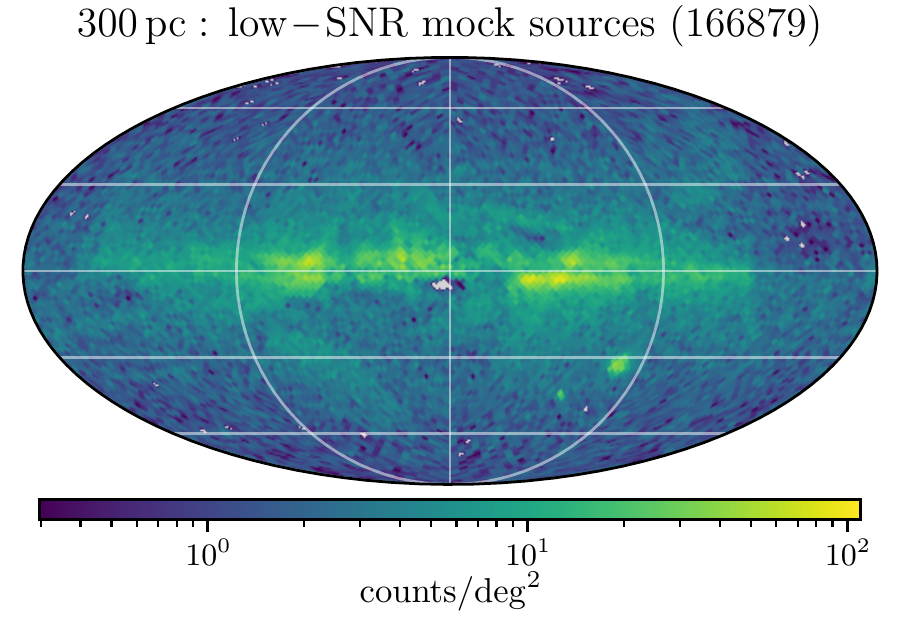}
	\hfill
	\includegraphics[width=0.29\linewidth]{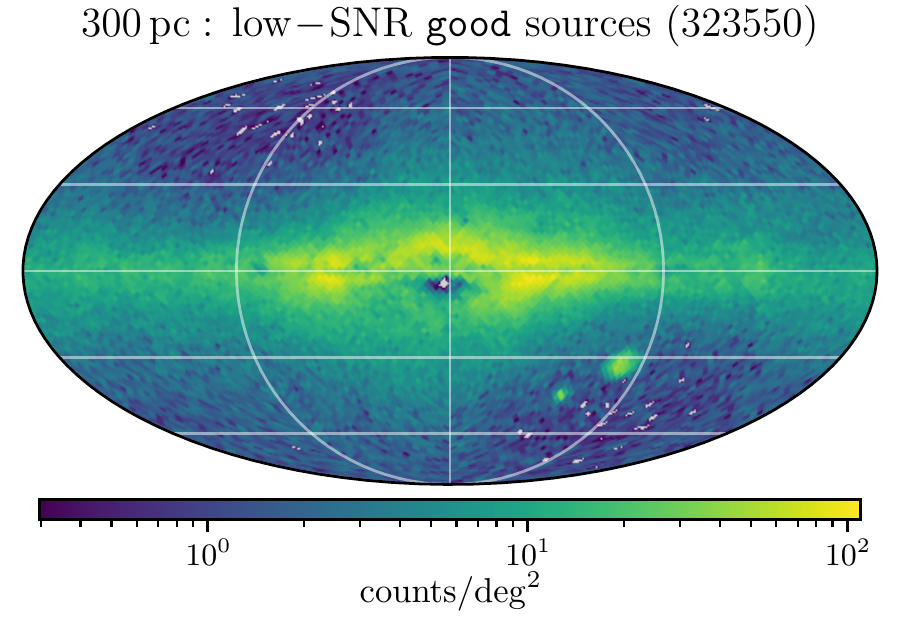}
	\hfill
	\includegraphics[width=0.29\linewidth]{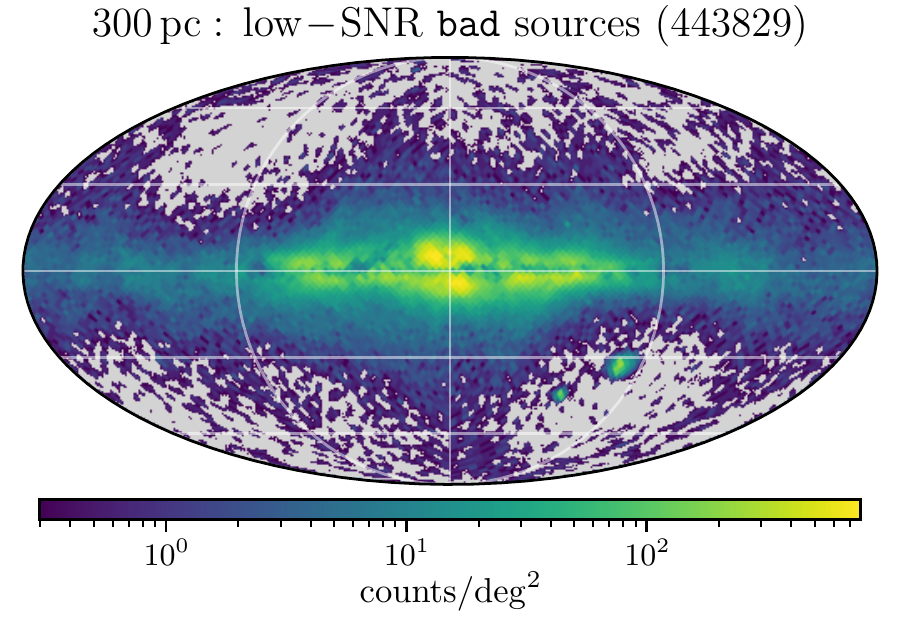}
	\includegraphics[width=0.29\linewidth]{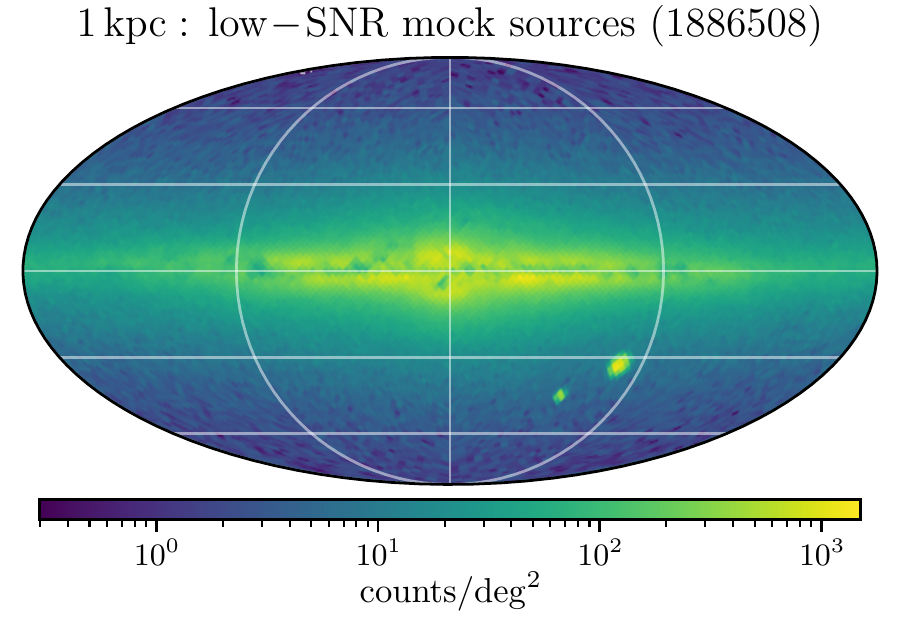}
	\hfill
	\includegraphics[width=0.29\linewidth]{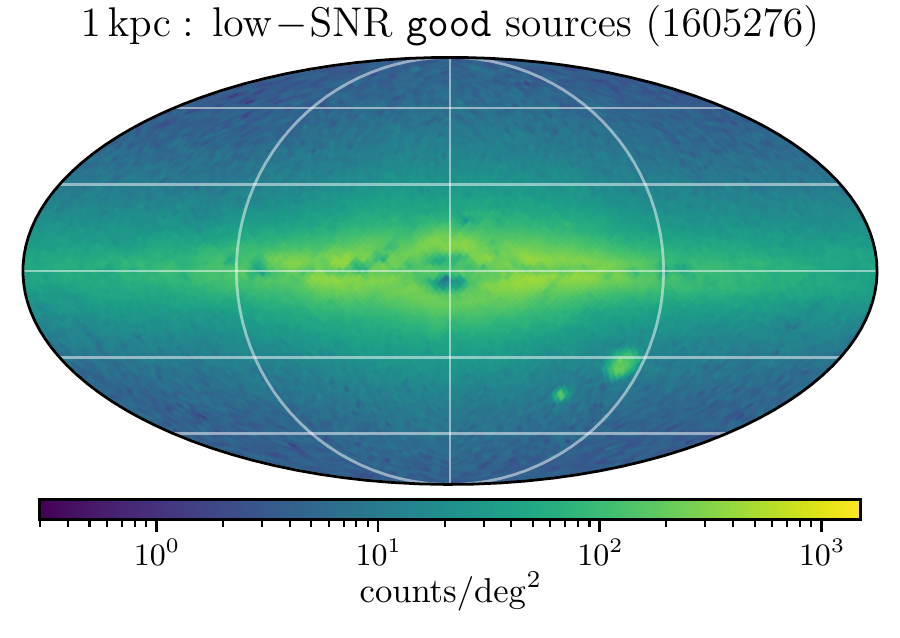}
	\hfill
	\includegraphics[width=0.29\linewidth]{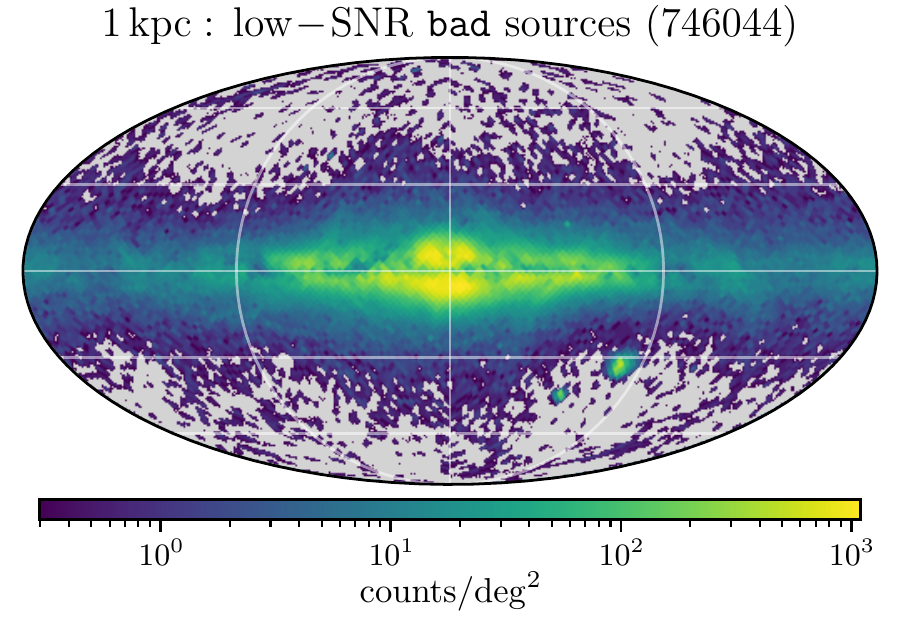}
	\includegraphics[width=0.29\linewidth]{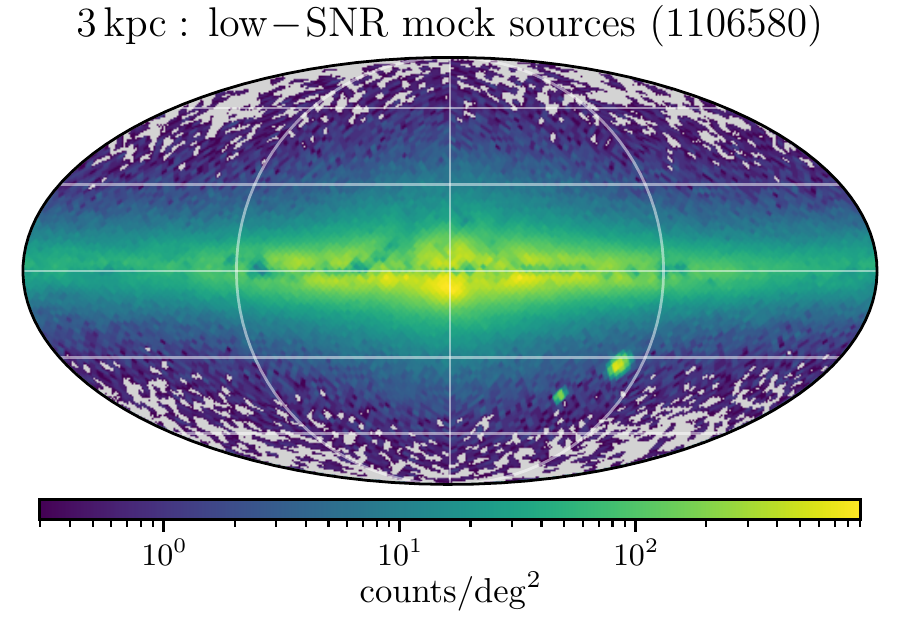}
	\hfill
	\includegraphics[width=0.29\linewidth]{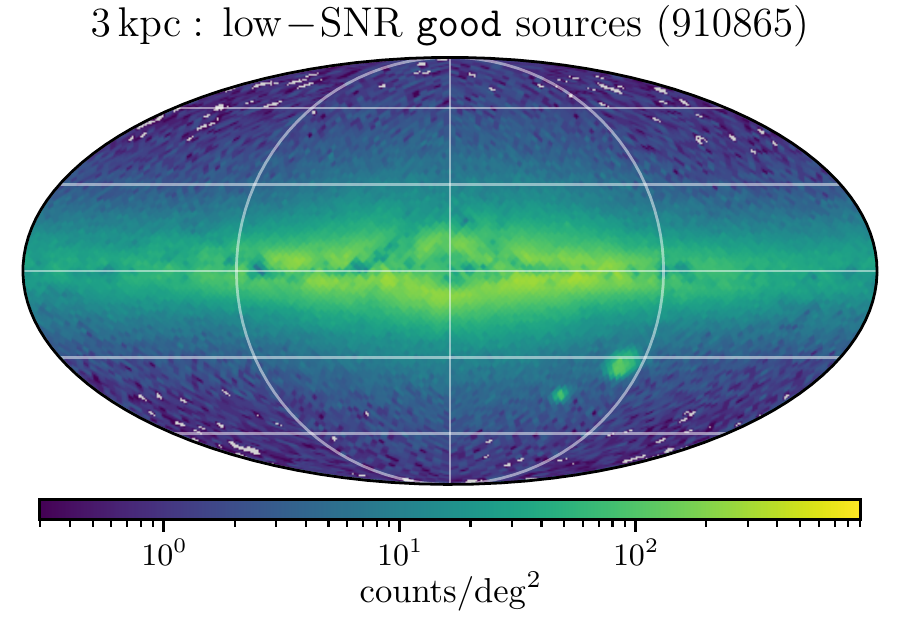}
	\hfill
	\includegraphics[width=0.29\linewidth]{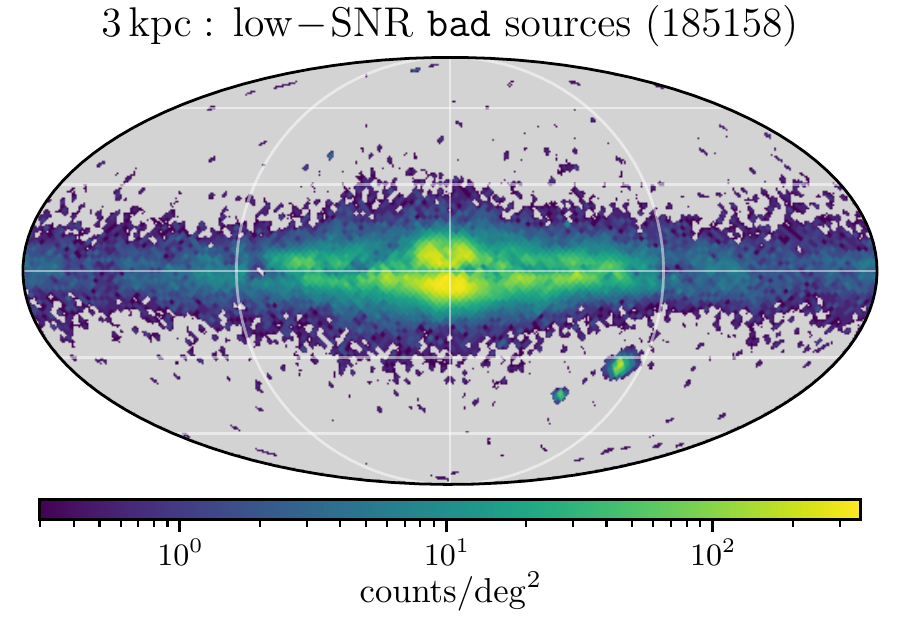}
	\includegraphics[width=0.29\linewidth]{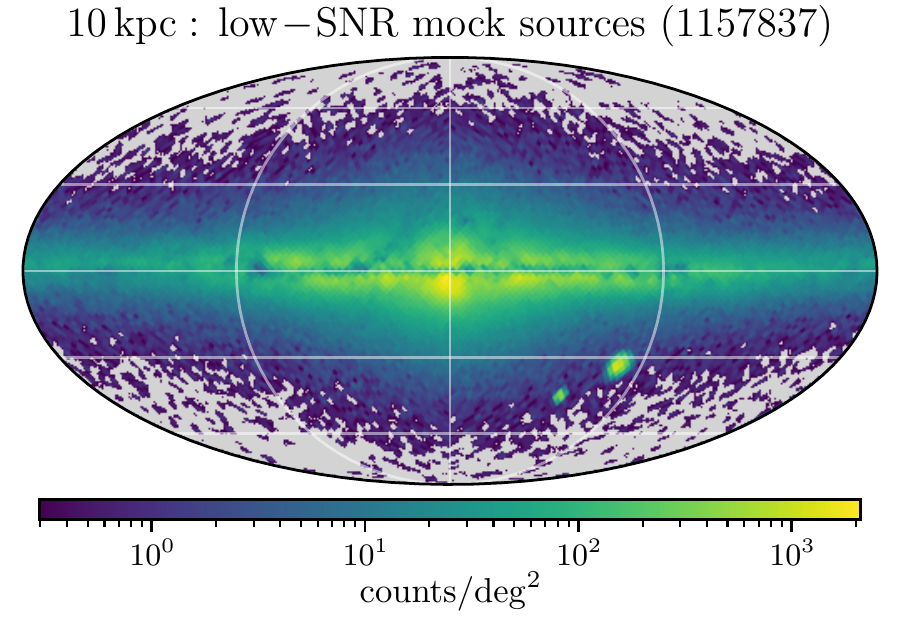}
	\hfill
	\includegraphics[width=0.29\linewidth]{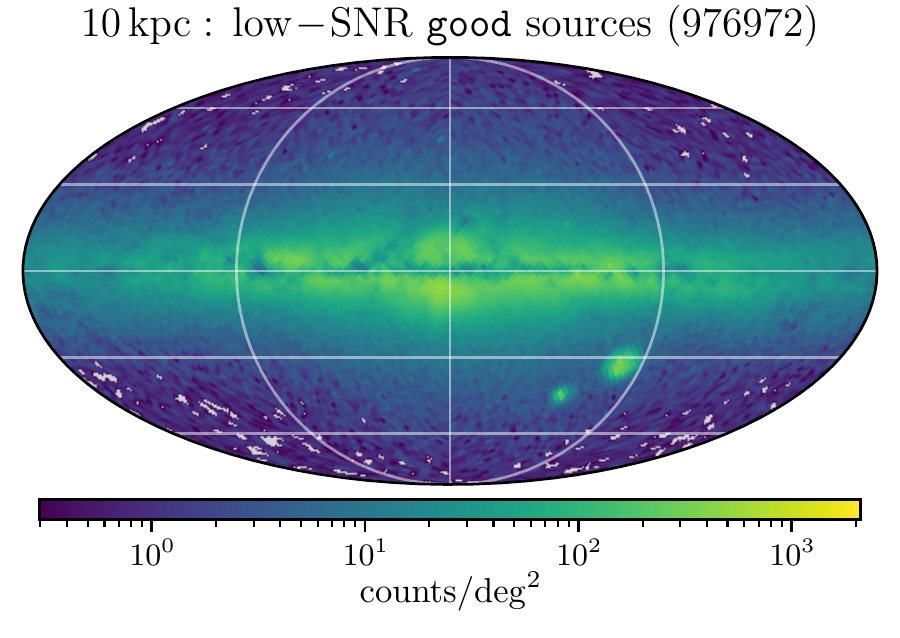}
	\hfill
	\includegraphics[width=0.29\linewidth]{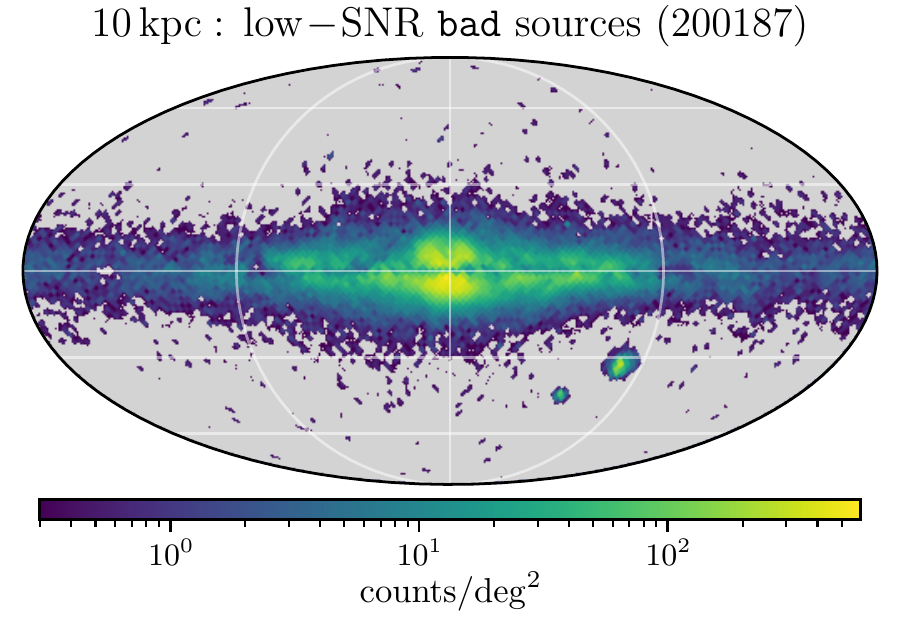}
	\includegraphics[width=0.29\linewidth]{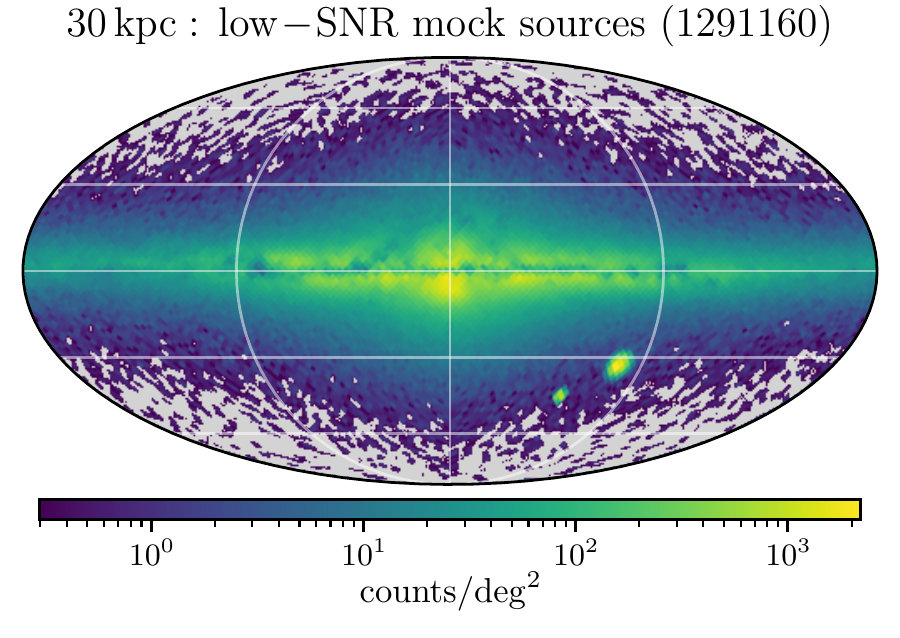}
	\hfill
	\includegraphics[width=0.29\linewidth]{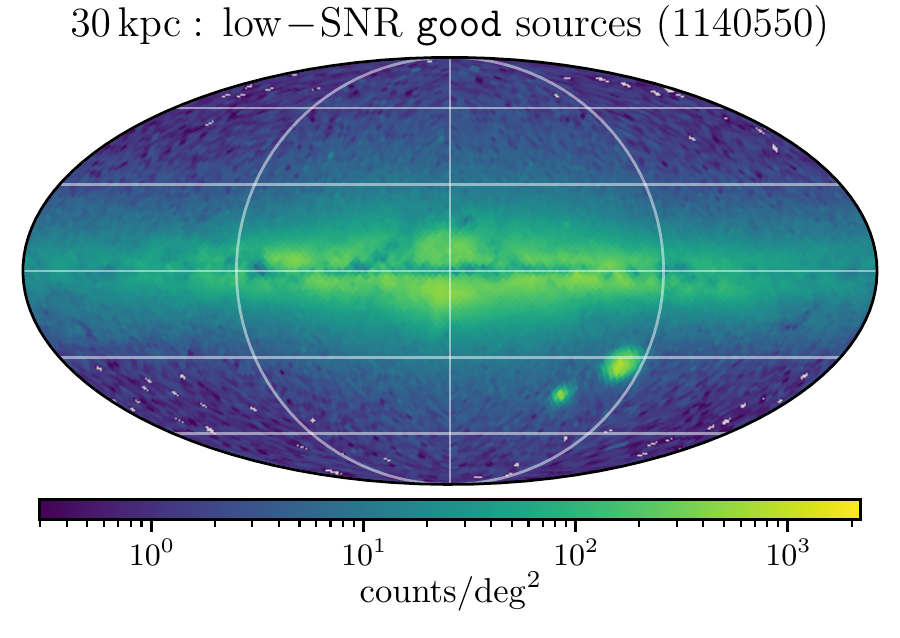}
	\hfill
	\includegraphics[width=0.29\linewidth]{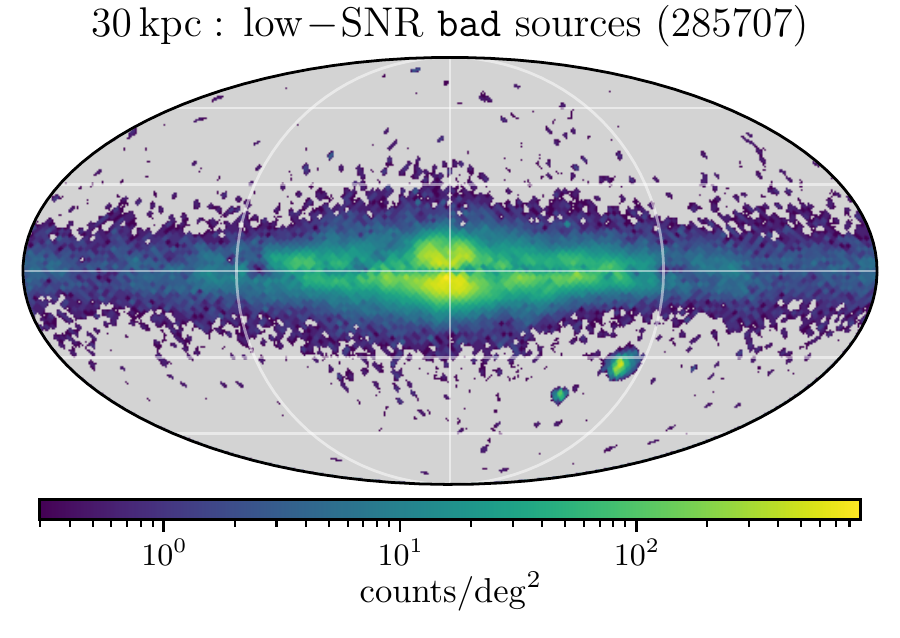}
	\caption{Sky distribution of low-SNR sources in GeDR3mock (left), in eDR3 classified as \texttt{good} (middle), and classified as \texttt{bad} (right column). We show the sky density of sources on a Mollweide projection of Galactic coordinates, with a logarithmic color scale (which is the same for the left and middle panels in each column). From top to bottom, the distance increases logarithmically from 100~pc to 30~kpc. We also list the number of sources in each panel, to illustrate the ratio of \texttt{good} to \texttt{bad} sources at the respective measured parallax and the number of expected sources according to GeDR3mock.}
	\label{fig:skydist_lowsnr}
\end{figure*}

For the low-SNR sources in Fig.~\ref{fig:skydist_lowsnr}, we expect (based on GeDR3mock) very few \texttt{good} sources in the 100~pc sample\footnote{Before adding the bad training sample from the GCNS comparison, as described in Section~\ref{sec:iterative-identification-spurious}, we would have more sources classified as \texttt{good} in the 100~pc sample.}. Interestingly, we already see the imprint of the central parts of the Galaxy and of the Magellanic clouds in the \texttt{good} sources in the 300~pc slice. This is expected: GeDR3mock also shows this imprint at 300~pc, due to poorly constrained (but valid) parallax measurements of sources that are, in reality, much farther away. In GeDR3mock, the visible scanning pattern is not well matched to that of eDR3. This is due to GeDR3mock uncertainties being trained on GDR2, and only roughly scaled to match eDR3 uncertainties. At 300~pc, the number of \texttt{good} sources we recover in eDR3 sources is almost twice as large as the number of GeDR3mock sources. This may be due to the true eDR3 uncertainties being larger than those modeled in GeDR3mock, scattering a greater number of distant sources into the 300~pc slice. The subsequent distance bins are then relatively similar in number and actual distribution. One prominent exception is the hole in the Galactic Center that is apparent to larger distances in the eDR3 data, and which results from the lower $G$-magnitude limit of the real survey, which is not fully reflected in the GeDR3mock $G$-magnitude limit approximation (90th percentile of the eDR3 $G$-magnitude distribution, see Appendix\,\ref{sec:gedr3mock_query}). The \texttt{bad} sources trace Galactic overdensities well, again with decreasing fractions for increasing distance bins. At 100~pc, the scanning law is still visible, but the sparsity is mainly due to most sources having a high SNR in this distance bin.

\subsubsection{CAMD}
\label{sec:camd_slices}
\begin{figure*}
	\includegraphics[width=0.25\linewidth]{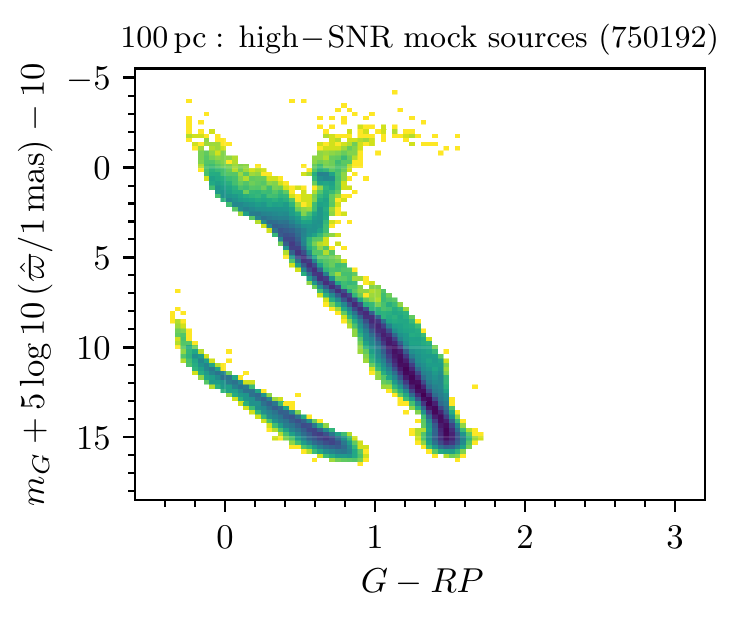}
	\hspace{0.02\linewidth}
	\includegraphics[width=0.25\linewidth]{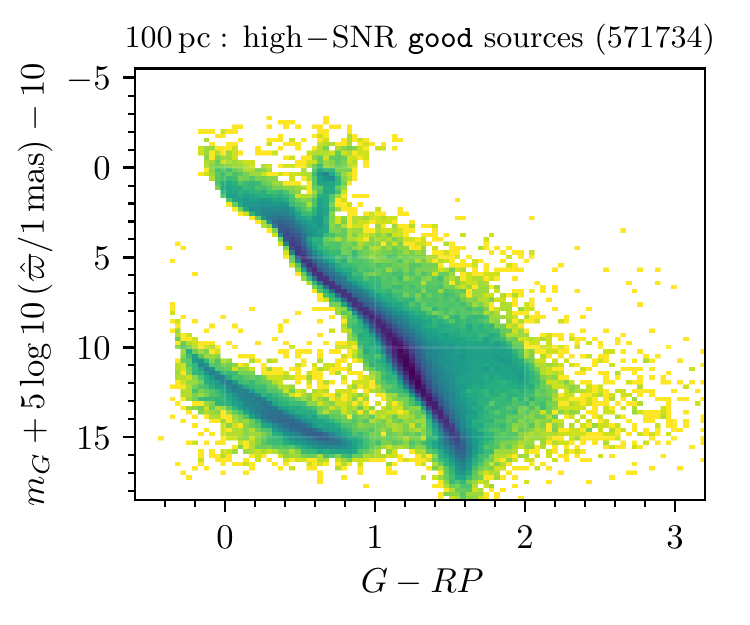}
	\hspace{0.02\linewidth}
	\includegraphics[width=0.25\linewidth]{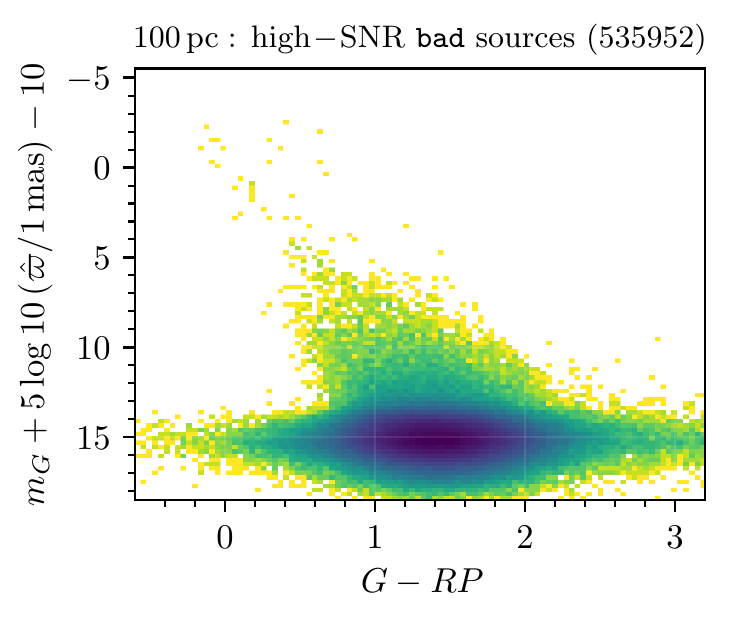}
	\includegraphics[width=0.25\linewidth]{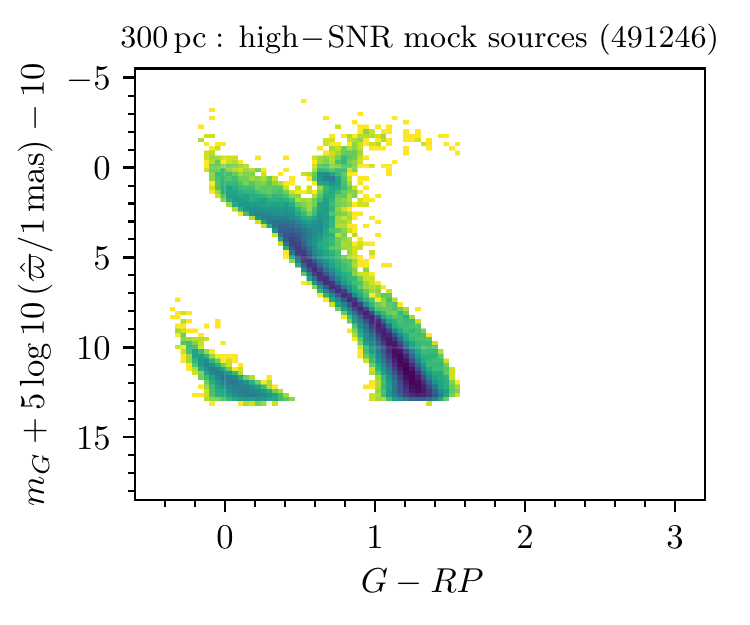}
	\hspace{0.02\linewidth}
	\includegraphics[width=0.25\linewidth]{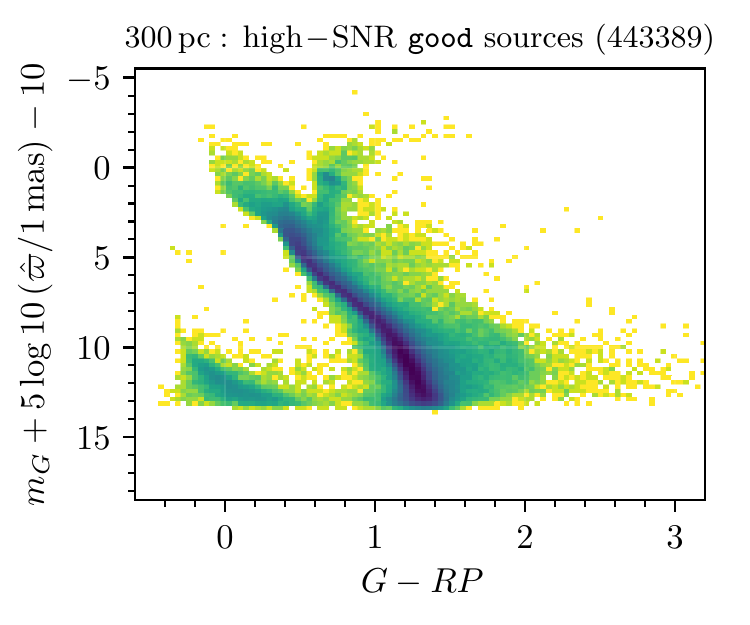}
	\hspace{0.02\linewidth}
	\includegraphics[width=0.25\linewidth]{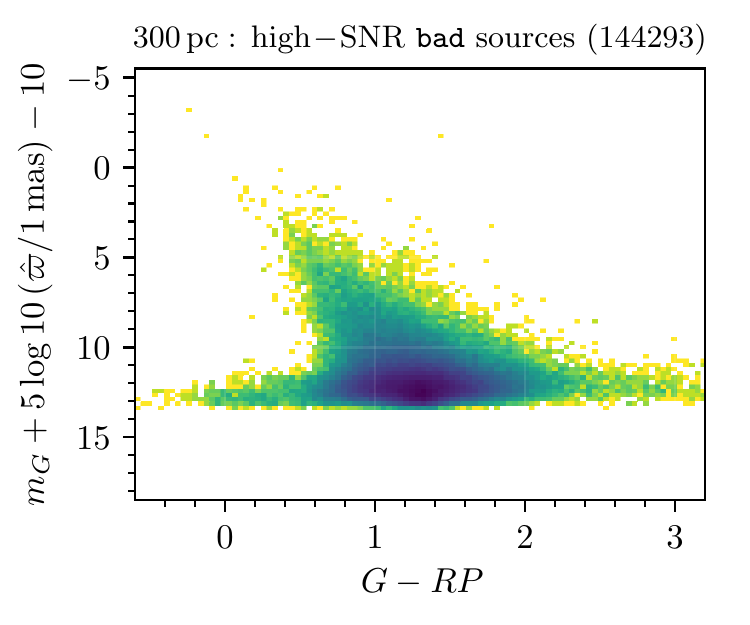}
	\includegraphics[width=0.25\linewidth]{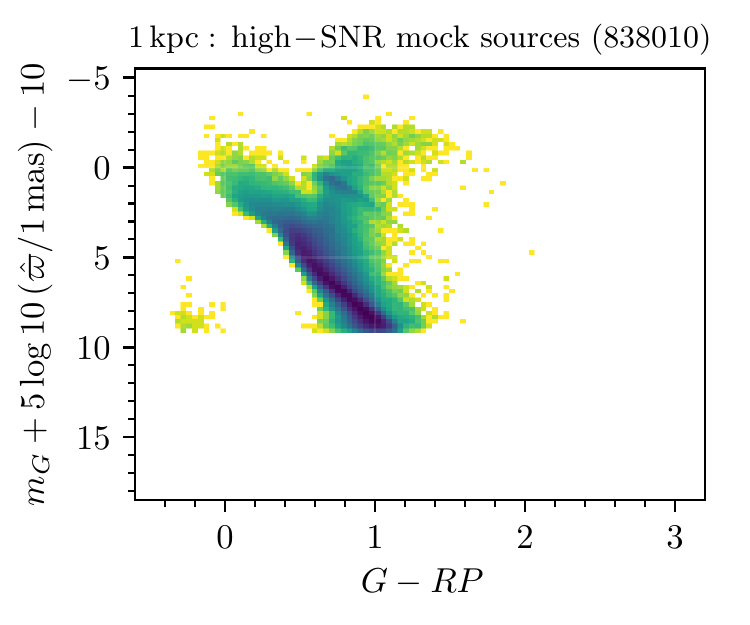}
	\hspace{0.02\linewidth}
	\includegraphics[width=0.25\linewidth]{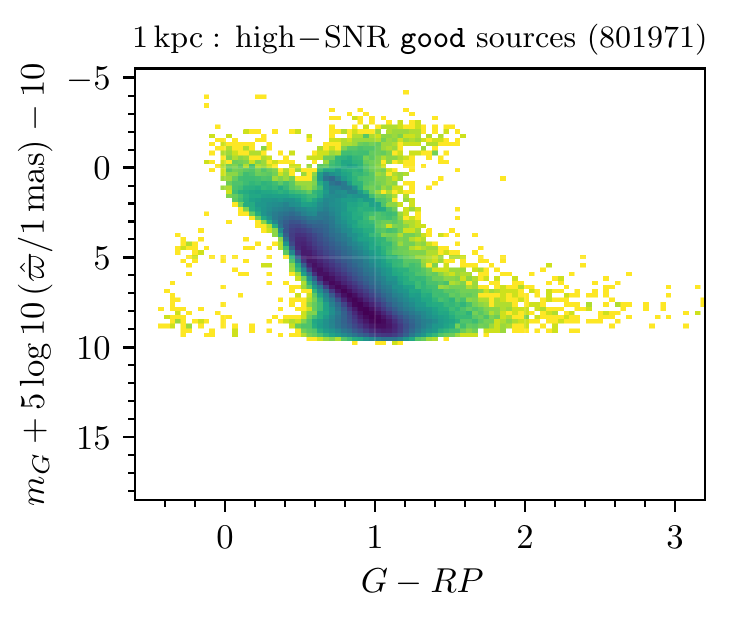}
	\hspace{0.02\linewidth}
	\includegraphics[width=0.25\linewidth]{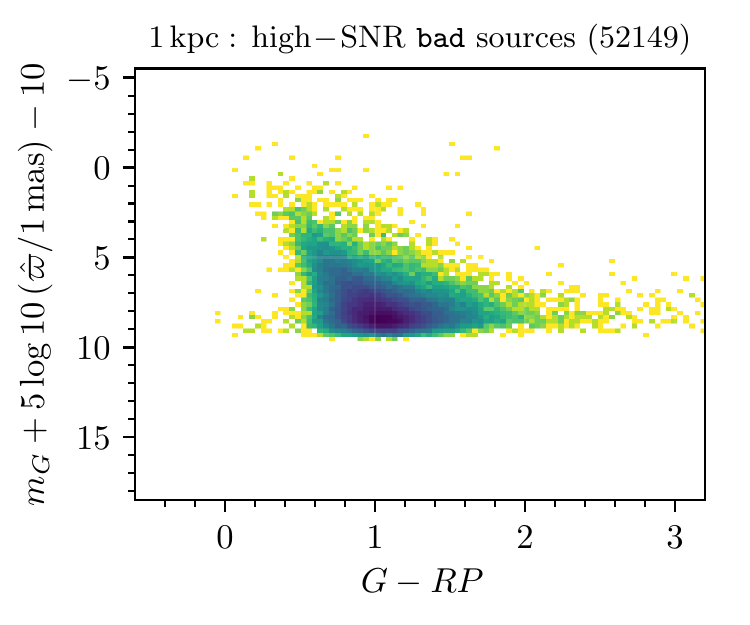}
	\includegraphics[width=0.25\linewidth]{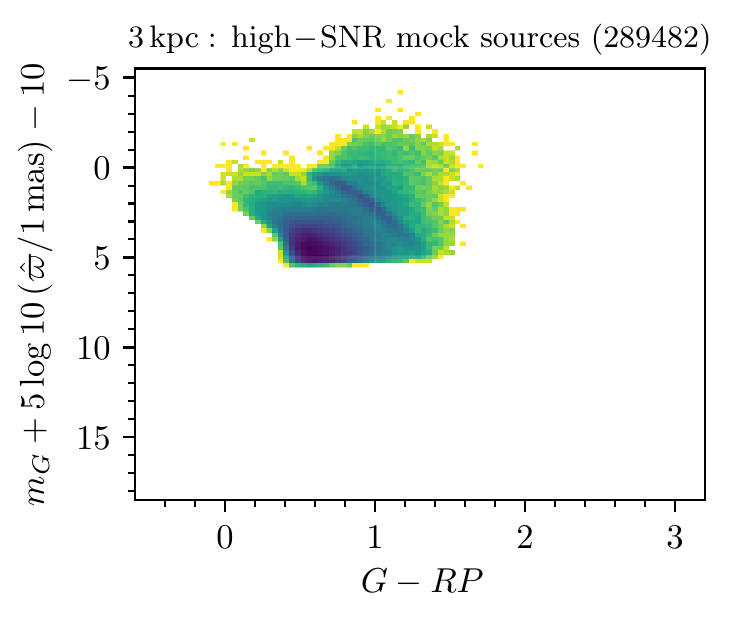}
	\hspace{0.02\linewidth}
	\includegraphics[width=0.25\linewidth]{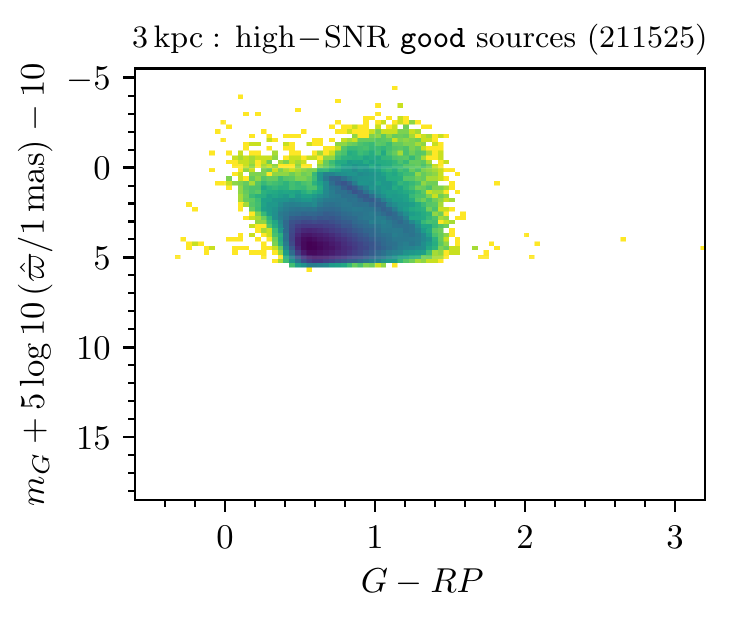}
	\hspace{0.02\linewidth}
	\includegraphics[width=0.25\linewidth]{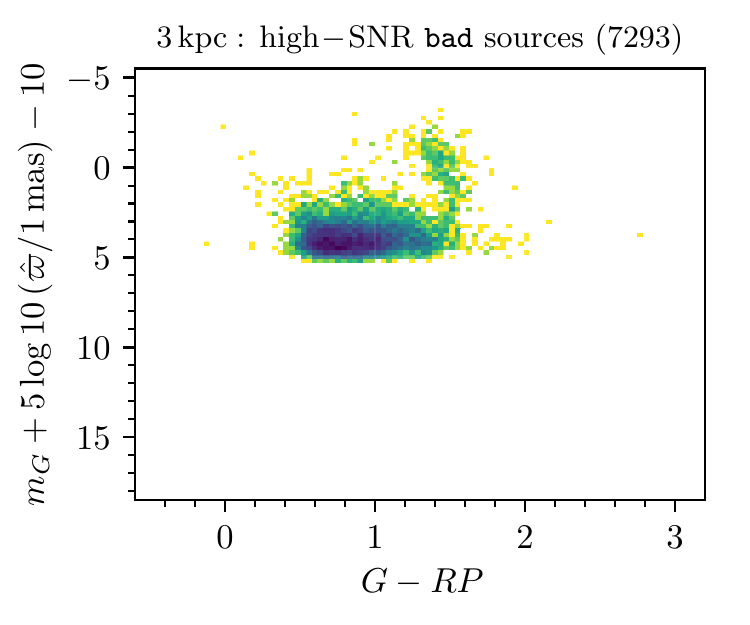}
	\includegraphics[width=0.25\linewidth]{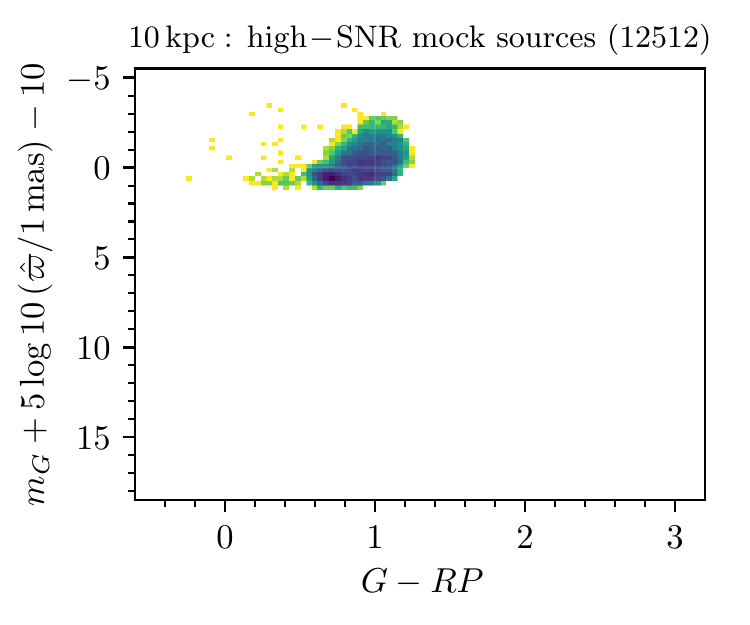}
	\hspace{0.02\linewidth}
	\includegraphics[width=0.25\linewidth]{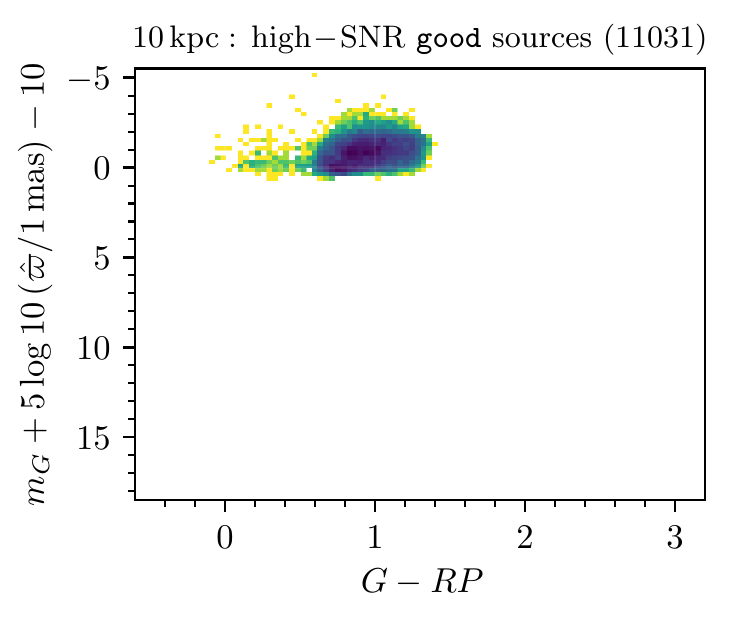}
	\hspace{0.02\linewidth}
	\includegraphics[width=0.25\linewidth]{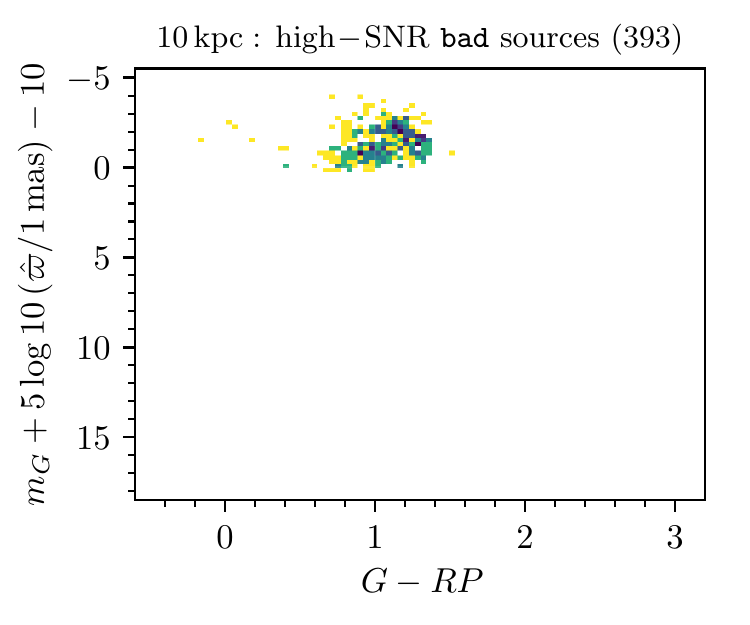}
	\caption{CAMD of high-SNR sources in GeDR3mock (left) in eDR3 classified as \texttt{good} (middle) and classified as \texttt{bad} (right column), on a logarithmic color scale. From top to bottom, the distance increases logarithmically from 100~pc to 10~kpc. We also list the number of sources, which illustrates the ratio of \texttt{good} to \texttt{bad} sources at the respective measured parallax, and the number of expected sources according to GeDR3mock.}
	\label{fig:camd_highsnr}
\end{figure*}

Now we turn at the CAMDs of the high-SNR sources in Fig.~\ref{fig:camd_highsnr}. The density is shown on a logarithmic color scale. In the 100~pc sample, a prominent difference between eDR3 \texttt{good} sources and GeDR3mock is the lack of WD-MS binaries and sources with photometric excess noise in the latter. However, in GeDR3mock, we see a spreading of the photometry in the brown dwarf regime, which comes from the photometric uncertainties, and which is also visible in the real data. The \texttt{bad} sources have an overdensity at $M_\mathrm{G} \approx 15$, which is due to the fact that the number of \texttt{bad} sources increases at smaller parallaxes, as well as the fact that most spurious astrometric solutions are for faint sources. There is also a triangle pointing into the direction of the MS, which may still contain a few good sources. When going further out in distance, the limiting absolute magnitude decreases and more and more extinction sets in. Increasing extinction is apparent in the elongation of the red clump with increasing distance. At 10~kpc, only the brightest stars can be seen in the high-SNR sample. In the \texttt{bad} sample, an interesting ``fountain'' pattern arises at 3~kpc, which might be an indication of misclassified red AGB stars.

\begin{figure*}
	\includegraphics[width=0.25\linewidth]{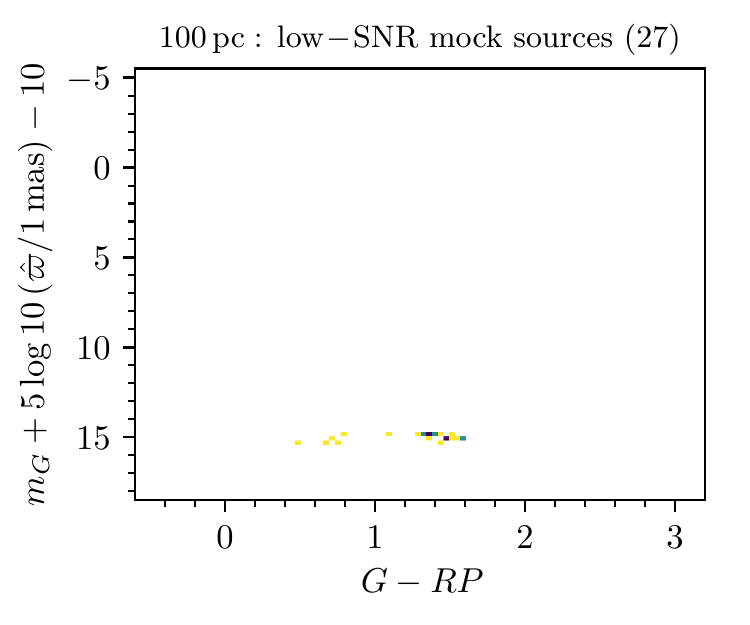}
	\hspace{0.02\linewidth}
	\includegraphics[width=0.25\linewidth]{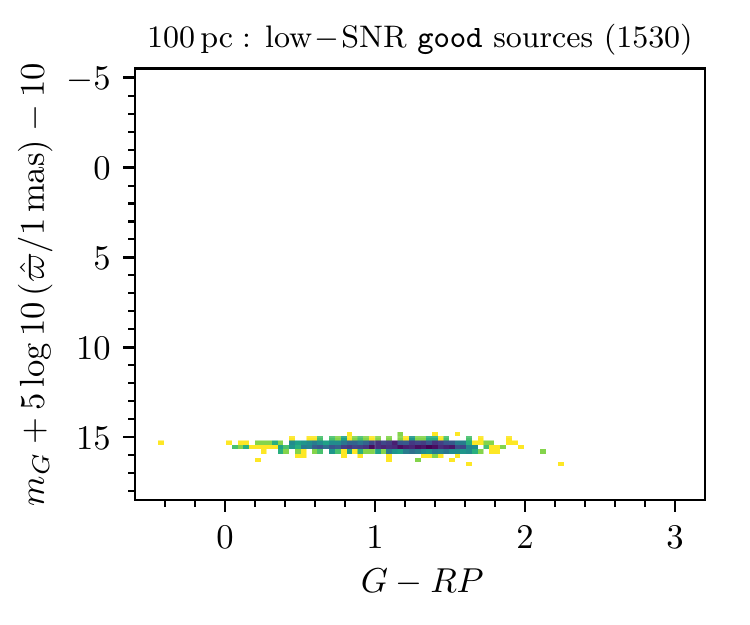}
	\hspace{0.02\linewidth}
	\includegraphics[width=0.25\linewidth]{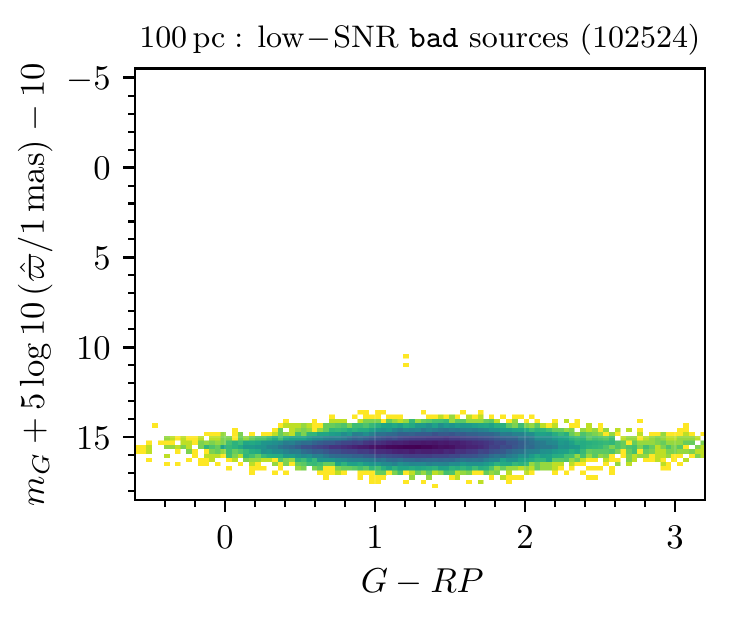}
	\includegraphics[width=0.25\linewidth]{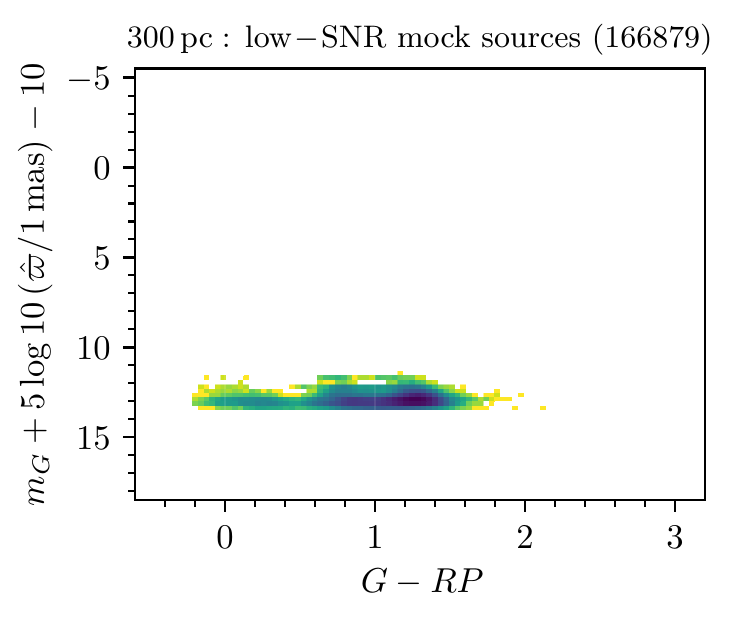}
	\hspace{0.02\linewidth}
	\includegraphics[width=0.25\linewidth]{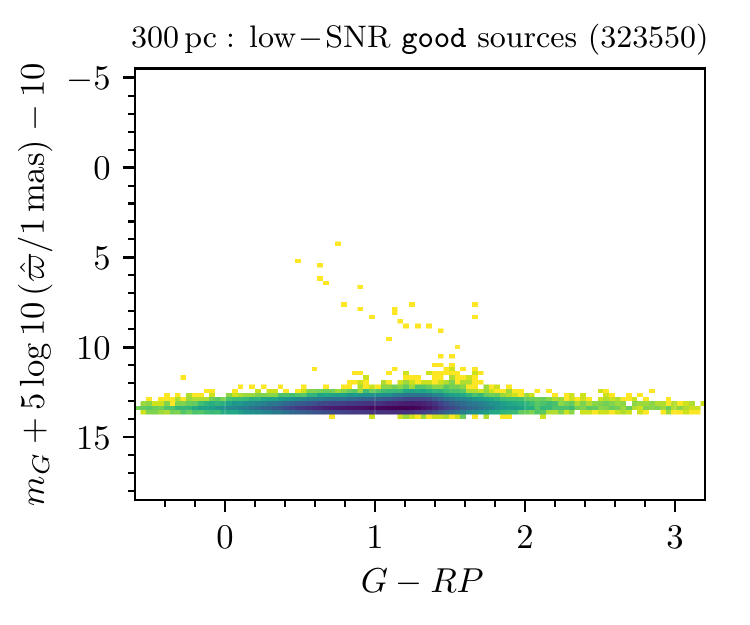}
	\hspace{0.02\linewidth}
	\includegraphics[width=0.25\linewidth]{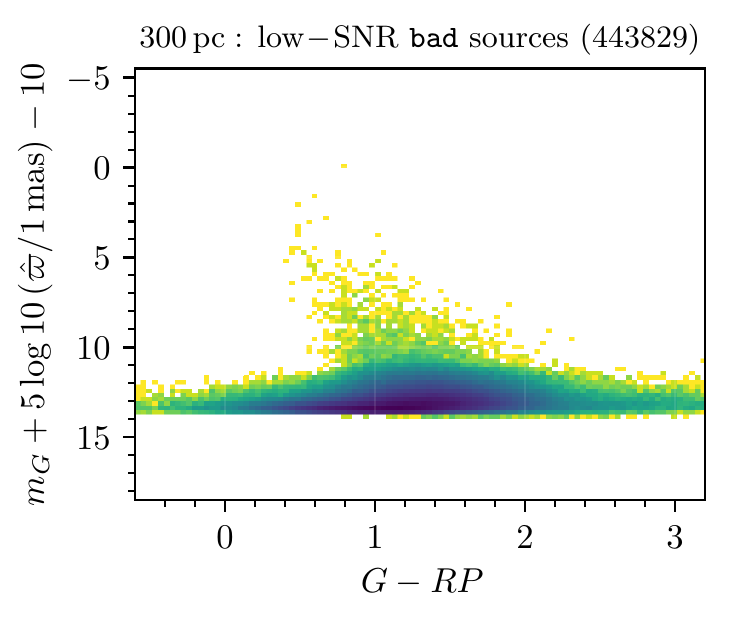}
	\includegraphics[width=0.25\linewidth]{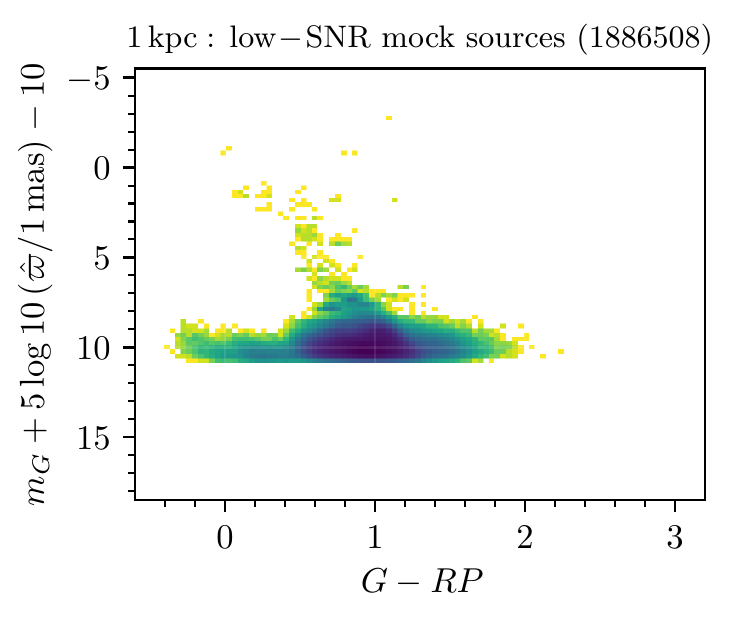}
	\hspace{0.02\linewidth}
	\includegraphics[width=0.25\linewidth]{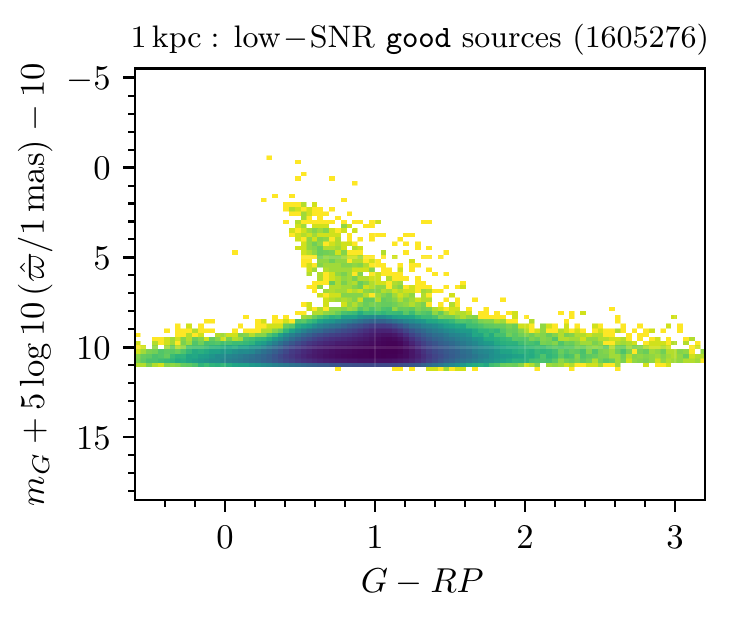}
	\hspace{0.02\linewidth}
	\includegraphics[width=0.25\linewidth]{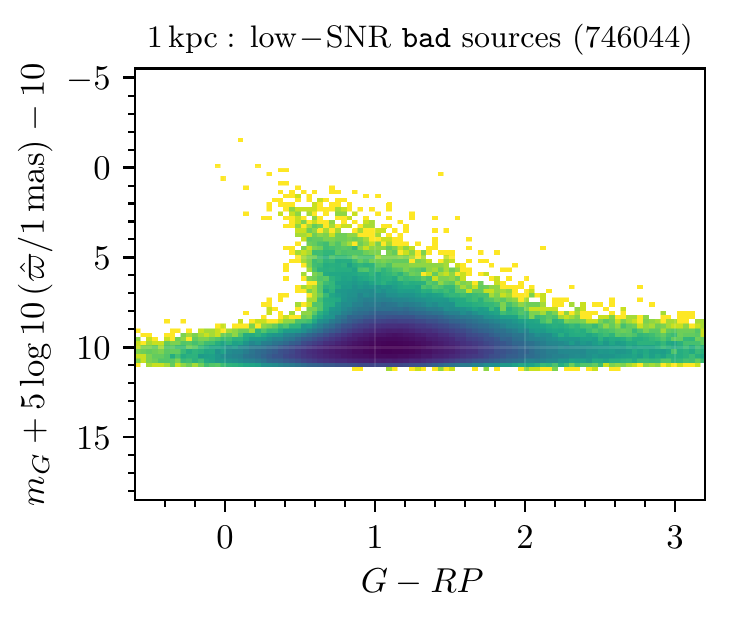}
	\includegraphics[width=0.25\linewidth]{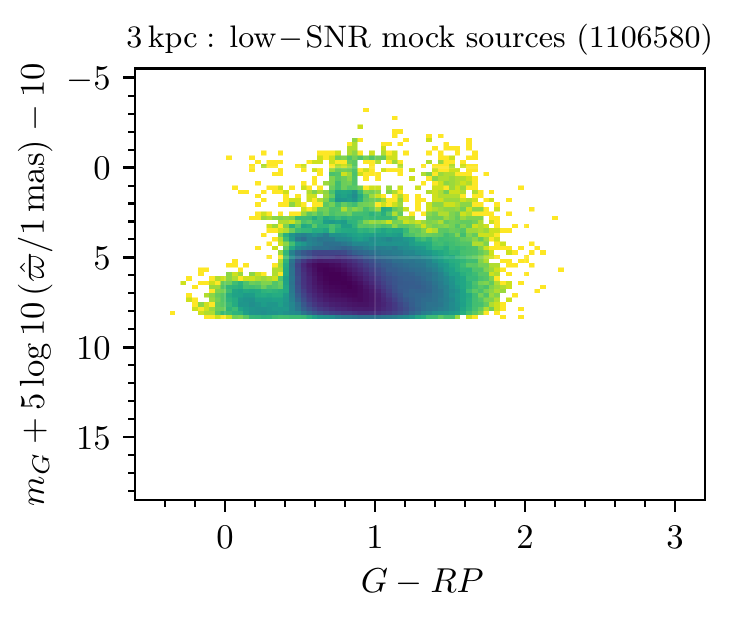}
	\hspace{0.02\linewidth}
	\includegraphics[width=0.25\linewidth]{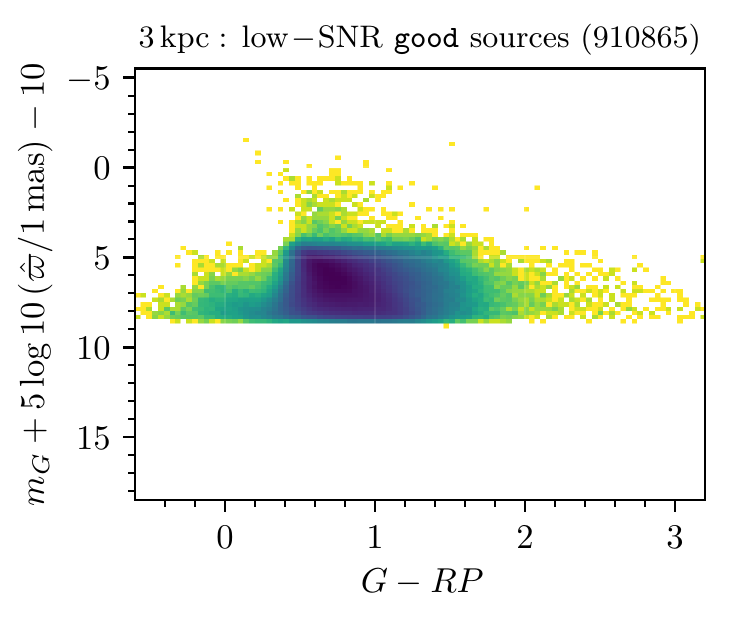}
	\hspace{0.02\linewidth}
	\includegraphics[width=0.25\linewidth]{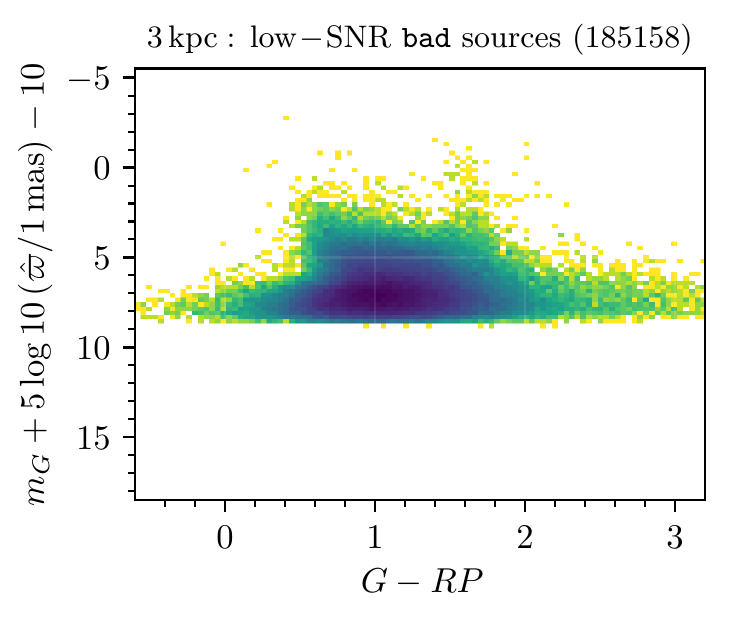}
	\includegraphics[width=0.25\linewidth]{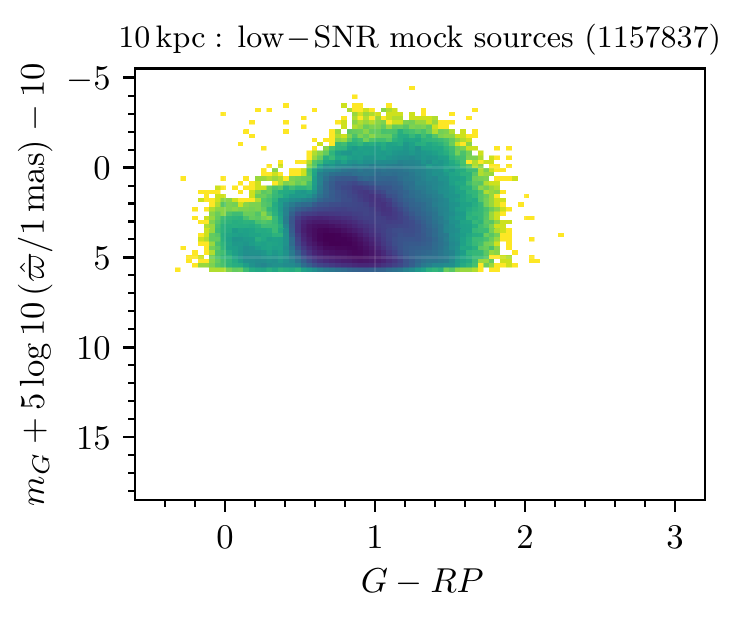}
	\hspace{0.02\linewidth}
	\includegraphics[width=0.25\linewidth]{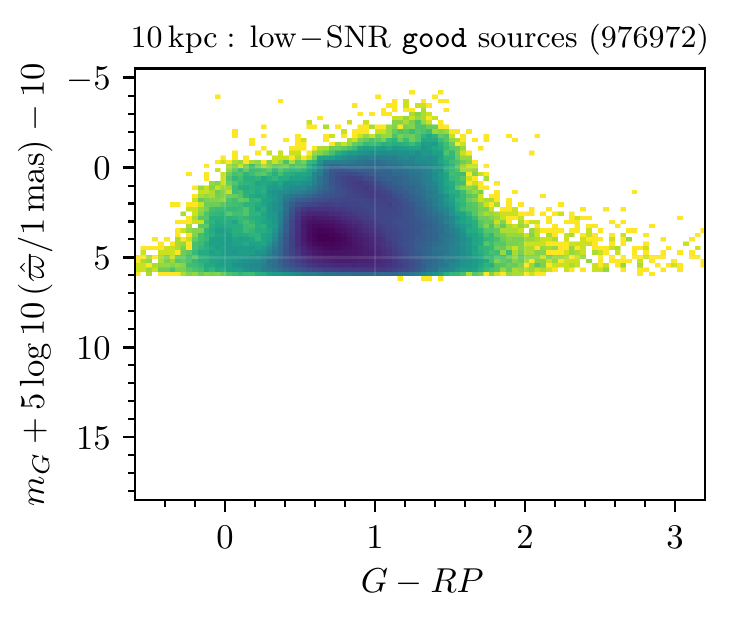}
	\hspace{0.02\linewidth}
	\includegraphics[width=0.25\linewidth]{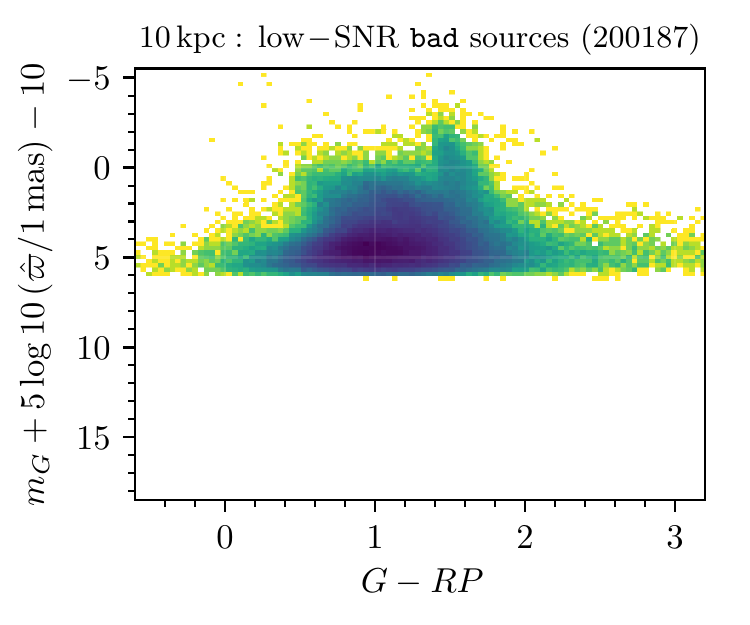}
	\includegraphics[width=0.25\linewidth]{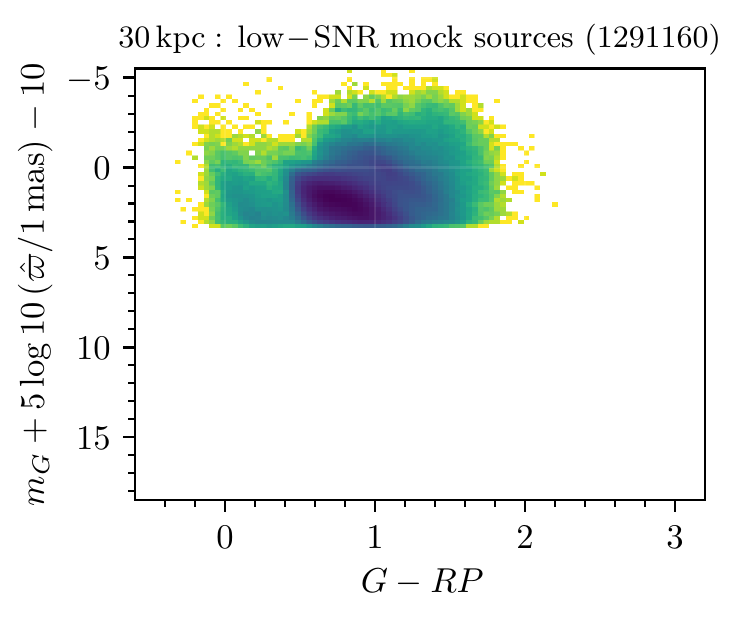}
	\hspace{0.02\linewidth}
	\includegraphics[width=0.25\linewidth]{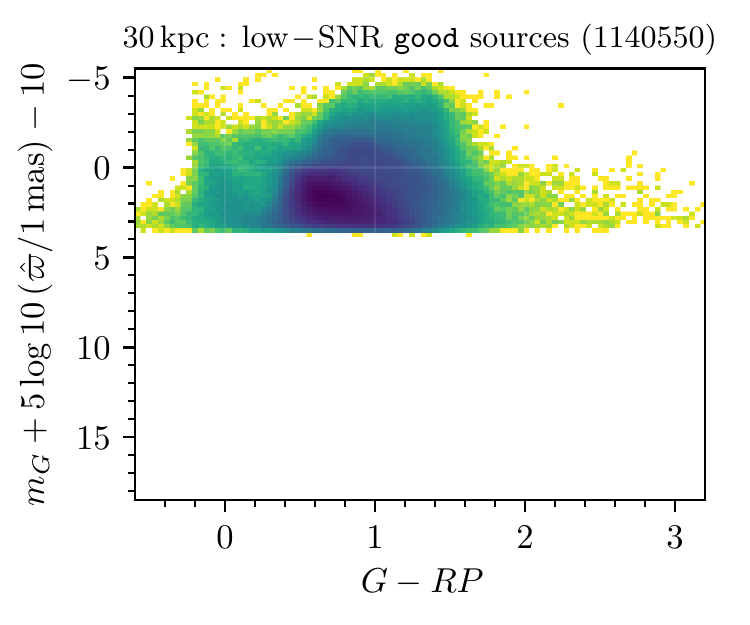}
	\hspace{0.02\linewidth}
	\includegraphics[width=0.25\linewidth]{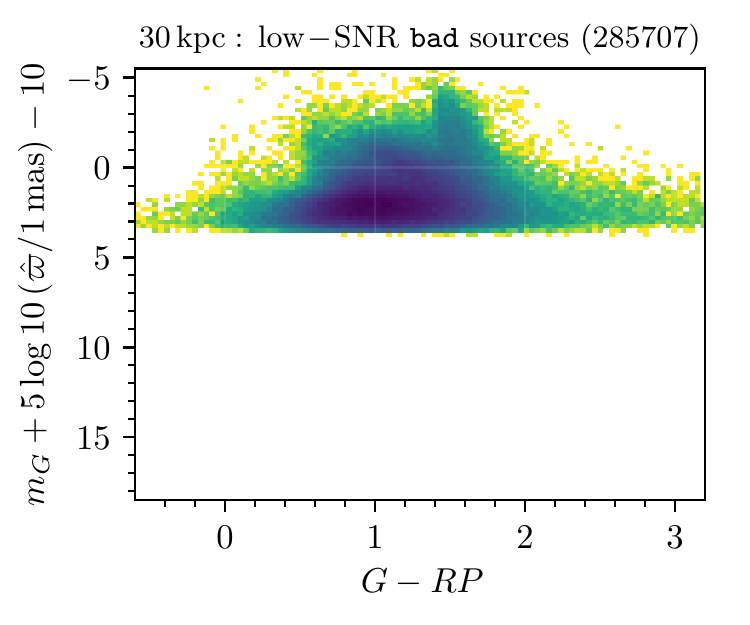}
	\caption{CAMD of low-SNR sources in GeDR3mock (left) in eDR3 classified as \texttt{good} (middle) and classified as \texttt{bad} (right column), on a logarithmic color scale. From top to bottom, the distance increases logarithmically from 100~pc to 30~kpc. We also list the number of sources, which illustrates the ratio of \texttt{good} to \texttt{bad} sources at the respective measured parallax, and the number of expected sources according to GeDR3mock.}
	\label{fig:camd_lowsnr}
\end{figure*}

Looking at the low-SNR CAMDs in Fig.~\ref{fig:camd_lowsnr}, we see that in the first distance bins we expect and see only very faint objects contributing and their $G-RP$ color is smeared out due to the high photometric uncertainty in that regime. Compared to GeDR3mock, the eDR3 data has even larger tails into the red and blue. Otherwise, the \texttt{good} sources closely mimick the GeDR3mock behavior, with a stronger qualitative deviation at 3~kpc, where the ``fountain'' pattern of red AGB stars is visible in GeDR3mock and in the \texttt{bad} sources. This might indicate a misclassification in that regime, as was the case for the high-SNR sources. 
At 10~kpc in the \texttt{good} sources, we can see massive main sequence stars and turn-off stars, as well as the red clump and perhaps a few sdB stars at the very blue end. For the bad sources, a weak signal of all those populations seems to be present, and again, some red AGB stars might have been wrongly predicted as \texttt{bad}. Overall, most of the physical structure can be seen in the \texttt{good} sample, making us confident that even at low-SNR our astrometric fidelity classifier is useful. Nevertheless, the ``fountain'' pattern might be a good candidate for further retraining, similar to Section~\ref{sec:iterative-identification-spurious}.

\section{Cleaning CAMDs of sources with spurious colors}
\label{sec:cmd_fidelity}
We have thus far focused on deriving a purely astrometric fidelity estimator. Here, we generalize this goal somewhat to estimating the reliability of a source's position in the {\it Gaia} CAMD. 

The fraction of sources with spurious astrometry and/or photometry is likely particularly high in regions of the CAMD that are expected to be sparsely populated astrophysically, such as the region between the main sequence (MS) and WD cooling tracks. Sources in this region of the CAMD are often of particular interest precisely because they are rare. When contaminants with poor astrometry and/or photometry can be efficiently filtered out, sources below the MS and brighter than normal WDs of similar color represent a mix of unresolved WD+MS binaries \citep[][]{Rebassa-Mansergas2021}, cataclysmic variables \citep[e.g.][]{Abrahams2020}, extremely low-mass WDs \citep[e.g.][]{Pelisoli2019}, and related transitional binary evolution products \citep[e.g.][]{El-Badry2021, El-Badry2021b}. It is thus desirable to construct a (relatively) clean sample of objects that genuinely fall in this part of color-absolute magnitude space, separating them from normal MS contaminants that scatter into it due to spurious absolute magnitudes and/or colors. 

\begin{figure*}
    \centering
    \includegraphics[width=\textwidth]{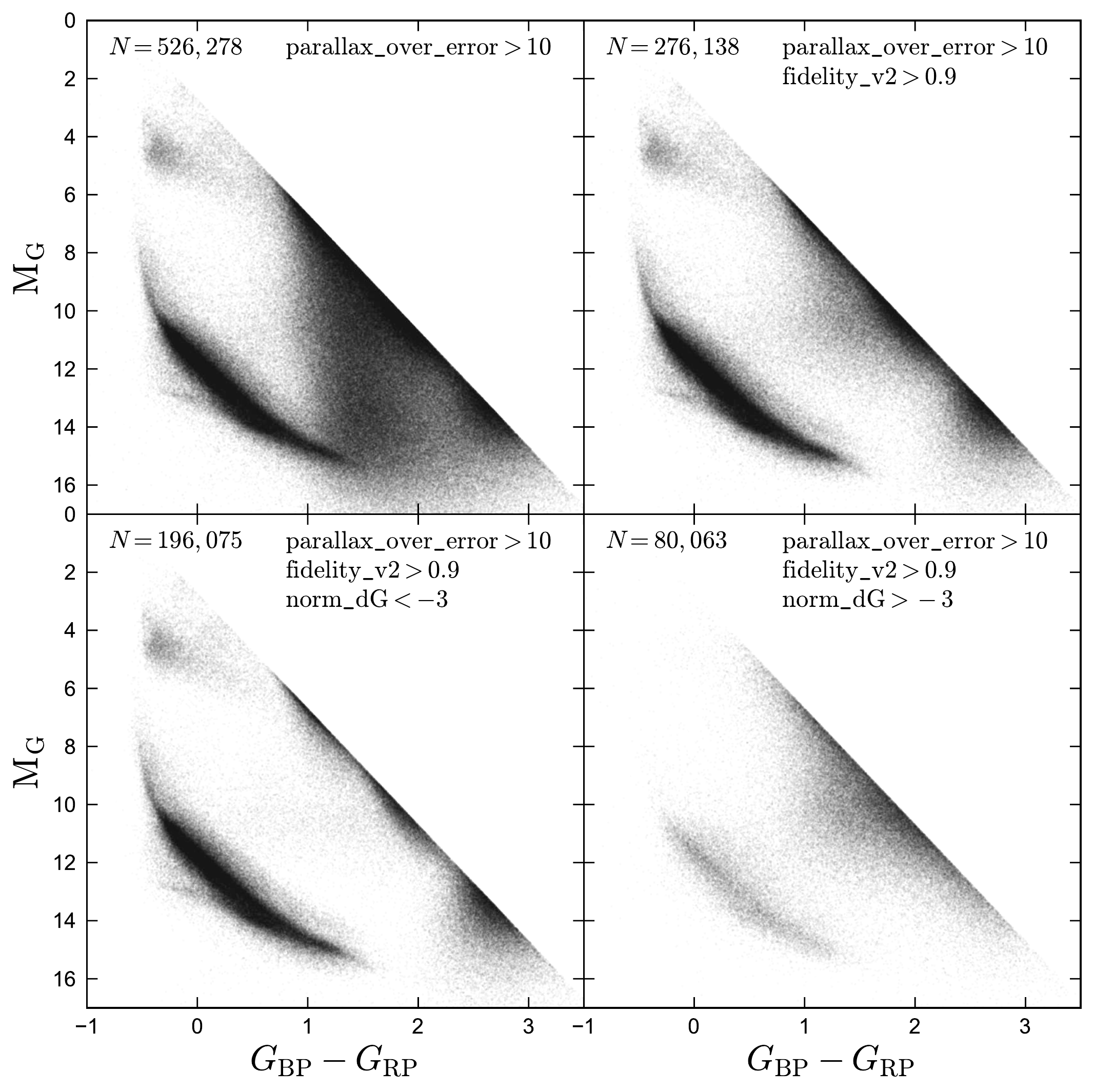}
    \caption{Cleaning of the CAMD using cuts on astrometric fidelity and photometric contamination. Upper left panel shows all sources in {\it Gaia} eDR3 with \texttt{parallax\_over\_error} > 10 and reported colors and magnitudes that fall well below the main sequence. Upper right panel shows the same sample after sources with \texttt{fidelity\_v2} < 0.9 have been removed. This eliminates a majority of sources between the main sequence and white dwarf cooling track. Lower left shows the sample after an additional cut of \texttt{norm\_dG} < -3 (Equation~\ref{eq:delta_Gi}), which filters out objects with potentially contaminated photometry. A majority of the surviving sources are on the MS or WD cooling track, and the density of objects between them is significantly reduced. Lower right shows the sources with good astrometric fidelity but potentially contaminated colors; most of these fall between the MS and WD cooling tracks and likely have unreliable colors. }
    \label{fig:cmd_below_ms}
\end{figure*}

Fig.~\ref{fig:cmd_below_ms} illustrates how the astrometric fidelity and color contamination parameters can be used to clean such a sample. We begin by selecting from {\it Gaia} eDR3 all sources with precise reported parallaxes that fall below the main sequence:
\begin{lstlisting}
SELECT * from gaiaedr3.gaia_source 
WHERE parallax_over_error > 10 AND 
phot_g_mean_mag + 5 * log10(parallax / 100) > (4*bp_rp + 2.7) 
\end{lstlisting}

Here we crudely define the line $M_{G} = 4\times \left(BP - RP\right ) + 2.7$ as the lower boundary of the main sequence. This query yields 526,278 sources, which are shown in the upper left panel of Fig.~\ref{fig:cmd_below_ms}. Almost half of the selected sources fall squarely between the MS and WD cooling track, but we expect that at this stage, a large majority of these sources are spurious. The upper right panel shows the result of applying a cut \texttt{fidelity\_v2} > 0.9. This reduces the total number of objects by  factor of $\sim$2 and preferentially removes sources farther away from the main sequence, significantly reducing the number of sources likely to be spurious. 

We inspected Pan-STARRS~1 images of a random subset of the remaining sources and found that a large majority of them are faint sources that are within $\lesssim 10$ arcsec of a much brighter source. Given that $BP-RP$ colors are calculated from spectra dispersed over a $2\times 3$ arcsec region \citep[][]{2021A&A...649A...3R}, one expects such sources to have unreliable colors. Their parallaxes may also be less reliable than normal due to centroiding errors, though this has only been found to be severe for sources within $\lesssim 4$ arcsec of a brighter source \citep[][]{El-Badry2021c}.

We experimented with a variety of cuts to filter out sources with unreliable colors. One possible approach is to remove all sources with a neighbor that is brighter than the source itself within, say, 5 arcsec (e.g., \texttt{dist\_nearest\_neighbor\_at\_least\_0\_brighter} > 5). Such a cut would be overly aggressive for neighbors only slightly brighter than the source itself, which are not expected to cause significant photometric contamination at 5 arcsec separation. On the other hand, it would not be aggressive enough for sources with very bright neighbors, which can cause significant photometric contamination even at $>10$ arcsec separations. To account for the farther-reaching contamination from brighter neighbors, we calculate a modified contamination metric as follows.

For each source in {\it Gaia} eDR3, we consider all neighboring sources within 30~arcsec as potential contaminants. For each neighbor $i$, we calculate $\Delta G_{i} = G_{\rm source} - G_{i}$, where $G_i$ is the $G-$band magnitude of the neighbors. We then subtract from $\Delta G_i$ the angular separation, in arcsec, between the two sources: 

\begin{align}
    \label{eq:delta_Gi}
    \Delta\tilde{G}_{i}
    =
    \frac{\Delta G_{i}}{\mathrm{mag}}
    -
    \frac{\theta_{i}}{\mathrm{arcsec}}
    \, .
\end{align}
Finally, we record as \texttt{norm\_dG} the {\it largest} value of $\Delta\tilde{G}_{i}$ among all neighbors within 30~arcsec. Large values of \texttt{norm\_dG} correspond to sources with significant contamination. We find that most sources with \texttt{norm\_dG} $< -3$ (e.g, an equally bright neighbor more than 3 arcsec away, or a 2-magnitudes-brighter neighbor more than 5 arcsec away) have unproblematic colors. Sources with no neighbors within 30 arcsec are assigned \texttt{norm\_dG} = \texttt{nan}; these are also expected to be uncontaminated. The lower left panel of Fig.~\ref{fig:cmd_below_ms} shows the effects of applying a cut of \texttt{norm\_dG} $< -3$, and the lower right panel shows the sources that are filtered out by it. Most of the removed sources fall between the WD cooling tracks and the MS. There is no reason to expect e.g. cataclysmic variables to preferentially fall close to bright neighbors, so we suspect that these sources are mostly MS stars that have spurious colors or parallaxes.

It is not immediately obvious whether the spurious sources that fall between the MS and WD sequence and are removed by a cut on \texttt{norm\_dG} have spurious colors, spurious parallaxes, or both. To distinguish between these possibilities, we investigated the reported parallaxes of sources at high Galactic latitude that (a) are between the MS and WD cooling track, (b) have high astrometric fidelity but are removed by a cut on \texttt{norm\_dG} (e.g. bottom panel of Fig.~\ref{fig:cmd_below_ms}), and (c) have a bright companion within 5 arcsec. At high latitudes, most such pairs are gravitationally bound wide binaries \citep[e.g.][]{El-Badry2021c}, and thus the parallaxes of the two stars should be consistent. We found that the reported parallaxes are indeed consistent within $3\sigma$ in a large majority of pairs, even when one component falls far from the MS in the {\it Gaia} CMD. We conclude that for most sources with high reported astrometric fidelity but spurious CMD position, it is the colors, not the parallaxes, that are incorrect. 

In addition to the astrometric fidelity parameter, we report \texttt{norm\_dG} values for all sources in {\it Gaia} eDR3. We have found that \texttt{norm\_dG} is strongly correlated with the \texttt{phot\_bp\_rp\_excess\_factor} parameter or the more elaborate photometric excess measure, C* \citep[eq.6]{2021A&A...649A...3R}. However, because \texttt{norm\_dG} is a geometric quantity, cuts on it are more straightforward to account for in selection function modeling than cuts on \texttt{phot\_bp\_rp\_excess\_factor} and other image parameter diagnostics. In addition, \texttt{norm\_dG} can be calculated for all {\it Gaia} sources, while \texttt{phot\_bp\_rp\_excess\_factor} is only reported for sources with measured {\it Gaia} colors. We thus recommend filtering on \texttt{norm\_dG} as well as astrometric fidelity when a sample with reliable colors and parallaxes is required. Cuts on \texttt{norm\_dG} will of course remove sources with close neighbors, so cuts on this parameter will preferentially remove objects in wide binaries and dense regions of the sky. Cuts on the astrometric fidelity parameter likely also discriminate against dense regions of the sky, since information about the distance to bright neighbors is used in calculating it. 

\section{Access to the catalog}

\begin{figure}
\includegraphics[width=\linewidth]{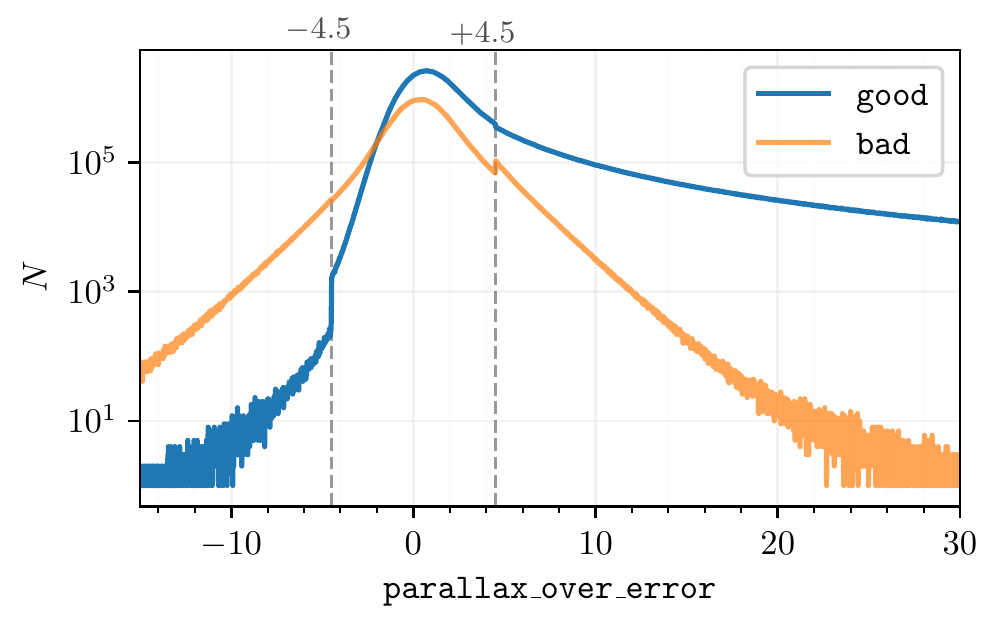}
	\caption{SNR (\texttt{parallax\_over\_error}) distribution for \texttt{good} and \texttt{bad} sources in Gaia eDR3. The break at $\left|\mathrm{SNR}\right| = 4.5$, where we switch between the low-SNR and the high-SNR classifier, is shown by dashed gray lines.}
	\label{fig:SNR_histogram}
\end{figure}

Our catalog is hosted at the German Astrophysical Virtual Observatory (GAVO),\footnote{\url{https://dc.zah.uni-heidelberg.de/}} in the table \texttt{gedr3spur.main}.\footnote{\url{https://dc.zah.uni-heidelberg.de/tableinfo/gedr3spur.main}} The simplest way to access the astrometric fidelities for a sample of stars is to crossmatch directly via a Table Access Protocol (TAP) upload join in TOPCAT.\footnote{The TOPCAT program is described at \url{http://www.star.bris.ac.uk/~mbt/topcat/}.} If the local table has Gaia eDR3 \texttt{source\_id}s, one can simply query:
\begin{lstlisting}
SELECT src.*
FROM gedr3spur.main as src
JOIN TAP_UPLOAD.t1 AS target
-- TAP_UPLOAD.tX needs to be the table number in TOPCAT
  USING (source_id)
\end{lstlisting}
Because there is a 100~MB upload limit, one can increase the number of sources queried at a time by hiding all columns except \texttt{source\_id}.
GAVO hosts a light version of Gaia eDR3, containing only the most commonly used columns. One can directly query this light version of Gaia eDR3 and simultaneously cross-match to our astrometric fidelities. For example, the following query returns a histogram of the parallax SNR distribution for the 1,111 million sources in eDR3 classified as \texttt{good}:
\begin{lstlisting}
SELECT COUNT(*) AS ct, 
ROUND(parallax/parallax_error,2) AS bin  
FROM gaia.edr3lite -- only contains most important rows 
JOIN gedr3spur.main using (source_id)  
WHERE fidelity_v2 >= 0.5  GROUP BY bin
\end{lstlisting}
Requiring $\mathtt{fidelity\_v2} < 0.5$ returns the SNR distribution of the 356 million \texttt{bad} sources. These are depicted in Fig.~\ref{fig:SNR_histogram}. We see a jump in the number of sources at the $\left| \mathrm{SNR} \right| = 4.5$ transition between our high- and low-SNR classifier. At the transition point, the number of \texttt{bad} sources increases by 65\,\%, i.e. the high-SNR classifier appears to obtain higher purity. This is unsurprising, as the high-SNR classifier takes $\left| \mathtt{parallax\_over\_error} \right|$ into account, which is one of the most informative features. Due to imbalances in the training data, the low-SNR classifier does not take this feature into account (see Section~\ref{sec:two-SNR-regimes}). \citet{EDR3CatalogueValidation} estimates the total contamination of the sources with SNR $>$ 5 to be of the order of $\sim$ 3 million (equal to the number of sources with SNR $<$ -5). Our classifier finds 11.3 million \texttt{bad} sources in this regime, which constitute 6\,\% of the 192 million sources with SNR $>$ 5.

\begin{figure}
\includegraphics[width=\linewidth]{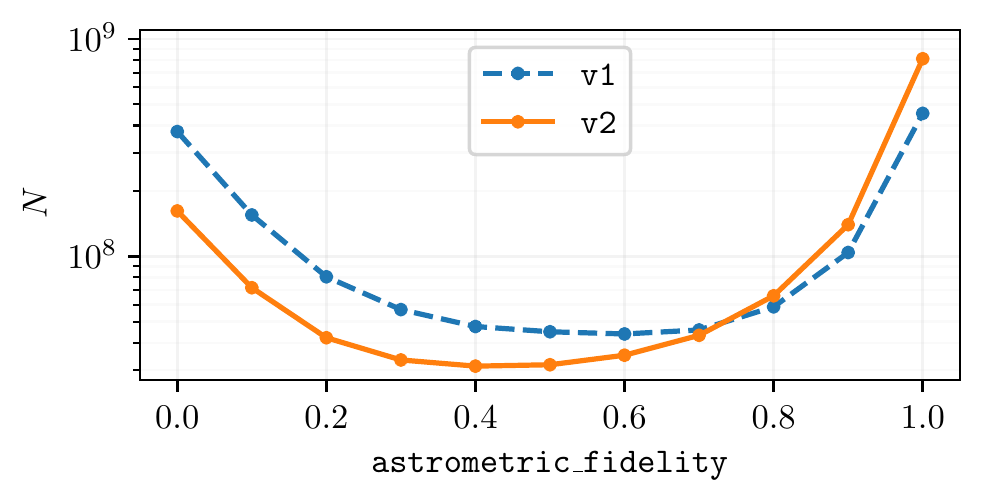}
	\caption{Distribution of astrometric fidelities for eDR3, in bins of 0.1, for both our previous (\texttt{v1}) and our updated (\texttt{v2}) classifier. The \texttt{v2} classifier marks fewer sources as \texttt{bad}, particularly in dense regions of the sky, due to our addition of \texttt{good} training datasets covering this regime. We expect our \texttt{v2} classifier to perform better (in particular, to achieve higher completeness) in dense regions of the sky.}
	\label{fig:fidelities}
\end{figure}

In Fig.~\ref{fig:fidelities}, we compare \texttt{v1} with \texttt{v2} fidelities on all\footnote{Due to an ingestion error, we missed nearest-neighbor columns of 18 sources. Due to these missing features, these sources did not get assigned final fidelity values.} eDR3 sources with astrometric solutions. We can see that many more sources are classified as \texttt{good} in \texttt{v2}, i.e. 1.1 billion compared to 0.7 billion in \texttt{v1}. This is presumably due to the new high-density training data.

\begin{figure*}
	\includegraphics[width=0.3\linewidth]{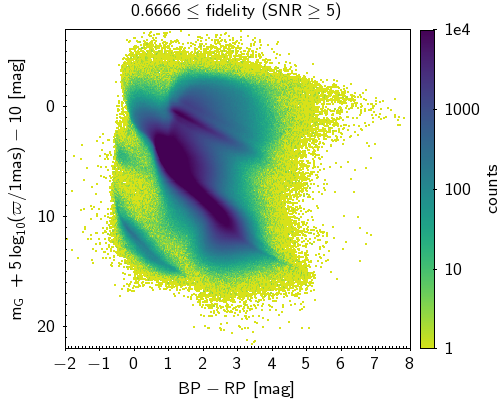}
	\hspace{0.02\linewidth}
	\includegraphics[width=0.3\linewidth]{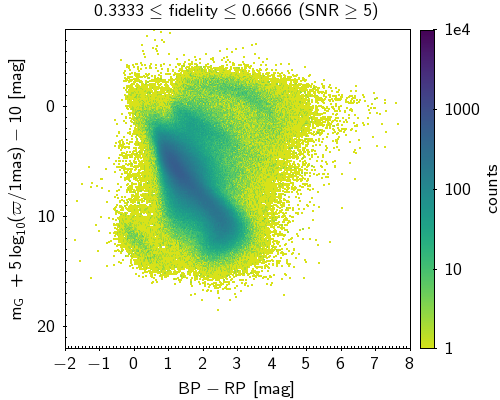}
	\hspace{0.02\linewidth}
	\includegraphics[width=0.3\linewidth]{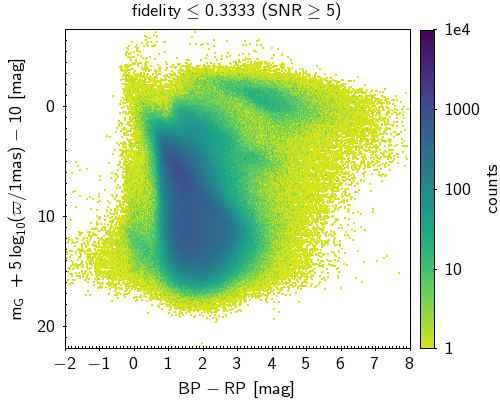}
	\includegraphics[width=0.3\linewidth]{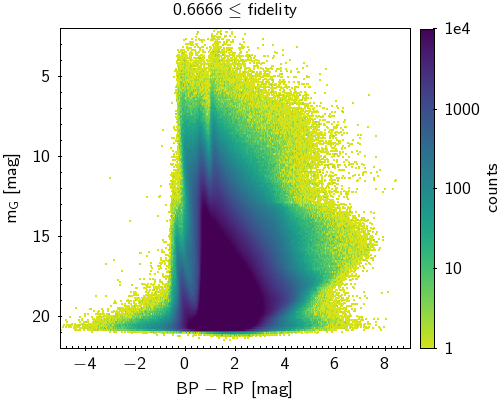}
	\hspace{0.02\linewidth}
	\includegraphics[width=0.3\linewidth]{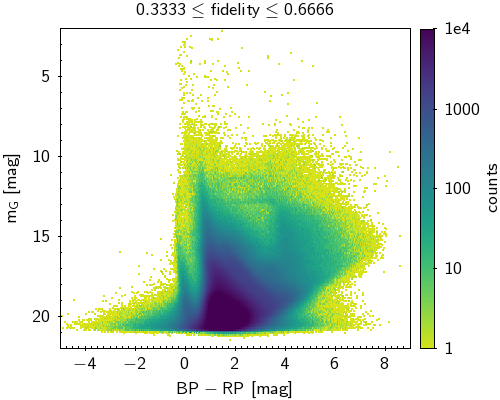}
	\hspace{0.02\linewidth}
	\includegraphics[width=0.3\linewidth]{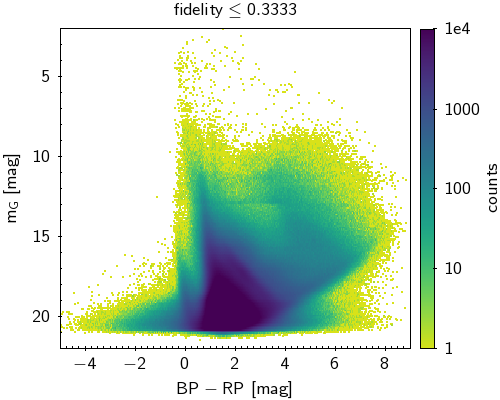}
	\caption{Upper panels: CAMD of SNR$\ge5$ sources split into \texttt{fidelity\_v2} bins, decreasing from left to right. All three panels share the same color bar. Lower panels: as upper panels, but for CMD and all sources.}
	\label{fig:camd_with_fidelity}
\end{figure*}

The upper panels of Fig.~\ref{fig:camd_with_fidelity} show the CAMD of eDR3 sources with $\mathrm{SNR} \ge 5$, while the lower panels show the CMD of all eDR3 sources, in three bins of decreasing fidelities, from left to right. For the $\mathrm{SNR} \ge 5$ sources in the upper panels, most sources have high fidelities (180 million have fidelity $\ge\nicefrac{2}{3}$, 10 million have fidelity $\le\nicefrac{1}{3}$, and 3 million lie in-between). For all eDR3 sources in the lower panel, we have 1,054 million with fidelity $\ge\nicefrac{2}{3}$, 304 million with fidelity $\le\nicefrac{1}{3}$, and 110 million in between. 

The CAMD in the upper left panel shows the richness of Gaia stellar populations visible at high parallax SNR. However, even at lower fidelities (middle and right panel), MS, WD and RC overdensities are visible, together with a fan at very bright and red sources, which might be indicative of good sources that are still misclassified by our model. In the very low fidelity regime (upper right panel), the space between MS and WD has many sources, presumably MS or turn-off stars scattered due to spurious astrometric solutions. It is important to remark, however, that we are only interested in finding spurious astrometric solutions, and not in finding incorrect CAMD positions, which can also result from spurious photometry. One class of sources with spurious photometry is apparent in the upper-left and upper-middle panels for the faint MS, where the $BP-RP$ color is underestimated due to a calibration error \citep{2021A&A...649A...3R}, such that these sources appear too blue. Another class of source with spurious colours arises due to contaminating flux from nearby sources, which can be filtered out using \texttt{phot\_bp\_rp\_excess\_factor} or our \texttt{norm\_dG} parameter, as described in Section~\ref{sec:cmd_fidelity}.

For the CMDs of all eDR3 sources in the lower panels, we see that bright sources are usually assigned high fidelities (left panel) whereas the lower fidelity sources (middle and right panel) are deficient in bright sources. However, at low fidelities we can see overdensities at the change of window functions from WC0b to WC1 at $m_G = 13\,\mathrm{mag}$ and at 11.5~mag which could correspond to the first gate of WC0b (i.e. gate 12, cf. Fig.~4 of \citealt{2020arXiv201203380L}). We also see that low-fidelity sources dominate at the red end.

\begin{figure}
	\includegraphics[width=\linewidth]{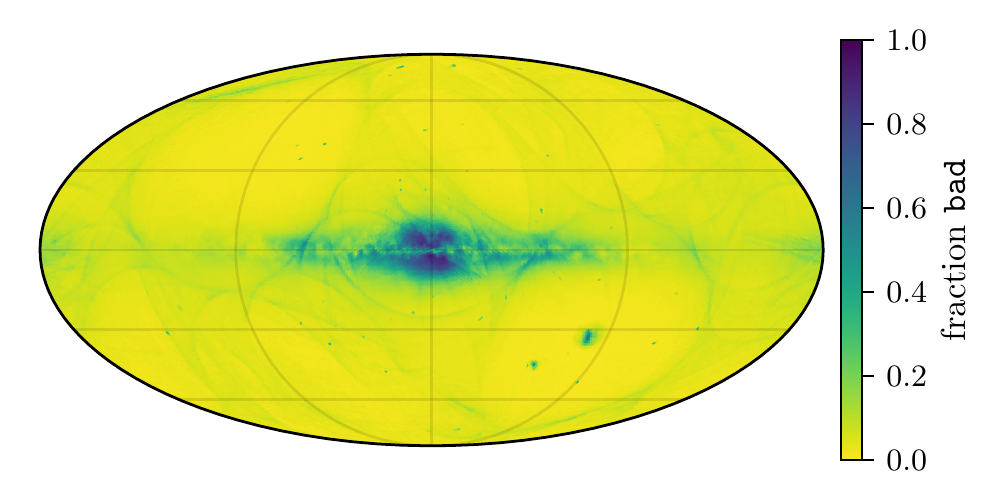}
	\caption{Fractions of \texttt{bad} sources with respect to all eDR3 sources with astrometric solutions over the sky, using a Mollweide projection of Galactic coordinates.}
	\label{fig:sky_fraction_bad}
\end{figure}

Finally, in Fig.~\ref{fig:sky_fraction_bad} we show the fraction of bad astrometric solutions, as classified by our \texttt{v2} model, compared to all astrometric solutions for eDR3. We can see that the densest areas are most vulnerable to spurious solutions, i.e., up to 90\,\% of sources in the bulge and LMC and can be \texttt{bad}. Dense star clusters in the field also have higher fractions of spurious astrometric solutions. The scanning law is clearly visible as well, and drives up the fraction of \texttt{bad} sources by a few percent.

\section{Conclusion}

We have extended the classification of valid and spurious astrometric solutions from the Gaia Catalogue of nearby stars \citep{EDR3GCNS} to all 1.47 billion sources in Gaia eDR3 with astrometry. Our training sample of spurious sources is obtained by taking all sources with $\mathtt{parallax\_over\_error} < -4.5$, as well as a set of low-SNR sources identified by the GCNS as \texttt{bad}. The bulk of our training sample of \texttt{good} astrometric solutions is obtained by taking all sources with a PS1 crossmatch in parts of the sky where no sources with $\mathtt{parallax\_over\_error} < -3.6$ exist. We additionally identify \texttt{good} sources that lie on the main sequence on a PS1 CAMD, as well as sources in globular clusters with measured parallaxes consistent with the cluster parallax \citep{VasilievBaumgardt2021-GCs}.

We train two neural network models: one for high-parallax-SNR and one for the low-parallax-SNR regime (divided at $\left| \mathrm{SNR} \right| = 4.5$), which take astrometric quality parameters from eDR3 and information about distances to nearby sources as inputs. 

Our validation shows that we outperform simple cuts but also logistic models that take into account a linear combination of our features as well as ExtraTrees. Across a wide range of parallaxes, sources that our classifier identifies as \texttt{good} are more plausibly distributed both on the sky and on a CAMD than sources that our classifier identifies as \texttt{bad}. Similarly, the spatial distribution (based on measured parallaxes) of OBA stars in the Milky Way midplane that our classifier marks as having \texttt{good} astrometric solutions is plausible, while the spatial distribution of stars with \texttt{bad} astrometric solutions is highly unphysical. The parallaxes of sources that our classifier identifies as \texttt{good} in the LMC and globular/open clusters are normally distributed with respect to their host parallaxes, with smaller scatter than sources that our classifier identifies as \texttt{bad}. Sources classified as \texttt{good} also have significantly lower $\chi^2$ when comparing to OGLE proper motions. Sources classified as \texttt{bad} usually occur in high-density regions, e.g. in the bulge and disc region and in the Magellanic clouds.

The astrometric fidelities presented in this paper can be used in place of simpler cuts on Gaia catalog columns, such as \texttt{ruwe} and \texttt{astrometric\_excess\_noise}, to filter out sources with spurious astrometric solutions. In most regimes, astrometric fidelities should yield a purer and more complete sample of sources with reliable astrometric solutions. In addition to the astrometric fidelities, we also provide simple diagnostics of the level of photometric contamination from bright neighbors (Section~\ref{sec:cmd_fidelity}), which can be used to filter out sources with unreliable colors. 

We provide the community with all the data and an online python notebook that can rerun the whole training and validation procedure automatically. This allows for retraining with additional sources that the user identified as \texttt{good} or \texttt{bad}, e.g. from \mbox{(in-)compatible} photometry and astrometry of dwarf galaxy sources. Worthwhile improvements could also come from training special purpose models, e.g., for dense regions, binaries or for sources that have all photometric bands available (i.e. BP, RP which would allow for additional features like \texttt{phot\_bp\_rp\_excess\_factor}). One could also re-run for GDR2 data and for future releases. Two more improvements that come to mind are taking the parallax zero point and parallax error scalings into account. Creating a selection function for the \texttt{good} sources would be a worthwhile project on its own.

\section*{Acknowledgements}

We thank the anonymous referee for swift, constructive and thorough feedback on the manuscript.
It is a pleasure to thank Claus Fabricius, Jordi Portell for help in understanding the astrometric solution. We thank Anthony Brown, Eugene Vasiliev, Zephyr Penoyre, Lennart Lindegren, Andy Everall, Adam Riess and Ron Drimmel for valuable feedback at the EDR3 workshop. We are also thankful for input and discussions with the MPIA Gaia and MW group. Furthermore, the authors would like to thank Douglas P. Finkbeiner and Joshua S. Speagle for helpful discussions and suggestions.

This work has made use of data from the European Space Agency (ESA) mission Gaia, processed by the Gaia Data Processing and Analysis Consortium (DPAC). Funding for the DPAC has been provided by national institutions, in particular the institutions participating in the Gaia Multilateral Agreement.

This research or product makes use of public auxiliary data provided by ESA/Gaia/DPAC as obtained from the publicly accessible ESA Gaia SFTP.

This work was funded by the DLR (German space agency) via grant 50\,QG\,1403.

GG acknowledges funding from the Alexander von Humboldt Foundation, through the Sofja Kovalevskaja Award.

The OGLE project has received funding from the National Science
Centre, Poland, grant MAESTRO 2014/14/A/ST9/00121 to AU.

JR will not travel anywhere by aeroplane for the purpose of promoting this paper.

Software: \texttt{TOPCAT} \citep{2005ASPC..347...29T}, \texttt{Tensorflow} \citep{Abadi2016Tensorflow}, \texttt{Keras} \citep{chollet2015keras}, \texttt{HEALpix} \citep{2005ApJ...622..759G}, \texttt{astropy} \citep{Astropy}, \texttt{matplotlib} \citep{Matplotlib}.

\subsection*{Data availability} 
The data underlying this article are available in the article and in its online supplementary material.


\bibliographystyle{mnras}
\bibliography{spurious} 

\begin{thebibliography}{}
\makeatletter
\relax
\def\mn@urlcharsother{\let\do\@makeother \do\$\do\&\do\#\do\^\do\_\do\%\do\~}
\def\mn@doi{\begingroup\mn@urlcharsother \@ifnextchar [ {\mn@doi@}
  {\mn@doi@[]}}
\def\mn@doi@[#1]#2{\def\@tempa{#1}\ifx\@tempa\@empty \href
  {http://dx.doi.org/#2} {doi:#2}\else \href {http://dx.doi.org/#2} {#1}\fi
  \endgroup}
\def\mn@eprint#1#2{\mn@eprint@#1:#2::\@nil}
\def\mn@eprint@arXiv#1{\href {http://arxiv.org/abs/#1} {{\tt arXiv:#1}}}
\def\mn@eprint@dblp#1{\href {http://dblp.uni-trier.de/rec/bibtex/#1.xml}
  {dblp:#1}}
\def\mn@eprint@#1:#2:#3:#4\@nil{\def\@tempa {#1}\def\@tempb {#2}\def\@tempc
  {#3}\ifx \@tempc \@empty \let \@tempc \@tempb \let \@tempb \@tempa \fi \ifx
  \@tempb \@empty \def\@tempb {arXiv}\fi \@ifundefined
  {mn@eprint@\@tempb}{\@tempb:\@tempc}{\expandafter \expandafter \csname
  mn@eprint@\@tempb\endcsname \expandafter{\@tempc}}}

\bibitem[\protect\citeauthoryear{Abadi et~al.,}{Abadi
  et~al.}{2016}]{Abadi2016Tensorflow}
Abadi M.,  et~al., 2016, TensorFlow: Large-Scale Machine Learning on
  Heterogeneous Distributed Systems (\mn@eprint {arXiv} {1603.04467})

\bibitem[\protect\citeauthoryear{{Abrahams}, {Bloom}, {Mowlavi}, {Szkody},
  {Rix}, {Ventura}, {Brink}  \& {Filippenko}}{{Abrahams}
  et~al.}{2020}]{Abrahams2020}
{Abrahams} E.~S.,  {Bloom} J.~S.,  {Mowlavi} N.,  {Szkody} P.,  {Rix} H.-W.,
  {Ventura} J.-P.,  {Brink} T.~G.,   {Filippenko} A.~V.,  2020, arXiv e-prints,
  \href {https://ui.adsabs.harvard.edu/abs/2020arXiv201112253A} {p.
  arXiv:2011.12253}

\bibitem[\protect\citeauthoryear{{Anders} et~al.,}{{Anders}
  et~al.}{2019}]{2019A&A...628A..94A}
{Anders} F.,  et~al., 2019, \mn@doi [\aap] {10.1051/0004-6361/201935765}, \href
  {https://ui.adsabs.harvard.edu/abs/2019A&A...628A..94A} {628, A94}

\bibitem[\protect\citeauthoryear{{Bailer-Jones}}{{Bailer-Jones}}{2015}]{2015PASP..127..994B}
{Bailer-Jones} C. A.~L.,  2015, \mn@doi [\pasp] {10.1086/683116}, \href
  {https://ui.adsabs.harvard.edu/abs/2015PASP..127..994B} {127, 994}

\bibitem[\protect\citeauthoryear{{Bailer-Jones}, {Rybizki}, {Fouesneau},
  {Mantelet}  \& {Andrae}}{{Bailer-Jones} et~al.}{2018}]{2018AJ....156...58B}
{Bailer-Jones} C.~A.~L.,  {Rybizki} J.,  {Fouesneau} M.,  {Mantelet} G.,
  {Andrae} R.,  2018, \mn@doi [\aj] {10.3847/1538-3881/aacb21}, \href
  {https://ui.adsabs.harvard.edu/abs/2018AJ....156...58B} {156, 58}

\bibitem[\protect\citeauthoryear{{Chambers} et~al.,}{{Chambers}
  et~al.}{2016}]{2016arXiv161205560C}
{Chambers} K.~C.,  et~al., 2016, arXiv e-prints, \href
  {https://ui.adsabs.harvard.edu/abs/2016arXiv161205560C} {p. arXiv:1612.05560}

\bibitem[\protect\citeauthoryear{Chollet et~al.}{Chollet
  et~al.}{2015}]{chollet2015keras}
Chollet F.,  et~al., 2015, Keras, \url{https://keras.io}

\bibitem[\protect\citeauthoryear{{El-Badry}, {Rix}, {Quataert}, {Kupfer}  \&
  {Shen}}{{El-Badry} et~al.}{2021a}]{El-Badry2021b}
{El-Badry} K.,  {Rix} H.-W.,  {Quataert} E.,  {Kupfer} T.,   {Shen} K.~J.,
  2021a, arXiv e-prints, \href
  {https://ui.adsabs.harvard.edu/abs/2021arXiv210804255E} {p. arXiv:2108.04255}

\bibitem[\protect\citeauthoryear{{El-Badry} et~al.,}{{El-Badry}
  et~al.}{2021b}]{El-Badry2021}
{El-Badry} K.,  et~al., 2021b, \mn@doi [\mnras] {10.1093/mnras/stab1318}, \href
  {https://ui.adsabs.harvard.edu/abs/2021MNRAS.505.2051E} {505, 2051}

\bibitem[\protect\citeauthoryear{{El-Badry}, {Rix}  \& {Heintz}}{{El-Badry}
  et~al.}{2021c}]{El-Badry2021c}
{El-Badry} K.,  {Rix} H.-W.,   {Heintz} T.~M.,  2021c, \mn@doi [\mnras]
  {10.1093/mnras/stab323}, \href
  {https://ui.adsabs.harvard.edu/abs/2021MNRAS.506.2269E} {506, 2269}

\bibitem[\protect\citeauthoryear{{Fabricius} et~al.,}{{Fabricius}
  et~al.}{2021}]{EDR3CatalogueValidation}
{Fabricius} C.,  et~al., 2021, \mn@doi [\aap] {10.1051/0004-6361/202039834},
  \href {https://ui.adsabs.harvard.edu/abs/2021A&A...649A...5F} {649, A5}

\bibitem[\protect\citeauthoryear{{Gaia Collaboration} et~al.,}{{Gaia
  Collaboration} et~al.}{2018}]{2018A&A...616A..10G}
{Gaia Collaboration} et~al., 2018, \mn@doi [\aap]
  {10.1051/0004-6361/201832843}, \href
  {https://ui.adsabs.harvard.edu/abs/2018A&A...616A..10G} {616, A10}

\bibitem[\protect\citeauthoryear{{Gaia Collaboration} et~al.,}{{Gaia
  Collaboration} et~al.}{2021a}]{EDR3Summary}
{Gaia Collaboration} et~al., 2021a, \mn@doi [\aap]
  {10.1051/0004-6361/202039657}, \href
  {https://ui.adsabs.harvard.edu/abs/2021A&A...649A...1G} {649, A1}

\bibitem[\protect\citeauthoryear{{Gaia Collaboration} et~al.,}{{Gaia
  Collaboration} et~al.}{2021b}]{EDR3GCNS}
{Gaia Collaboration} et~al., 2021b, \mn@doi [\aap]
  {10.1051/0004-6361/202039498}, \href
  {https://ui.adsabs.harvard.edu/abs/2021A&A...649A...6G} {649, A6}

\bibitem[\protect\citeauthoryear{{Gentile Fusillo} et~al.,}{{Gentile Fusillo}
  et~al.}{2021}]{2021arXiv210607669G}
{Gentile Fusillo} N.~P.,  et~al., 2021, arXiv e-prints, \href
  {https://ui.adsabs.harvard.edu/abs/2021arXiv210607669G} {p. arXiv:2106.07669}

\bibitem[\protect\citeauthoryear{Geurts, Ernst  \& Wehenkel}{Geurts
  et~al.}{2006}]{10.1007/s10994-006-6226-1}
Geurts P.,  Ernst D.,   Wehenkel L.,  2006, \mn@doi [Mach. Learn.]
  {10.1007/s10994-006-6226-1}, 63, 3–42

\bibitem[\protect\citeauthoryear{Goodfellow, Bengio  \& Courville}{Goodfellow
  et~al.}{2016}]{Goodfellow-et-al-2016}
Goodfellow I.,  Bengio Y.,   Courville A.,  2016, Deep Learning.
MIT Press, \url {http://www.deeplearningbook.org}

\bibitem[\protect\citeauthoryear{{G{\'o}rski}, {Hivon}, {Banday}, {Wandelt},
  {Hansen}, {Reinecke}  \& {Bartelmann}}{{G{\'o}rski}
  et~al.}{2005}]{2005ApJ...622..759G}
{G{\'o}rski} K.~M.,  {Hivon} E.,  {Banday} A.~J.,  {Wandelt} B.~D.,  {Hansen}
  F.~K.,  {Reinecke} M.,   {Bartelmann} M.,  2005, \mn@doi [\apj]
  {10.1086/427976}, \href {http://adsabs.harvard.edu/abs/2005ApJ...622..759G}
  {622, 759}

\bibitem[\protect\citeauthoryear{{Green}}{{Green}}{2018}]{Green2018dustmaps}
{Green} G.,  2018, \mn@doi [The Journal of Open Source Software]
  {10.21105/joss.00695}, \href
  {https://ui.adsabs.harvard.edu/abs/2018JOSS....3..695G} {3, 695}

\bibitem[\protect\citeauthoryear{{Green}, {Schlafly}, {Zucker}, {Speagle}  \&
  {Finkbeiner}}{{Green} et~al.}{2019}]{Green2019-Bayestar19}
{Green} G.~M.,  {Schlafly} E.,  {Zucker} C.,  {Speagle} J.~S.,   {Finkbeiner}
  D.,  2019, \mn@doi [\apj] {10.3847/1538-4357/ab5362}, \href
  {https://ui.adsabs.harvard.edu/abs/2019ApJ...887...93G} {887, 93}

\bibitem[\protect\citeauthoryear{Hunter}{Hunter}{2007}]{Matplotlib}
Hunter J.~D.,  2007, \mn@doi [Computing in Science \& Engineering]
  {10.1109/MCSE.2007.55}, 9, 90

\bibitem[\protect\citeauthoryear{{Kingma} \& {Ba}}{{Kingma} \&
  {Ba}}{2014}]{Kingma2014}
{Kingma} D.~P.,  {Ba} J.,  2014, arXiv e-prints, \href
  {https://ui.adsabs.harvard.edu/abs/2014arXiv1412.6980K} {p. arXiv:1412.6980}

\bibitem[\protect\citeauthoryear{{Lindegren} et~al.,}{{Lindegren}
  et~al.}{2021a}]{2020arXiv201203380L}
{Lindegren} L.,  et~al., 2021a, \mn@doi [\aap] {10.1051/0004-6361/202039709},
  \href {https://ui.adsabs.harvard.edu/abs/2021A&A...649A...2L} {649, A2}

\bibitem[\protect\citeauthoryear{{Lindegren} et~al.,}{{Lindegren}
  et~al.}{2021b}]{2020arXiv201201742L}
{Lindegren} L.,  et~al., 2021b, \mn@doi [\aap] {10.1051/0004-6361/202039653},
  \href {https://ui.adsabs.harvard.edu/abs/2021A&A...649A...4L} {649, A4}

\bibitem[\protect\citeauthoryear{{Luri} et~al.,}{{Luri}
  et~al.}{2018}]{2018A&A...616A...9L}
{Luri} X.,  et~al., 2018, \mn@doi [\aap] {10.1051/0004-6361/201832964}, \href
  {https://ui.adsabs.harvard.edu/abs/2018A&A...616A...9L} {616, A9}

\bibitem[\protect\citeauthoryear{{Pelisoli} \& {Vos}}{{Pelisoli} \&
  {Vos}}{2019}]{Pelisoli2019}
{Pelisoli} I.,  {Vos} J.,  2019, \mn@doi [\mnras] {10.1093/mnras/stz1876},
  \href {https://ui.adsabs.harvard.edu/abs/2019MNRAS.488.2892P} {488, 2892}

\bibitem[\protect\citeauthoryear{{Rebassa-Mansergas}
  et~al.,}{{Rebassa-Mansergas} et~al.}{2021}]{Rebassa-Mansergas2021}
{Rebassa-Mansergas} A.,  et~al., 2021, \mn@doi [\mnras]
  {10.1093/mnras/stab2039}, \href
  {https://ui.adsabs.harvard.edu/abs/2021MNRAS.506.5201R} {506, 5201}

\bibitem[\protect\citeauthoryear{{Riello} et~al.,}{{Riello}
  et~al.}{2021}]{2021A&A...649A...3R}
{Riello} M.,  et~al., 2021, \mn@doi [\aap] {10.1051/0004-6361/202039587}, \href
  {https://ui.adsabs.harvard.edu/abs/2021A&A...649A...3R} {649, A3}

\bibitem[\protect\citeauthoryear{{Rybizki} et~al.,}{{Rybizki}
  et~al.}{2020}]{Rybizki2020-GeDR3mock}
{Rybizki} J.,  et~al., 2020, \mn@doi [\pasp] {10.1088/1538-3873/ab8cb0}, \href
  {https://ui.adsabs.harvard.edu/abs/2020PASP..132g4501R} {132, 074501}

\bibitem[\protect\citeauthoryear{{Schlegel}, {Finkbeiner}  \&
  {Davis}}{{Schlegel} et~al.}{1998}]{SFD1998}
{Schlegel} D.~J.,  {Finkbeiner} D.~P.,   {Davis} M.,  1998, \mn@doi [\apj]
  {10.1086/305772}, \href
  {https://ui.adsabs.harvard.edu/abs/1998ApJ...500..525S} {500, 525}

\bibitem[\protect\citeauthoryear{{Taylor}}{{Taylor}}{2005}]{2005ASPC..347...29T}
{Taylor} M.~B.,  2005, in {Shopbell} P.,  {Britton} M.,   {Ebert} R.,  eds,
  Astronomical Society of the Pacific Conference Series Vol. 347, Astronomical
  Data Analysis Software and Systems XIV. p.~29

\bibitem[\protect\citeauthoryear{{The Astropy Collaboration} et~al.,}{{The
  Astropy Collaboration} et~al.}{2018}]{Astropy}
{The Astropy Collaboration} et~al., 2018, \mn@doi [\aj]
  {10.3847/1538-3881/aabc4f}, \href
  {https://ui.adsabs.harvard.edu/abs/2018AJ....156..123T} {156, 123}

\bibitem[\protect\citeauthoryear{{Torra} et~al.,}{{Torra}
  et~al.}{2021}]{2020arXiv201206420T}
{Torra} F.,  et~al., 2021, \mn@doi [\aap] {10.1051/0004-6361/202039637}, \href
  {https://ui.adsabs.harvard.edu/abs/2021A&A...649A..10T} {649, A10}

\bibitem[\protect\citeauthoryear{{Udalski}, {Szyma{\'n}ski}  \&
  {Szyma{\'n}ski}}{{Udalski} et~al.}{2015}]{OGLE-IV}
{Udalski} A.,  {Szyma{\'n}ski} M.~K.,   {Szyma{\'n}ski} G.,  2015, \actaa,
  \href {https://ui.adsabs.harvard.edu/abs/2015AcA....65....1U} {65, 1}

\bibitem[\protect\citeauthoryear{{Vasiliev} \& {Baumgardt}}{{Vasiliev} \&
  {Baumgardt}}{2021}]{VasilievBaumgardt2021-GCs}
{Vasiliev} E.,  {Baumgardt} H.,  2021, \mn@doi [\mnras]
  {10.1093/mnras/stab1475}, \href
  {https://ui.adsabs.harvard.edu/abs/2021MNRAS.505.5978V} {505, 5978}

\bibitem[\protect\citeauthoryear{{Zari}, {Rix}, {Frankel}, {Xiang}, {Poggio},
  {Drimmel}  \& {Tkachenko}}{{Zari} et~al.}{2021}]{Zari2021-hot-stars}
{Zari} E.,  {Rix} H.~W.,  {Frankel} N.,  {Xiang} M.,  {Poggio} E.,  {Drimmel}
  R.,   {Tkachenko} A.,  2021, \mn@doi [\aap] {10.1051/0004-6361/202039726},
  \href {https://ui.adsabs.harvard.edu/abs/2021A&A...650A.112Z} {650, A112}

\bibitem[\protect\citeauthoryear{{Zinn}}{{Zinn}}{2021}]{2021AJ....161..214Z}
{Zinn} J.~C.,  2021, \mn@doi [\aj] {10.3847/1538-3881/abe936}, \href
  {https://ui.adsabs.harvard.edu/abs/2021AJ....161..214Z} {161, 214}

\makeatother
\end{thebibliography}




\appendix

\section{features}
\begin{figure*}
	\includegraphics[width=\linewidth]{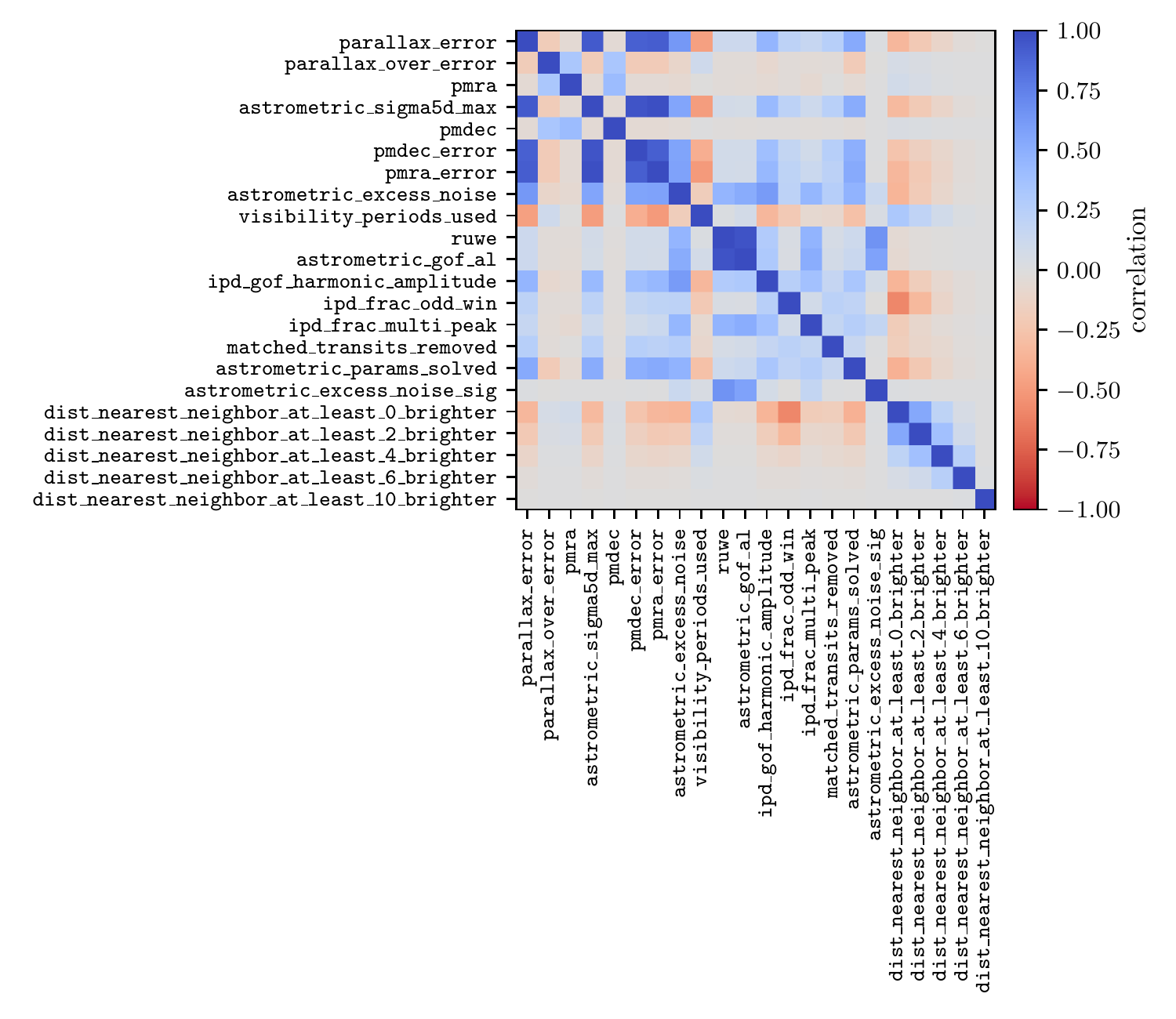}
	\caption{Spearman's cross correlation matrix of the features in the training dataset. Positive correlations are shown in blue and anti-correlations in red. Gray indicates no correlation.}
	\label{fig:feature_correlation}
\end{figure*}

In order to provide a bit more insight into the features used, Fig.~\ref{fig:feature_correlation} shows their correlation matrix in the training dataset.

Below, we also show the ExtraTree importance of each feature (averaged over 3 random determinations and normalised to the most important). These importance estimates do not necessarily carry over to the neural network model, and are only shown in order to give a rough idea of the discriminatory power of each feature:\\
1.000 astrometric\_excess\_noise\\
0.743 dist\_nearest\_neighbor\_at\_least\_0\_brighter\\
0.367 parallax\_over\_error\\
0.339 ipd\_frac\_multi\_peak\\
0.323 astrometric\_sigma5d\_max\\
0.274 ruwe\\
0.250 astrometric\_params\_solved\\
0.213 astrometric\_gof\_al\\
0.200 visibility\_periods\_used\\
0.139 parallax\_error\\
0.136 ipd\_gof\_harmonic\_amplitude\\
0.082 pmdec\_error\\
0.060 pmra\_error\\
0.060 astrometric\_excess\_noise\_sig\\
0.048 pmra\\
0.048 pmdec\\
0.038 dist\_nearest\_neighbor\_at\_least\_2\_brighter\\
0.020 ipd\_frac\_odd\_win\\
0.007 matched\_transits\_removed\\
0.002 dist\_nearest\_neighbor\_at\_least\_4\_brighter\\
0.000 dist\_nearest\_neighbor\_at\_least\_6\_brighter\\
0.000 dist\_nearest\_neighbor\_at\_least\_10\_brighter\\


\section{GeDR3mock}
\label{sec:gedr3mock_query}
In order to compare the distance slices with expected distributions in the CMD and sky we query the GeDR3mock \citep{Rybizki2020-GeDR3mock} using the following query:
\begin{lstlisting}
SELECT g.* 
FROM (
    SELECT source_id,
    parallax_obs/parallax_error_scaled as parallax_over_error,    
    GAVO_NORMAL_RANDOM( phot_g_mean_mag,
      phot_g_mean_mag_error) AS phot_g_mean_mag,
    GAVO_NORMAL_RANDOM( phot_rp_mean_mag,
      phot_rp_mean_mag_error) AS phot_rp_mean_mag,
    GAVO_NORMAL_RANDOM( phot_bp_mean_mag,
      phot_bp_mean_mag_error) AS phot_bp_mean_mag,
-- Using pseudo-observed instead of true photometry
    parallax_error_scaled, parallax_obs as parallax
    FROM (
        SELECT y.*,
        GAVO_NORMAL_RANDOM( parallax,
        parallax_error_scaled) AS parallax_obs
-- Pseudo-observed parallaxes by sampling the error
        FROM (
            SELECT z.*, 
  POWER(10, ((LOG(parallax_error)/LOG(10)+1)*1.3)-1) AS       parallax_error_scaled
-- Rescaling the error according to the GCNS paper
            FROM gedr3mock.main as z
-- WHERE source_id BETWEEN 6001046503471185920 AND 6001609453424607231
-- uncomment above line in order to run with synchronous mode otherwise takes 1-2 hours
            ) as y 
        ) as x
    WHERE
    (x.parallax_obs > 8) OR
    (x.parallax_obs > 1
    AND x.parallax_obs < 1.01) OR
    (x.parallax_obs > 3.3
    AND x.parallax_obs < 3.4) OR
    (x.parallax_obs > 0.1
    AND x.parallax_obs < 0.101) OR
    (x.parallax_obs > 0.333
    AND x.parallax_obs < 0.334) OR
    (x.parallax_obs > 0.0325
    AND x.parallax_obs < 0.034)
-- specifying the different parallax slices from our validation set
    ) as g
JOIN gcns.maglims6 AS lim
ON (g.source_id/140737488355328=lim.hpx)
WHERE phot_g_mean_mag<lim.magnitude_90
-- dropping sources in gedr3mock that are below Gaia's magnitude limit
\end{lstlisting}
This yields pseudo-observed quantities as the photometry and parallaxes are sampled with their respective uncertainties. We also apply the GeDR3 G magnitude limits per level 6 HEALpix which we approximate with the 90th percentile of the GeDR3 G magnitude distribution per HEALpix for sources which have a parallax measurement \citep{EDR3GCNS}.
\\
\bsp	
\label{lastpage}
\end{document}